\documentclass[fleqn,usenatbib]{mnras}

\usepackage{newtxtext,newtxmath}

\usepackage[T1]{fontenc}

\DeclareRobustCommand{\VAN}[3]{#2}
\let\VANthebibliography\thebibliography
\def\thebibliography{\DeclareRobustCommand{\VAN}[3]{##3}\VANthebibliography}

\usepackage{graphicx}	
\usepackage{amsmath}	

\usepackage{amssymb}	

\usepackage{float}
\usepackage{mathtools}
\usepackage{enumitem}
\usepackage{bm}
\usepackage{hyperref}
\usepackage{natbib}
\usepackage{makecell,multirow,diagbox}
\usepackage{array}
\usepackage{booktabs}

\title[HSC-SSP tomographic peak analyses]{Cosmological Studies from HSC-SSP Tomographic Weak Lensing Peak Abundances}

\author[Liu et al.]{
Xiangkun Liu,$^{1}$\thanks{E-mail: liuxk@ynu.edu.cn}
Shuo Yuan,$^{2}$
Chuzhong Pan,$^{3}$
Tianyu Zhang,$^{1}$
Qiao Wang$^{2}$
and Zuhui Fan$^{1}$\thanks{E-mail: zuhuifan@ynu.edu.cn}
\\
$^{1}$South-Western Institute for Astronomy Research, Yunnan University, Kunming 650500, People's Republic of China\\
$^{2}$National Astronomical Observatories, Chinese Academy of Sciences, Beijing 100101, People's Republic of China\\
$^{3}$Department of Astronomy, School of Physics, Peking University, Beijing 100871, People's Republic of China
}

\date{Accepted XXX. Received YYY; in original form ZZZ}

\pubyear{2022}

\begin{document}
\label{firstpage}
\pagerange{\pageref{firstpage}--\pageref{lastpage}}
\maketitle

\begin{abstract}
We perform weak lensing tomographic peak studies using the first-year shear data from Hyper Suprime-Cam Subaru Strategic Program (HSC-SSP) survey. The effective area used in our analyses after field selection, mask and boundary exclusions is $\sim 58 \deg^2$. 
The source galaxies are divided into low- and high-redshift bins with $0.2 \leq z_{\rm p} \leq 0.85$ and $0.85 \leq z_{\rm p} \leq 1.5$, respectively. We utilize our halo-based theoretical peak model including the projection effect of large-scale structures to derive cosmological constraints from the observed tomographic high peak abundances with the signal-to-noise ratio in the range of $\nu_\mathrm{N}=[3.5,5.5]$. These high peaks are closely associated with the lensing effects of massive clusters of galaxies. Thus the inclusion of their member galaxies in 
the shear catalog can lead to significant source clustering and dilute their lensing signals. We account for this systematic effect in our theoretical modelling. Additionally, the impacts of baryonic effects, galaxy intrinsic alignments, 
as well as residual uncertainties in shear and photometric redshift calibrations are also analyzed. Within the flat $\Lambda$CDM model, the derived constraint is $S_8=\sigma_8(\Omega_{\rm m}/0.3)^{0.5} =0.758_{-0.076}^{+0.033}$ and $0.768_{-0.057}^{+0.030}$ with the source clustering information measured from the 
two cluster catalogs, CAMIRA and WZL, respectively. The asymmetric uncertainties are due to the different degeneracy direction of $(\Omega_{\rm m}, \sigma_8)$ from high peak abundances comparing to that from the cosmic shear two-point correlations which give rise approximately 
the power index $\alpha=0.5$. Fitting to our constraints, we obtain $\alpha\approx 0.38$ and $\Sigma_8=\sigma_8(\Omega_{\rm m}/0.3)^{\alpha}=0.772_{-0.032}^{+0.028}$ (CAMIRA) and $0.781_{-0.033}^{+0.028}$ (WZL). In comparison with the results from non-tomographic peak analyses, the $1\sigma$ uncertainties on $\Sigma_8$ 
are reduced by a factor of $\sim 1.3$.  
\end{abstract}




\section{Introduction}

\label{sec:intro}

Weak gravitational lensing (WL) effects, small observed image distortions and flux changes of background galaxies arising from the gravitational light deflection by large-scale structures, are unique means in probing the matter distribution in the Universe and thus uncovering the nature of dark matter and dark energy \citep{BS2001, HJ2008, FuFan2014, Kilbinger2015}. With the great efforts devoted to and the advance in accurate shape measurements for far-way faint galaxies \citep{KSB1995, BJ2002, Massey2005, Miller2007, Kitching2008, Zhang2008, Zhang2011, Tewes2012, Miller2013, Zuntz2013, Hoekstra2015, Mandelbaum2015, Zhang2015, FC2017, Sheldon2017}, WL surveys with ever growthing data quality and quantity, such as Canada-France Hawaii Telescope Lensing Survey \citep[CFHTLenS\footnote{\url{http://cfhtlens.org/}}]{Heymans2012}, the Dark Energy Survey \citep[DES\footnote{\url{http://www.darkenergysurvey.org/}}]{DES2016}, 
the Kilo-Degree survey \citep[KiDS\footnote{\url{http://kids.strw.leidenuniv.nl/}}]{deJong2013}, and the Hyper Suprime-Cam Subaru Strategic Program \citep[hereafter the HSC-SSP\footnote{\url{http://hsc.mtk.nao.ac.jp/ssp/}} survey]{Aihara2018}, have resulted in cosmological constraints with 
continuously improved precisions \citep{Kilbinger2013, Hildebrandt2017, Hikage2019, Asgari2021, Amon2022}. Furthermore, the next generation of large surveys, both ground-based and space-based ones, are already around the corner, and will further lead to an order of magnitude increase of data volume, including Vera C. Rubin Observatory Legacy Survey of Space and Time \citep[LSST\footnote{\url{http://www.lsst.org/}}]{LSST2019}, {\it Euclid}\footnote{\url{http://sci.esa.int/euclid/}} \citep{Laureijs2011}, Roman Space Telescope \citep[Roman\footnote{\url{https://roman.gsfc.nasa.gov}}]{Roman2018} and the Chinese Space Station Survey Space Telescope \citep[CSST\footnote{\url{http://nao.cas.cn/csst/}}]{Gong2019}.

Weak lensing cosmological studies rely on statistical analyses from a large sample of galaxies with accurate shear measurements. The cosmic shear two-point correlation function (2PCF) or the power spectrum is the second order statistics, and tomographic 2PCFs (power spectra) have been widely applied to derive cosmological constraints from different WL surveys. To further extract cosmological information from nonlinear structures, higher-order analyses have been performed \citep{Semboloni2011, Fu2014}. Recently, weak lensing peak statistics have been utilized in WL studies as an important complement to the 2PCFs \citep{KS1999, vanWaerbeke2000, DH2010, Hamana2004, Fan2010, Shan2012, Shan2014, Hamana2015, LK2015, LiuJ2015, LiuX2015, Liu2016, Kacprzak2016, Shan2018, Martinet2018, Zrcher2022}.  

Similarly to cosmic shear correlations, theoretical studies have shown that adding redshift information in peak analyses, i.e., tomographic peak statistics, can effectively enhance the cosmological gains in comparison with that from non-tomographic, or 2D peak statistics \citep{Yang2011,Martinet2015,Petri2016,Abruzzo2019,Yuan2019}.
Limited by the number density of galaxies in the shear samples from the current surveys, however, most of the observational peak studies to date are 2D without tomography. Very recently, \citet{Zrcher2022} carry out tomographic peak analyses using the first three years of data of DES \citep{DES2021}. 
They construct smoothed mass maps from the shear data in different redshift bins with several smoothing scales. Tomographic peak counts including the cross-peaks \citep {Martinet2021a}, i.e., peaks detected from maps generated by taking a harmonic-space product of the convergence of two different redshift bins, 
are then used for cosmological constraints employing mock templates generated by an emulator trained on a suite of PKDGRAV3 N-Body simulations \citep{Potter2017}. To avoid the bias resulting from the strong source clustering effects on high peaks for which their mocks cannot model well, they limit their analyses to peaks with the signal-to-noise ratio SNR$\le 4$.

In this study, we perform tomographic peak studies using the first-year shear sample from the Hyper Suprime-Cam Subaru Strategic Program (HSC-SSP) survey \citep{Mandelbaum2018a}. The sample is relatively deep with the effective number density of galaxies about $20\hbox{ arcmin}^{-2}$, and thus very suitable for tomographic peak statistics. We particularly concentrate on high peaks that are physically originated from massive halos along lines of sight. For cosmological constraints, we adopt our halo-based theoretical peak model that includes the impacts of the shape noise from the intrinsic ellipticities of galaxies and the projection effect of large-scale structures \citep{Fan2010, Yuan2018}. 
Comparing to the numerical-template approach, the theoretical modelling analyses can provide good insights of the physical information embedded in WL high peaks, and allow us to extract cosmological constraints from observed peak counts efficiently without the need of a large number of simulations demanding a careful design to 
sample fairly the multi-dimensional parameter space. On the other hand, a theoretical model always contains assumptions and approximations. We validate our model applicability using mock simulations before applying it to observational analyses. To mitigate the source clustering effect on high peaks, we perform careful analyses with the cluster catalogs identified in the same sky area \citep{Oguri2018Cluster, Wen2021} and take it into account in our modelling.  

In \citet{Oguri2021}, they construct shear-selected cluster samples using the latest HSC-SSP S19A shear catalog \citep{Li2021} and show that most of the high peaks indeed correspond well with clusters of galaxies. The physical assumption in our halo-based theoretical peak model is well in line with this observational result. On the other hand, in our cosmological analyses, instead of trying to use true cluster samples, we adopt the forward-modelling approach to include different effects into the theoretical model. This enables us to derive cosmological constraints directly from observed peaks without the need for cleaning up contaminations to the underlying cluster samples. 

This paper is structured as follows. In Section \ref{sec:data}, we describe the HSC-SSP data set used in our study. Section \ref{sec:WLpeak} contains the details of our WL tomographic peak analyses. Potential impacts of different systematic effects are discussed in Section \ref{sec:systematics}. 
In Section \ref{sec:constraints}, we show the cosmological constraints obtained from the tomographic peak counts. Summary and discussion are given in Section \ref{sec:summary}.

\section{The HSC-SSP shear sample} \label{sec:data}
HSC-SSP is a large wide-field multi-band imaging survey aiming to address a broad range of scientific questions from cosmology to solar bodies \citep{Aihara2018}. The weak lensing study from the Wide-layer observations with the targeted sky coverage of $\sim 1400\deg^2$ is one of the key objectives of the survey. 
In the latest third data release of HSC-SSP, the Wide-layer data reach the full depth of $\sim 26$ mag at $5\sigma$ in all the five bands ($grizy$) for about $670\deg^2$ \citep{Aihara2021}. The three-year shear catalog has also been completed \citep{Li2021}, which is however not publically released yet.  

\begin{figure}
\centering
\includegraphics[width=1.0\columnwidth]{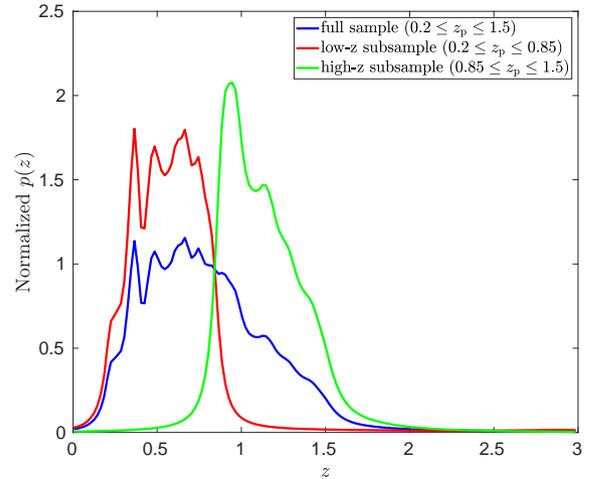}
\caption{Normalized redshift distributions with the blue, red and green lines for the full, low-z and high-z shear samples, respectively.}
\label{fig:pz_plot}
\end{figure}

Therefore in our analyses here, we use the first-year HSC-SSP shear catalog (S16A) \citep{Mandelbaum2018a}. The shapes of galaxies are measured on the $i$-band coadded images using the re-Gaussianization PSF correction method \citep{HS2003}. Only galaxies that pass the selection criteria listed in Table 4 of \citet{Mandelbaum2018a} are remained. The photometric redshift (photo-z) for each galaxy is determined based on the HSC five-band photometry \citep{Tanaka2018}, and it is proved to be most accurate with the scatter $\sigma[\Delta z_{\rm p}/(1+z_{\rm p})]\sim 0.05$ and the outlier rate of $\sim 15\%$ for galaxies down to $i=25$ in the redshift range of $0.2 \leq z_{\rm p} \leq 1.5$, where $z_{\rm p}$ is the best fit value of photo-z from its probability distribution for a galaxy. The specific photo-z catalog derived from Direct Empirical Photometric code \citep[DEmP]{HY2014} is taken in our study \citep{Tanaka2018}.

In \citet{Yuan2019}, we investigate theoretically the gain from tomographic high peak analyses. In the case with the galaxy number density $n_{\rm g} \sim 40 \hbox{ arcmin}^{-2}$ and the
redshift distribution similar to that expected from LSST, 2 to 4 tomographic bins are the optimal. For S16A sample, $n_{\rm g}\sim 20\hbox{ arcmin}^{-2}$, and 2 bins should be appropriate so that each bin can have a large enough number density of galaxies. 
To test the gain from more redshift bins, we perform analyses with 4 tomographic bins using mock simulations in \S\ref{subsec:mocktest}. The results show that the improvements for the cosmological constraints comparing to the 2-bin case are insignificant. We thus focus on 2-bin analyses using HSC-SSP S16A data. Please note that in this paper, unless otherwise stated as `4-bin' tomography in the mock analyses, the words `tomography' and `low-z' and `high-z' tomography represent the 2-bin tomographic case.

We divide galaxies into two redshift bins, referred to as low-z and high-z bins, with $0.2 \leq z_{\rm p} \leq 0.85$ and $0.85 \leq z_{\rm p} \leq 1.5$, respectively. The division redshift $0.85$ is chosen to be the mid point of the considered redshift range $[0.2,1.5]$. Because our peak analyses involve convergence reconstructions by employing
Fourier transformations, we select regular fields as much as possible from S16A whose original total coverage is $136.9\deg^2$. As a result, $52$ fields are chosen with 40 having an area of $1.5\times 1.5\deg^2$ each, and 12 with $1.0\times 1.0\deg^2$ each cropped from the central part of partially-irregular tracts. 
After this selection, the left-over sky area is $102\deg^2$.

The normalized redshift distributions of the full shear sample, and the low-z and high-z subsamples are shown, respectively, in blue, red and green in Figure \ref{fig:pz_plot}. They are calculated by stacking the normalized photo-z distributions of individual galaxies $P_j(z)$ with the shear weight $w_j$, i.e.,
$p(z)=\sum_j{w_jP_j(z)}/\sum_jw_j$, where the summation runs over all galaxies in a considered sample. The weighted number density is $n_{\rm g} \sim 11 $ and $ \sim 7.5 $ arcmin$^{-2}$ for the low-z and high-z subsamples, respectively.

\section{Weak Lensing Tomographic Peak Analyses} \label{sec:WLpeak}
\subsection{Brief Theory of Weak Lensing} \label{subsec:LB}
The WL effect can be described by the second derivatives of the lensing potential $\psi$ through the Jacobian matrix $\boldsymbol{A}$ given by \citep{BS2001},
\begin{equation}
\boldsymbol{A}=\bigg (\delta_{ij}-\frac {\partial^2 \psi(\boldsymbol{\theta})}{\partial \theta_i\partial \theta_j} \bigg )=
\left(\begin{array}{cc}
{1-\kappa-\gamma_{1}} & {-\gamma_{2}}\\
{-\gamma_{2}} & {1-\kappa+\gamma_{1}}
\end{array}\right)
\label{JacobianMatrix}
\end{equation}
where the convergence $\kappa$ and the shear $\boldsymbol{\gamma}=\gamma_1+i \gamma_2$ induce an isotropic size change and anisotropic shape distortions, respectively, on the observed image of a distant galaxy with respect to its unlensed image.

Under the Born approximation, the convergence $\kappa$ is directly related to the projected density fluctuation weighted by the lensing efficiency factor, i.e., 
\begin{equation}
\begin{split}
\kappa = &\frac{3H_{0}^{2}\Omega_\mathrm{m}}{2c^2} \int_{0}^{\chi_{H}}\mathrm{d}\chi' \int_{\chi'}^{\chi_{H}} \mathrm{d}\chi  \\ 
&\bigg[p_{\mathrm{s}}(\chi)\frac{f_{K}(\chi-\chi')f_{K}(\chi')}{f_{K}(\chi)a(\chi')}\bigg]{\delta[f_{K}(\chi')\boldsymbol{\theta},\chi']}
\end{split}
\label{born}
\end{equation}
where $H_0$, $c$, and $\chi$ are the Hubble constant, the speed of light, and the comoving radial distance, respectively, with $\chi_H=\chi(z=\infty)$. The function $p_{\rm s}$ is the source distribution in terms of $\chi$ calculated from the source redshift distribution, $f_K$ is the comoving angular-diameter distance, $a$ is the cosmic scale factor, and $\delta$ is the 3-dimensional density fluctuation. Peaks in a convergence field are physically intuitive. In particular, high peaks typically correspond to individual massive clusters of galaxies 
along their lines of sight \citep{Hamana2004, Fan2010, Yuan2018, Wei2018a, Oguri2021, Sabyr2022}.

On the other hand, for a WL survey, the observables are often the ellipticities of galaxies, which are directly related to WL shears, or more accurately the reduced shears $\boldsymbol{g}=\boldsymbol{\gamma}/(1-\kappa)$. Thus in all the peak analyses, procedures that generate scalar mass maps from the observed reduced shears are needed. As shown in Eq.(\ref{born}), one of the important scalar fields with a clear physical meaning is the covergence field that can be reconstructed from the reduced shears iteratively based on the relation between $\kappa$ and $\boldsymbol{\gamma}$ \citep{KS1993, Bartelmann1995, SS1995, SK1996, SS1997}. The aperture mass $M_{\rm ap}$ is another commonly used scalar field for peak studies. It can be constructed by applying a $Q$ filter to the tangential reduced shear field directly. In the limit of $\boldsymbol{g}\approx \boldsymbol{\gamma}$ at $\kappa\ll1$, $M_{\rm ap}$ is equivalent to the convergence field filtered by a compensated $U$ kernel with $Q(\theta)=(2/\theta^2)\int_0^\theta d\theta^{\prime} \theta^{\prime} U(\theta^{\prime})-U(\theta)$ and $\int d\theta \theta U(\theta)=0$ \citep{Schneider1998, Marian2012, Bard2013, Kacprzak2016, Martinet2018, Oguri2021, Zrcher2022}. For high peaks, however, this equivalence breaks down because the difference between $\boldsymbol{g}$ and $\boldsymbol{\gamma}$ is not negligible. To develop a theoretical model for the abundance of high aperture-mass peaks, we need to consider $M_{\rm ap}$ from the $Q$ filtering on the tangential reduced shears (Pan et al., in preparation).

In this study, we adopt the approach to reconstruct the convergence field from the observed reduced shears and apply our covergence peak model \citep{Fan2010, Yuan2018} to extract cosmological information from the tomographic high peak counts. 

\subsection{Map Making and Peak Identification} \label{subsec: MapPeak}
For the tomographic peak analyses, three sets of maps are generated for both low-z and high-z subsamples.

(1) Convergence maps. Following the details presented in \citet{LiuX2015}, we reconstruct the convergence fields from the shear catalog of each redshift bin. In summary, to reduce the shape noise from finite shear measurements, we first apply a smoothing to obtain the smoothed shear map on regular $1024\times1024$ grids for each of the 52 fields. It is calculated as follows \citep{Oguri2018Mass}
\begin{equation}
\begin{aligned}
\langle\bm{\epsilon}\rangle(\bm{\theta})
&=\frac{\sum_k w_{k}W_{\theta_\mathrm{G}}(\bm{\theta_k-\theta})\bm{e}_k}{2\sum_k w_{k}(1-e_{\mathrm{rms},k}^2)(1+m_k)W_{\theta_\mathrm{G}}(\bm{\theta_k-\theta})}\\
&-\frac{\sum_k w_{k}W_{\theta_\mathrm{G}}(\bm{\theta_k-\theta})\bm{c}_k}{\sum_k w_{k}(1+m_k)W_{\theta_\mathrm{G}}(\bm{\theta_k-\theta})}
\end{aligned}
\label{eq:smoothing}
\end{equation}
where $\bm{\theta}_k$ and $\bm{e}_k$ are the position and the two-component ellipticity of galaxy $k$, and $w_{k}$, $\bm{c}_k$, $m_k$ and $e_{\mathrm{rms},k}$ are the corresponding weight, additive and multiplicative biases of shear measurements, and the intrinsic ellipticity dispersion per component. We take the Gaussian smoothing kernel with  
\begin{equation}
W_{\theta_{\rm G}}(\boldsymbol{\theta})=\frac{1}{{\pi\theta_{\rm G}^2}}\exp{\left(-\frac{{|\boldsymbol{\theta}|^2}}{{\theta_{\rm G}^2}}\right)} 
\label{Gaussian} 
\end{equation}
and $\theta_{\rm G}=1.5$ arcmin, suitable for the galaxy number densities considered here and for the cluster-scale structures that are closely related to high WL peaks \citep{Hamana2004}. 

From $\langle\bm{\epsilon}\rangle(\bm{\theta})$, we reconstruct the convergence map iteratively using the nonlinear Kaiser-Squires inversion \citep{KS1993, SS1995, LiuX2015}. In brief, we start the iteration with $\kappa^{(0)}=0$ and $\gamma^{(0)}=\langle\epsilon\rangle$, and calculate $\kappa^{(n)}$ from $\gamma^{(n-1)}$ based on the relation between $\kappa$ and $\gamma$. Then $\gamma$ is updated to $\gamma^{(n)}=(1-\kappa^{(n)})\langle\epsilon\rangle$ for the next step. We truncate the process when the required converging accuracy of $10^{-6}$ is reached, which is defined as the maximum difference of the reconstructed convergence values in a field
between the two sequential iterations.

(2) Noise maps. We randomly rotate source galaxies to remove the correlated lensing signals and then follow the same procedures presented in (1) to generate the noise maps. For each field and each tomographic redshift bin, we perform 10 random rotations and obtain consequently 10 noise maps. 

(3) Filling factor maps. Bad data masking in a shear catalog can result in regions with unrealistic low galaxy numbers and thus bias the peak statistics \citep{Liu2014}. To identify the mask-affected areas, we construct the smoothed filling factor map $f$ for each field in each tomographic subsample by \citep{vW2013, LiuX2015}
\begin{equation}
f(\bm{\theta})=\frac{\Sigma_k W_{\theta_\mathrm{G}}(\bm{\theta_k-\theta})w_{k}}{\langle \Sigma_n W_{\theta_\mathrm{G}}(\bm{\theta_n-\theta})\tilde{w}_{n}\rangle}.
\label{filling}
\end{equation} 
Here the denominator represents the expected average value over a field if there are no masks. It is calculated by randomly populating galaxies (with their positions denoted as $\bm{\theta_n}$) over the full field area based on their average number density computed 
by dividing the total number of galaxies in the field in the considered shear sample by the area excluding the masks. The quantity $\tilde{w}_{n}$ is a randomly assigned weight according to the weight distribution in the shear sample. The nominator is calculated using 
the true galaxy position $\bm{\theta_k}$. To mitigate the mask effects in our studies, we exclude regions with $f<0.6$ in the peak counting \citep{Liu2014, LiuX2015}. In addition, we also remove the outer most $5\theta_\mathrm{G}$ regions in each of the four sides of a field to suppress the boundary effects. 
The final effective area for our tomographic peak studies is $\sim 58\deg^2$.
We note that the mask and boundary effects depend on the reconstruction methods. For the one used here, the exclusions described above are suitable to suppress the possible bias from masks and boundaries. 
As one of our future tasks, we will explore different reconstruction methods for better controling mask and boundary effects so that more areas can be used in cosmological analyses. 

To make a comparison between tomographic and 2D peak analyses, we also generate the above maps using the full shear sample in addition to that with the low-z and high-z subsamples.

\begin{figure*}
\centering
\includegraphics[width=1.0\columnwidth]{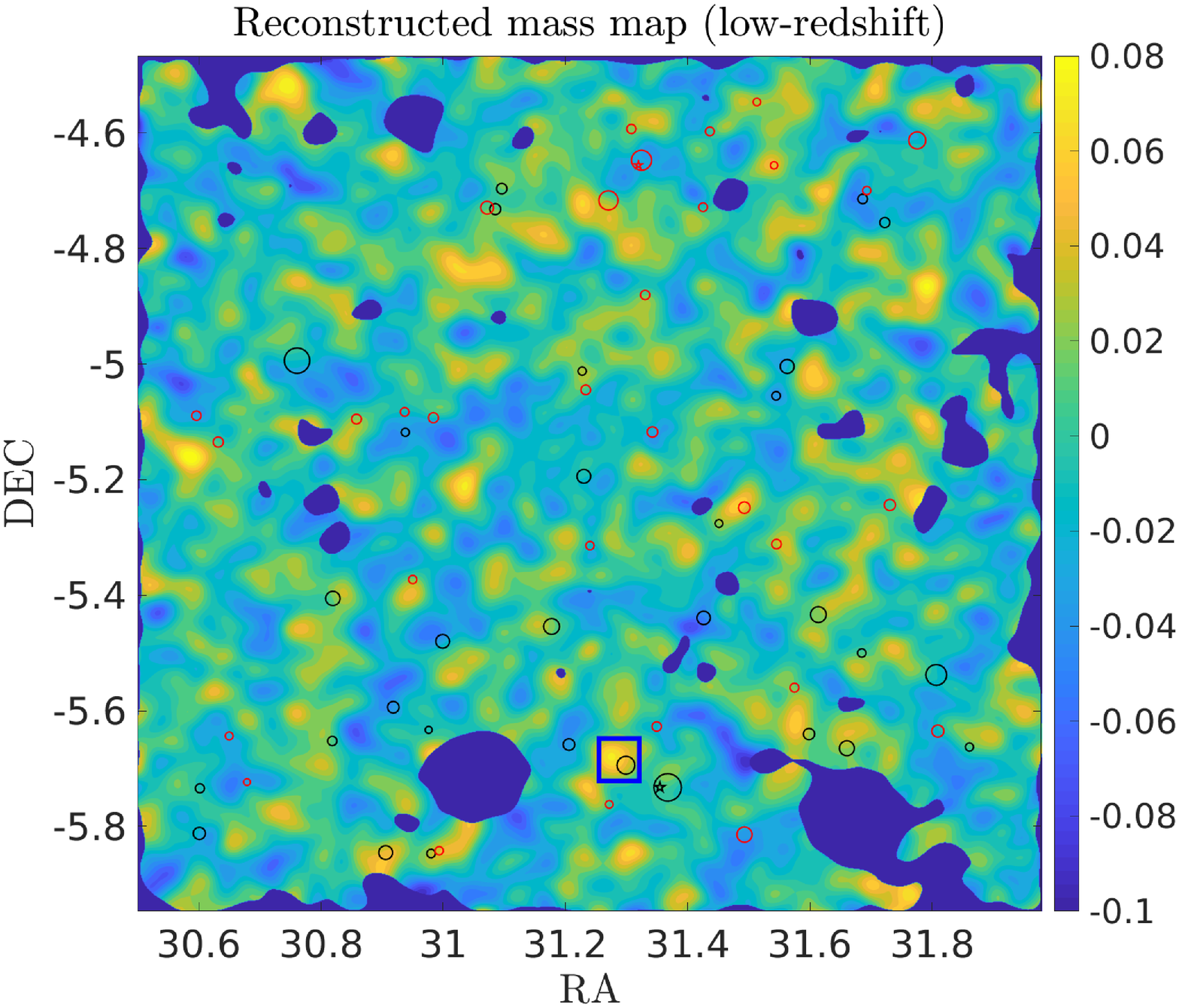}
\includegraphics[width=1.0\columnwidth]{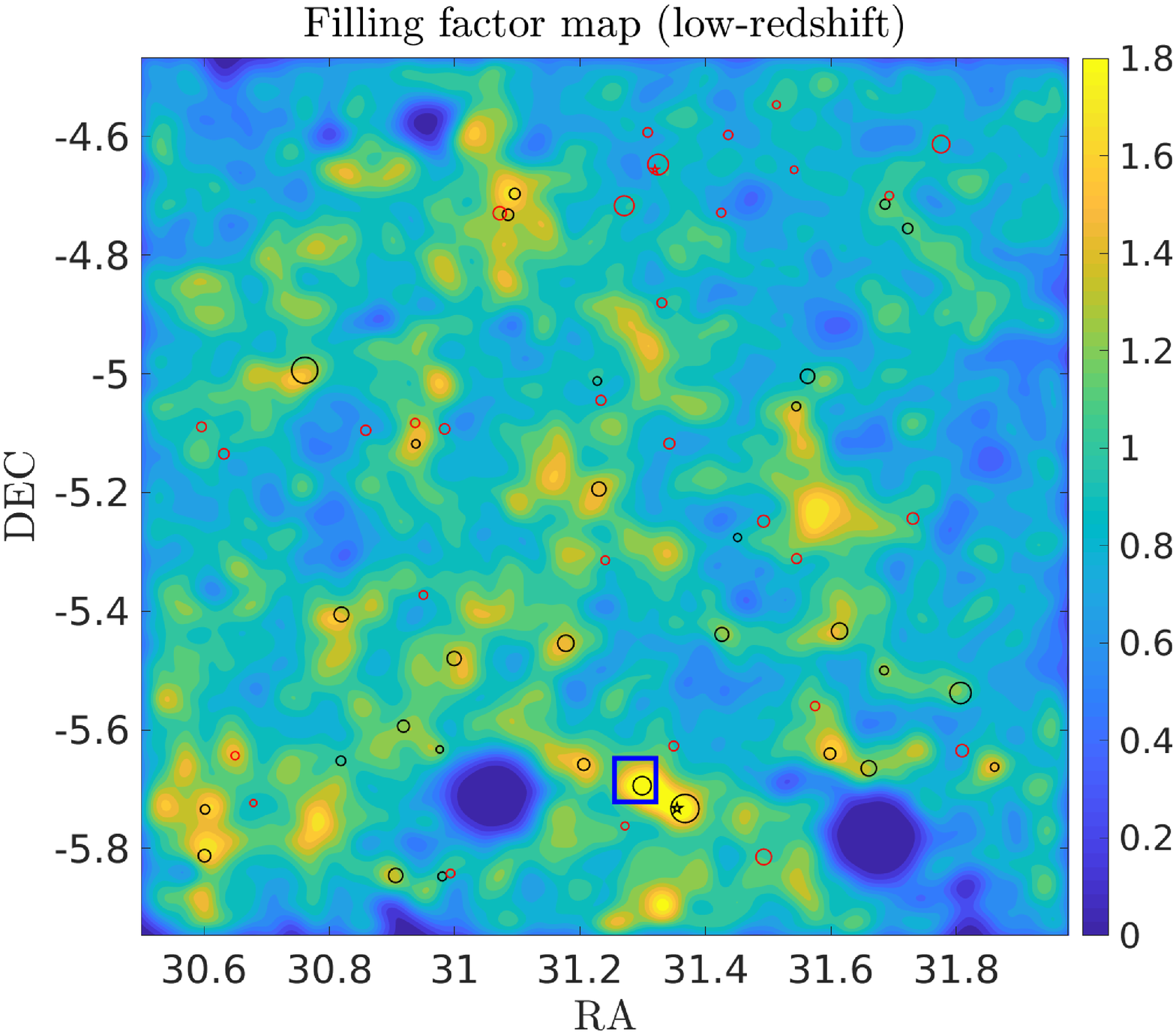}
\includegraphics[width=1.0\columnwidth]{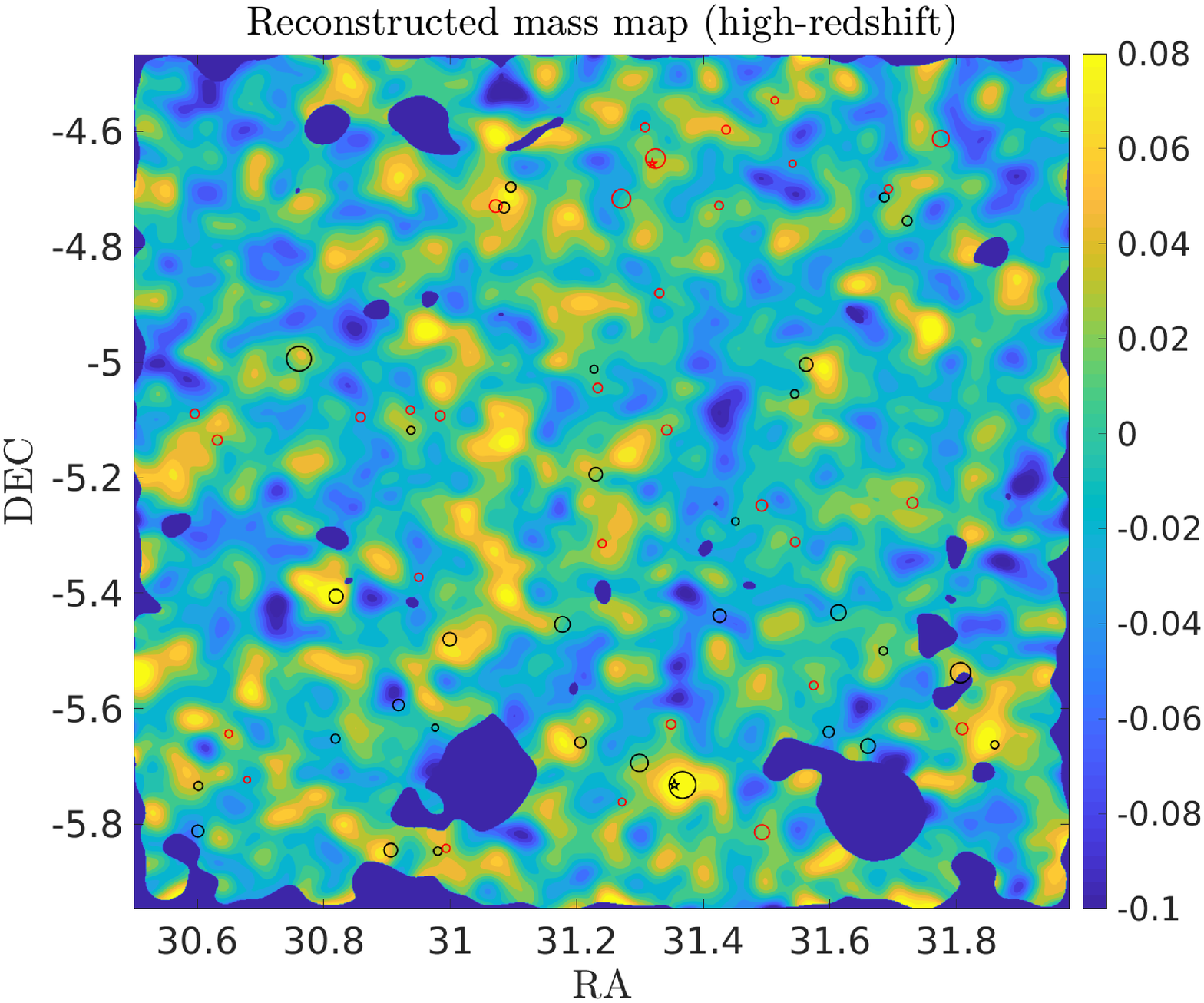}
\includegraphics[width=1.0\columnwidth]{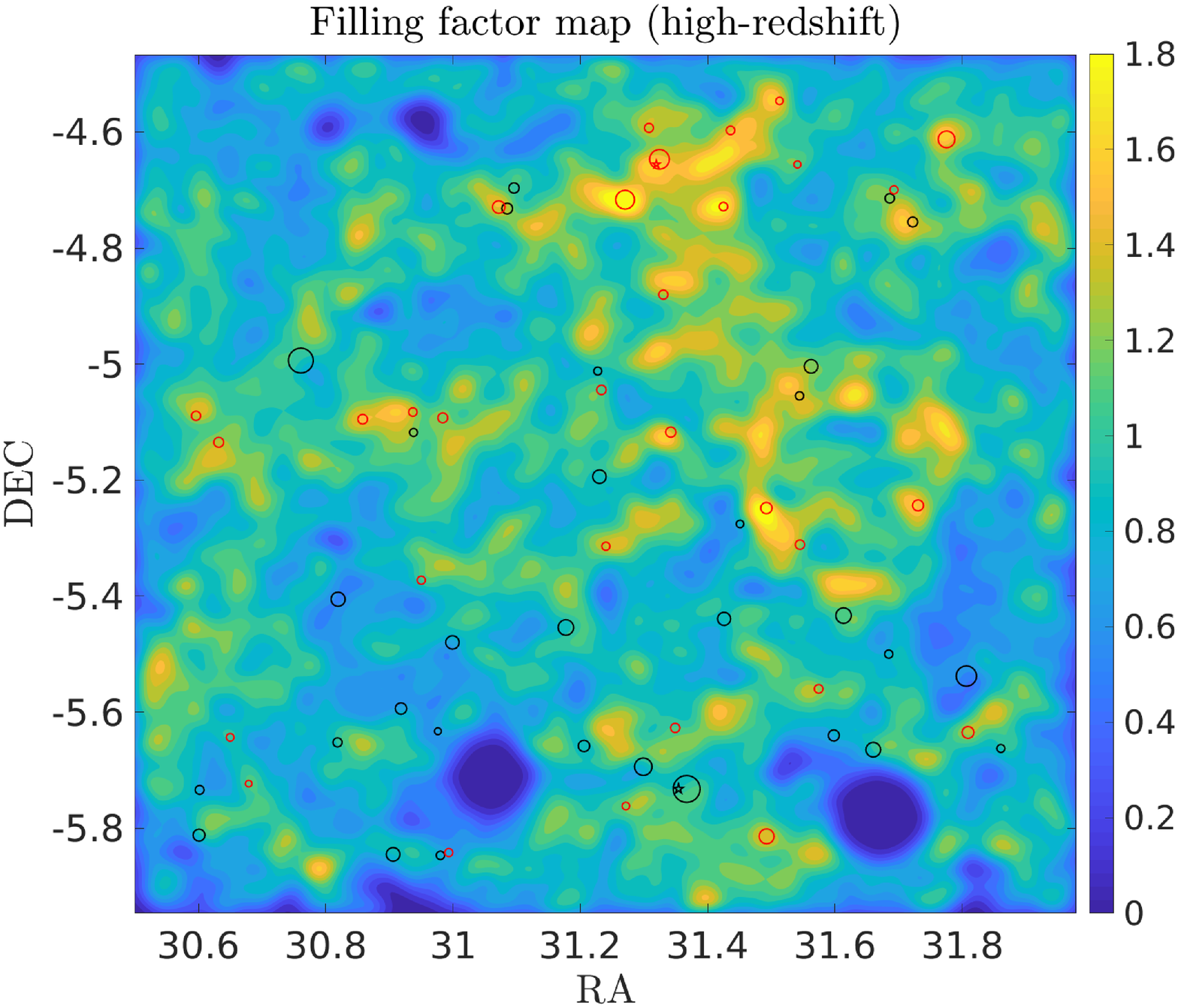}
\includegraphics[width=1.0\columnwidth]{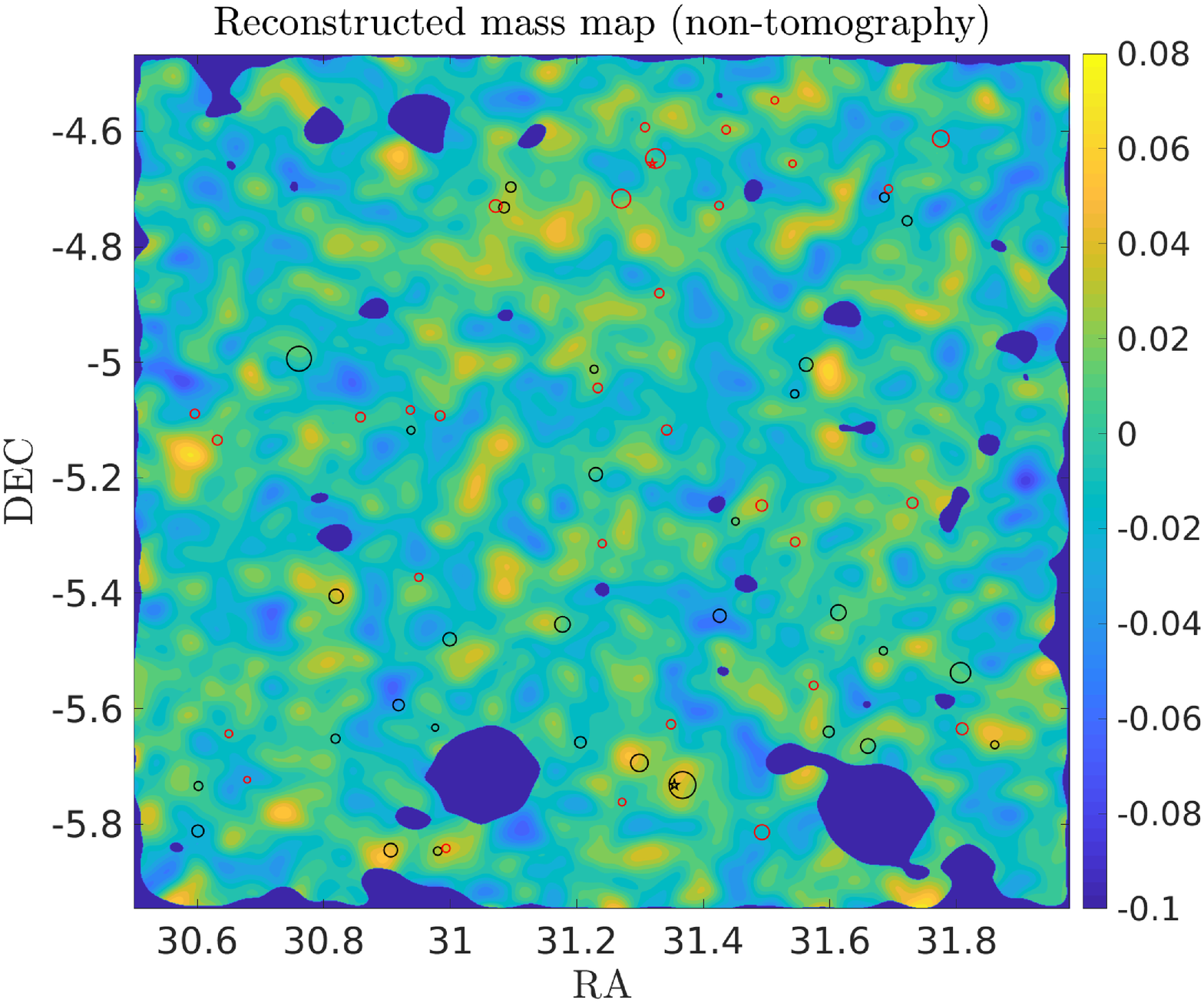}
\includegraphics[width=1.0\columnwidth]{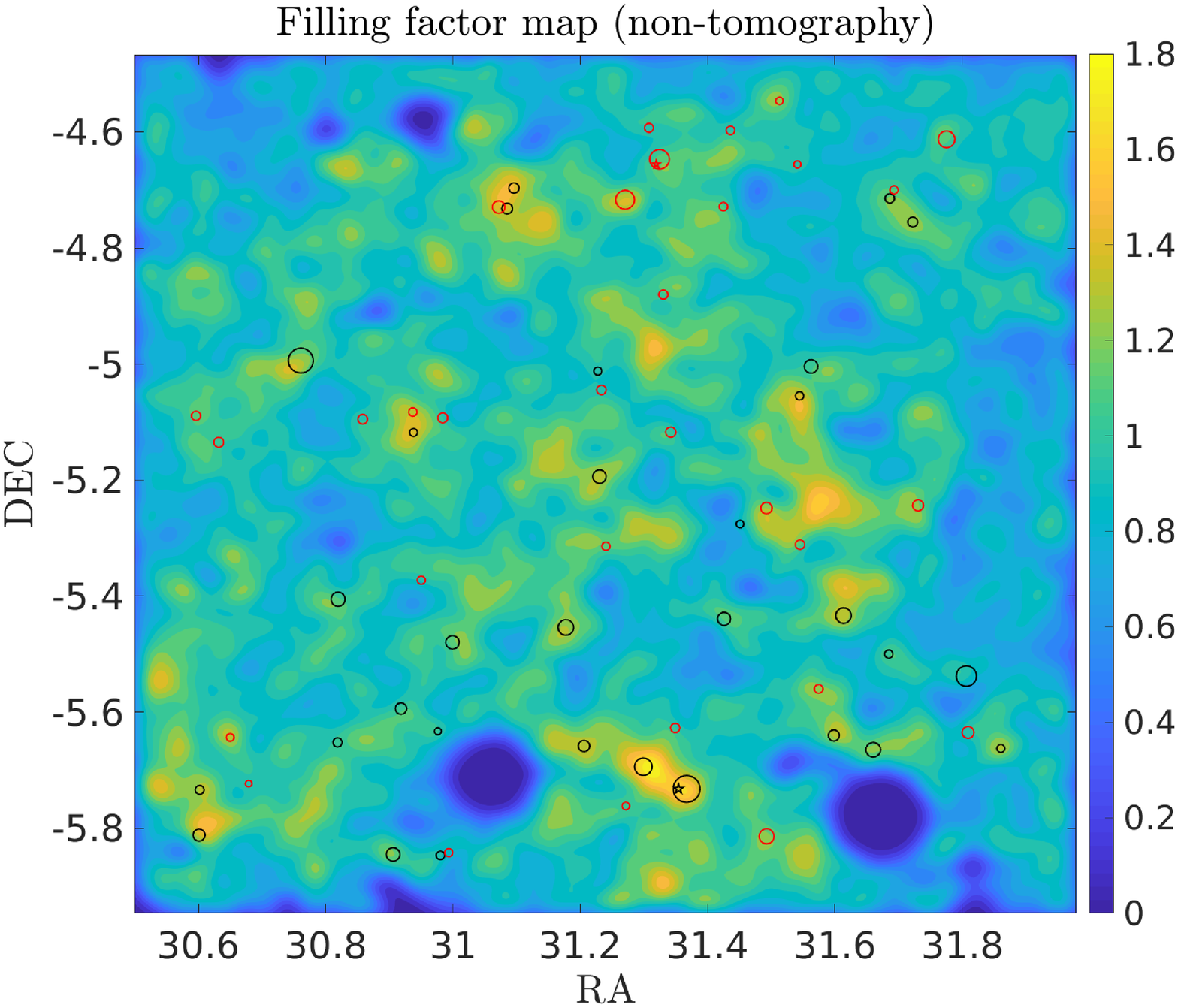}
\caption{Left panels: the reconstructed convergence maps of one field for low-z (top), high-z (middle) and non-tomographic (bottom) cases, respectively. The dark blue regions have the filling factor $f<0.6$. Right panels: the corresponding filling factor maps. The clusters from the CAMIRA and WZL catalogs 
are shown by the pentagrams and circles, with the size indicating the mass of the clusters. Black and red symbols are the ones with $0.2\le z_{\rm cluster} \le 0.85$ (low-zc) and $0.85\le z_{\rm cluster}\le 1.5$ (high-zc), respectively. The blue square highlighted in the top panels is a representative 
example of high peaks resulting mainly from individual massive clusters.
}
\label{fig:massREC_example}
\end{figure*}

In Figure \ref{fig:massREC_example}, we show the reconstructed convergence (mass) maps (left) and the filling factor maps (right) of one field for low-z (top), high-z (middle) and non-tomographic (bottom) cases, respectively. Regions with the filling factor $f<0.6$ are marked in dark blue in the convergence maps. The pentagrams and circles in the plots are the clusters in the field detected by the Cluster-finding Algorithm based on Multi-band Identification of Red-sequence gAlaxies (CAMIRA) \citep{Oguri2018Cluster} and by \citet{Wen2021} based on 3D distribution of galaxies (WZL) \citep{Wen2009, Wen2012, Wen2018}. The black and red symbols represent the clusters with their photometric redshifts in the range of $0.2\le z_{\rm cluster} \le 0.85$ (low-zc) and of $0.85\le z_{\rm cluster}\le 1.5$ (high-zc), respectively. The size of the pentagrams and circles indicates the mass of the clusters in CAMIRA and WZL. Only clusters with $M_{\rm vir}\ge 10^{14}h^{-1}M_{\odot}$ are shown (see \S\ref{subsec:boost} for the derivation of $M_{\rm vir}$). In general, peaks, particularly high peaks in the convergence maps associate well with clusters of galaxies as expected. However, the correspondence is not one to one because of the lensing efficiency, shape noise and the projection effects of large-scale structures. To be described in Appendix \ref{sec:appendix}, we take these effects into account in our WL peak model and therefore peaks can be used directly for our cosmological studies. 

From the right panels of Figure \ref{fig:massREC_example}, we see that the low-zc/high-zc clusters largely overlap with the high density regions in the filling factor maps constructed from the low-z/high-z shear samples. This shows that a significant fraction of their member galaxies are in the corresponding shear catalogs, which in turn can lead to a considerable dilution effect on their own lensing signals. As a visual example, the blue-squared peak in the lower part of the top left panel clearly arises from a low-zc cluster within the square. The concentration of its member galaxies in the low-z shear catalog shown in the top right panel results in a dilution bias for the peak height. This effect is significant for the low-z convergence peaks because of the overlap of the redshift range of the lensing clusters and the shear catalog. For the high-z bin, we can see that the peaks in the corresponding convergence map (middle left panel) associate more with low-zc clusters (black circles) than the high-zc clusters (red circles) because of the lensing efficiency factor. Most of the member galaxies of these low-zc clusters are not in the high-z shear catalog, and thus do not dilute the associated convergence peaks constructed from the high-z shear sample. We will present detailed statistical analyses of the dilution effect in \S\ref{subsec:boost}.

For peak identifications in a convergence map, if a pixel value is higher than that of its 8 neighbouring pixels, this pixel is defined as a peak. Its S/N is calculated by $\nu_\mathrm{N}=K/\sigma_{\mathrm{N},0}$, 
where $K$ is the reconstructed convergence value of the peak and $\sigma_{\mathrm{N},0}$ is the mean rms of the noise estimated from the noise maps excluding $f<0.6$ regions and the boundary regions as described above. 
For the 52 fields used in our analyses, the number densities of galaxies are rather similar with $n_{\rm g}=11.0\pm 0.76 \hbox{ arcmin}^{-2}$, $7.5\pm 0.62\hbox{ arcmin}^{-2}$, and $18.5\pm 1.26\hbox{ arcmin}^{-2}$ 
for the low-z, high-z and non-tomographic cases, respectively. The corresponding fluctuations of the shape noise over different fields are small with $\sigma_{\rm N, 0}=0.023\pm 0.0016$, $0.029\pm 0.0023$, and $0.018\pm 0.0013$ in the case of $\theta_{\rm G}=1.5\hbox { arcmin}$ for the three samples, respectively. It is therefore a good approximation to use the mean $\sigma_{\rm N,0}$ for each sample in our analyses. 

To apply our halo-based peak model \citep{Fan2010, Yuan2018} for cosmological constraints, here we consider high peaks with $\nu_\mathrm{N}\ge 3.5$, corresponding to a smoothed $K\ge 0.08$, $0.10$, and $0.06$ for the low-z, high-z and non-tomographic cases, respectively. We further limit to peaks with $\mathrm{S/N}\le 5.5$ because even higher peaks are very few for an effective HSC-SSP area of $\sim 58\deg^2$. 

For being self-contained, we describe concisely our theoretical peak model used in deriving cosmological constraints from the observed tomographic peak counts in Appendix \ref{sec:appendix}. More details of the model and previous applications can be found in \citet{Fan2010, Yuan2018, Yuan2019, Liu2014, LiuX2015, Liu2016, Shan2018}. 

\subsection{Derivation of Cosmological Constraints} \label{subsec:fitting}

For tomographic analyses, we combine the observed peak counts from low-z and high-z shear samples. As described in \S\ref{subsec: MapPeak}, observationally we can only obtain $\sigma_{\mathrm{N},0}$, thus the observed S/N of peaks
is defined by $\nu_\mathrm{N}=K/\sigma_{\mathrm{N},0}$. For each tomographic bin, peaks are divided into four $\nu_\mathrm{N}$ bins: [$3.5$, $4.0$], [$4.0$, $4.5$], [$4.5$, $5.0$], [$5.0$, $5.5$]. The peak counts from the $4$ $\nu_\mathrm{N}$ bins $\times$ $2$ tomographic redshift bins form the observed data set with $8$ data points. 

The following $\chi_{p}^2$ is minimized for cosmological parameter constraints,
\begin{equation}
 \chi_{p}^{2}=\sum_{i,j=1}^{N_{\rm bin}}\Delta N_{i}^{(p)}(\widehat{C_{ij}^{-1}})\Delta N_{j}^{(p)},
\label{chi2}
\end{equation}
where $\Delta N_i^{(p)}=N_{\rm peak}^{(p)}(\nu_{\mathrm{N},i})- N_{\rm peak}^{(d)}(\nu_{\mathrm{N},i})$ is the difference between the
theoretical prediction with cosmological model $p$ and the observed peak counts $d$. It is noted that in order to confront with the observed peak counts, in theoretical model predictions 
we need to convert the peak height $\nu=K/\sigma_0$ to $\nu_\mathrm{N}=K/\sigma_{\mathrm{N},0}$ using the ratio of $\sigma_0/\sigma_{\mathrm{N},0}$ and bin the calculated peaks based on $\nu_\mathrm{N}$, labelled as 
$N_{\rm peak}^{(p)}(\nu_{\mathrm{N},i})$ here. For tomographic analyses, $N_{\rm bin}=8$. 

The covariance matrix $C_{ij}$ is computed from our simulated HSC-SSP mocks to be detailed in \S\ref{subsec:mocktest}. We use the bootstrap analysis by resampling the 20 sets of HSC-SSP mocks ($52\times 20$ mock fields in total), and $C_{ij}$ is given by
\begin{equation}
\begin{split}
C_{ij}=\frac{1}{R-1}\sum_{r=1}^{R}[N^r_{\rm peak}(\nu_{\mathrm{N},i})-{\bar{N}}_{\rm peak}(\nu_{\mathrm{N},i})]\\
\times [N^r_{\rm peak}(\nu_{\mathrm{N},j})-{\bar{N}}_{\rm peak}(\nu_{\mathrm{N},j})],
\end{split}
\label{covar}
\end{equation}
where $R$ is the total number of bootstrap samples with $R=1000$, $N^r_{\rm peak}(\nu_{\mathrm{N},i})$ and ${\bar{N}}_{\rm peak}(\nu_{\mathrm{N},i})$ are the peak count from the bootstrap sample $r$ and 
the mean peak count averaged over the total $1000$ samples in the bin centred on $\nu_{\mathrm{N},i}$. 
As illustrated in Figure \ref{fig:massREC_example}, in real observations, the source clustering of the member galaxies of a cluster can dilute its lensing signal significantly. This effect cannot be correctly simulated in our mocks because the foreground matter distribution from the N-body simulations is different from that in the real Universe. In order to include the dilution effect in the covariance estimate for real observations, we calculate the theoretical peak counts with and without the dilution correction, $N_\mathrm{wb}$ and $N_\mathrm{wob}$, respectively, both under the cosmological model ($\Omega_{\rm m}$, $\sigma_8$)$=$($0.332$, $0.799$) from the marginalized constraints of the HSC shear correlation analyses in \citet{Hamana2020,Hamana2022}. We then use their ratio $D_i=N_{{\rm wb},i}/N_{{\rm wob},i}$ to perform element-based corrections for the covariance matrix. Specifically, $C^O_{ij}=C_{ij}D_iD_j$. The details to calculate $N_\mathrm{wb}$ is shown in \S\ref{subsec:boost}. 

With $C^O_{ij}$, the inverse of the covariance matrix is calculated in an unbiased way following \citet{Hartlap2007} 
\begin{equation}
\widehat{(\boldsymbol{C}^O)^{-1}}=\frac{R-N_{\mathrm{bin}}-2}{R-1}(\boldsymbol{C}^O)^{-1}.
\label{nobiascov}
\end{equation}
\citet{Sellentin2016} show that because the covariance matrix can only be estimated from data, either from simulations or bootstraps,  its difference from the true covariance matrix needs to be taken into account when the number of data sets used to estimate the covariance matrix is small. By marginalizing over the true covariance matrix conditioned on the estimated one, the likelihood function is better described by a multivariate t-distribution, instead of Gaussian. As a result, the approach adopting a debiased inverse covariance matrix as in Eq.(\ref{nobiascov}) can underestimate the errors of cosmological parameter inference. In our analyses, the number of simulations used to estimate the covariance is 1000, which is large. Thus we do not expect a very significant difference between the approach of \citet{Hartlap2007} and that of \citet{Sellentin2016}. Nevertheless, it is worth to investigate this issue in our future studies.

For the non-tomographic case, peaks are divided into the same $\nu_\mathrm{N}$ bins with the total $N_{\rm bin}=4$.

We derive cosmological constraints on $(\Omega_{\rm m}, \sigma_{\rm 8})$ using Markov Chain Monte Carlo (MCMC) under flat $\Lambda$CDM fixing the other parameters with the Hubble constant $h=0.7$, the power index of the initial density fluctuation power spectrum $n_{\rm s}=0.96$ and the present baryon matter density $\Omega_{\rm b}=0.046$. 

\begin{figure*}
\centering
\includegraphics[width=1.0\columnwidth]{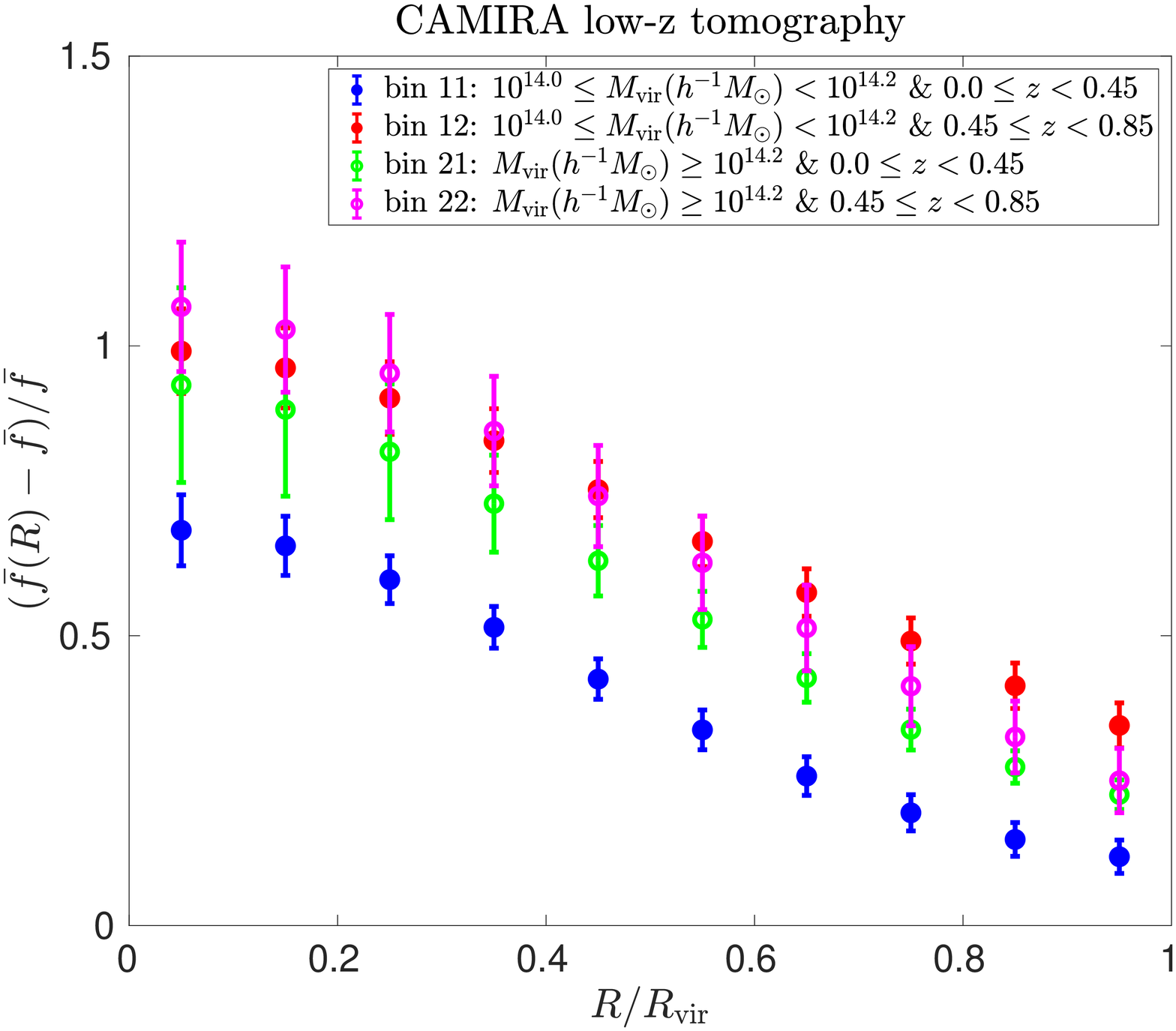}
\includegraphics[width=1.0\columnwidth]{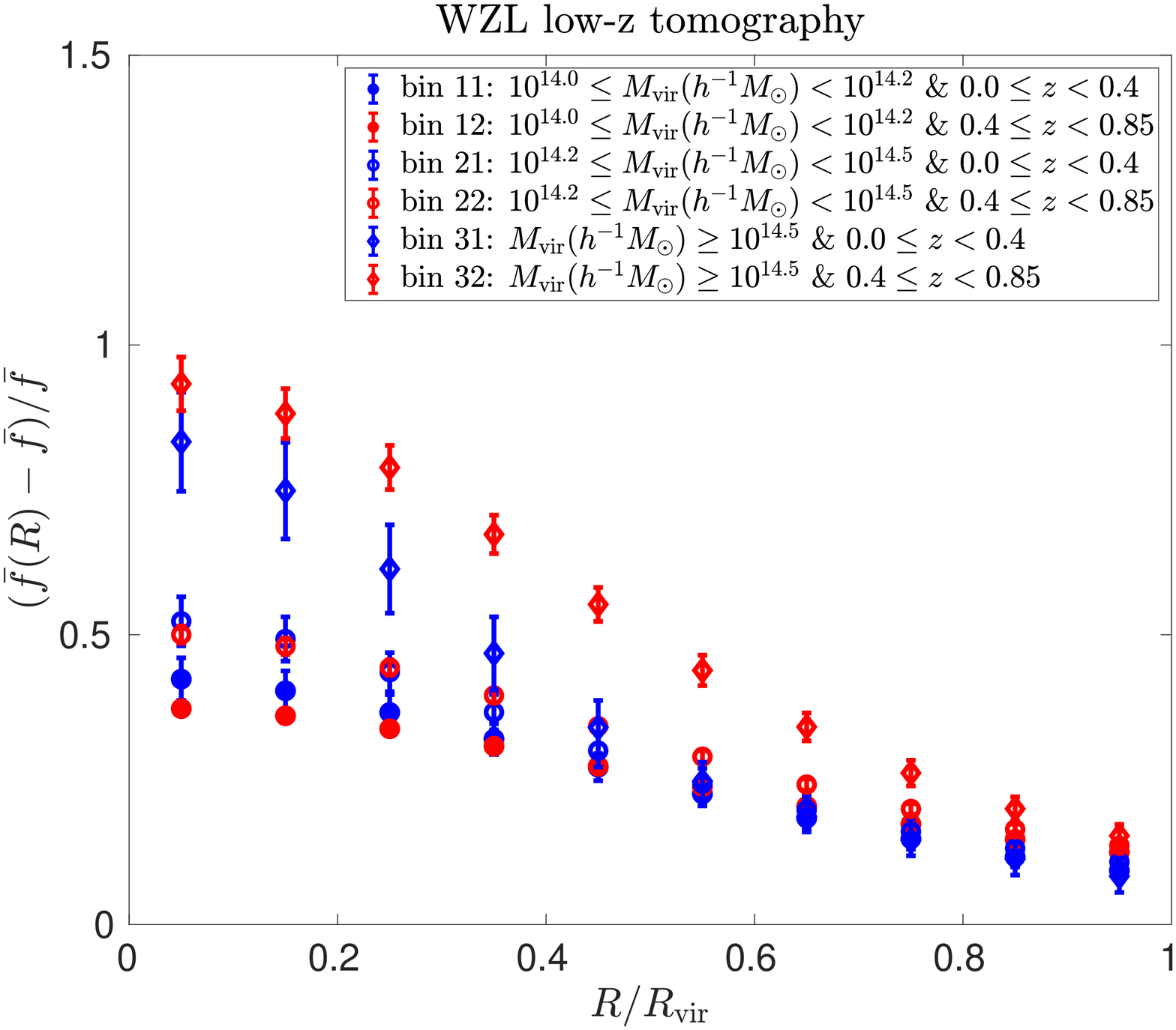}
\includegraphics[width=1.0\columnwidth]{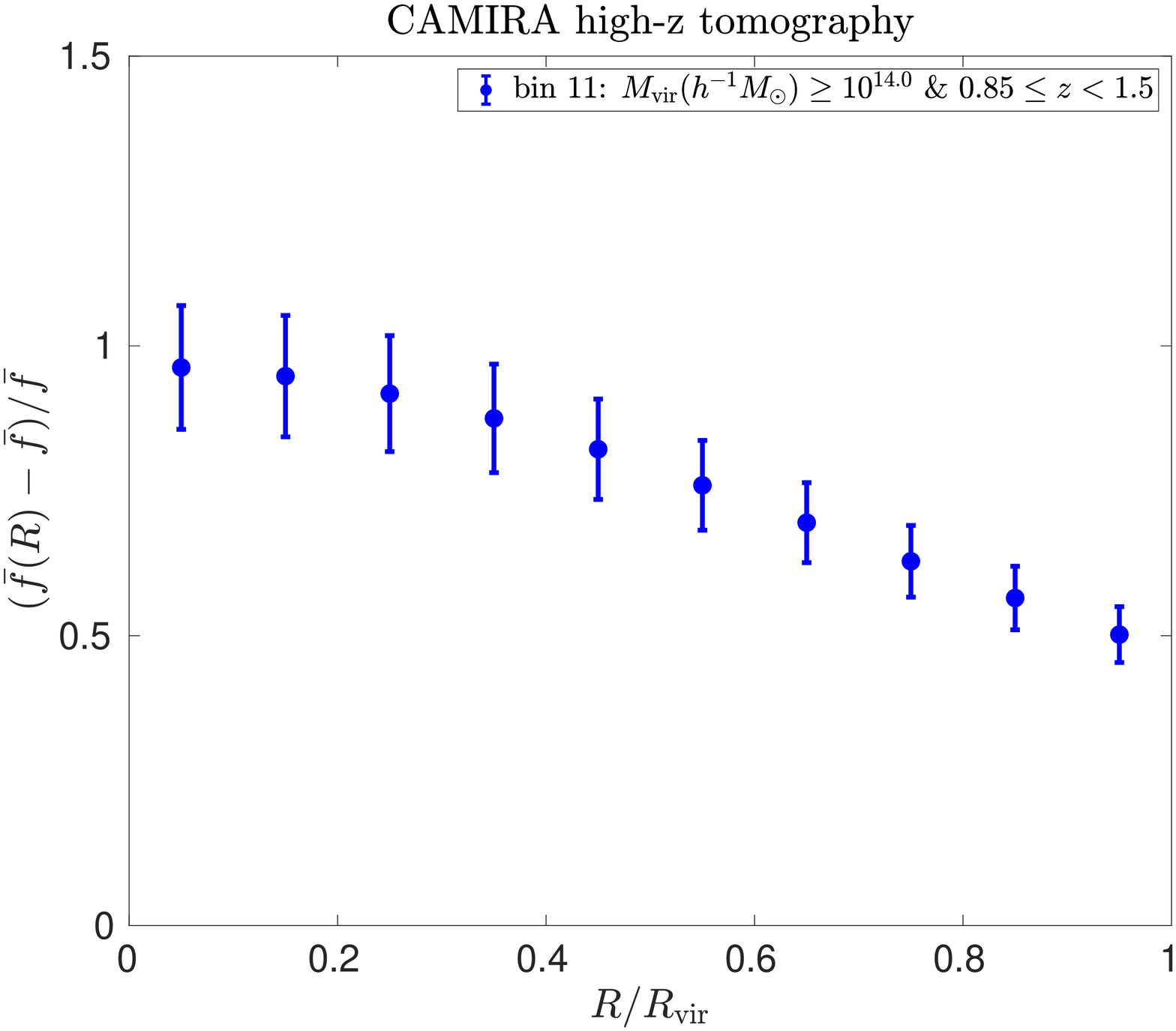}
\includegraphics[width=1.0\columnwidth]{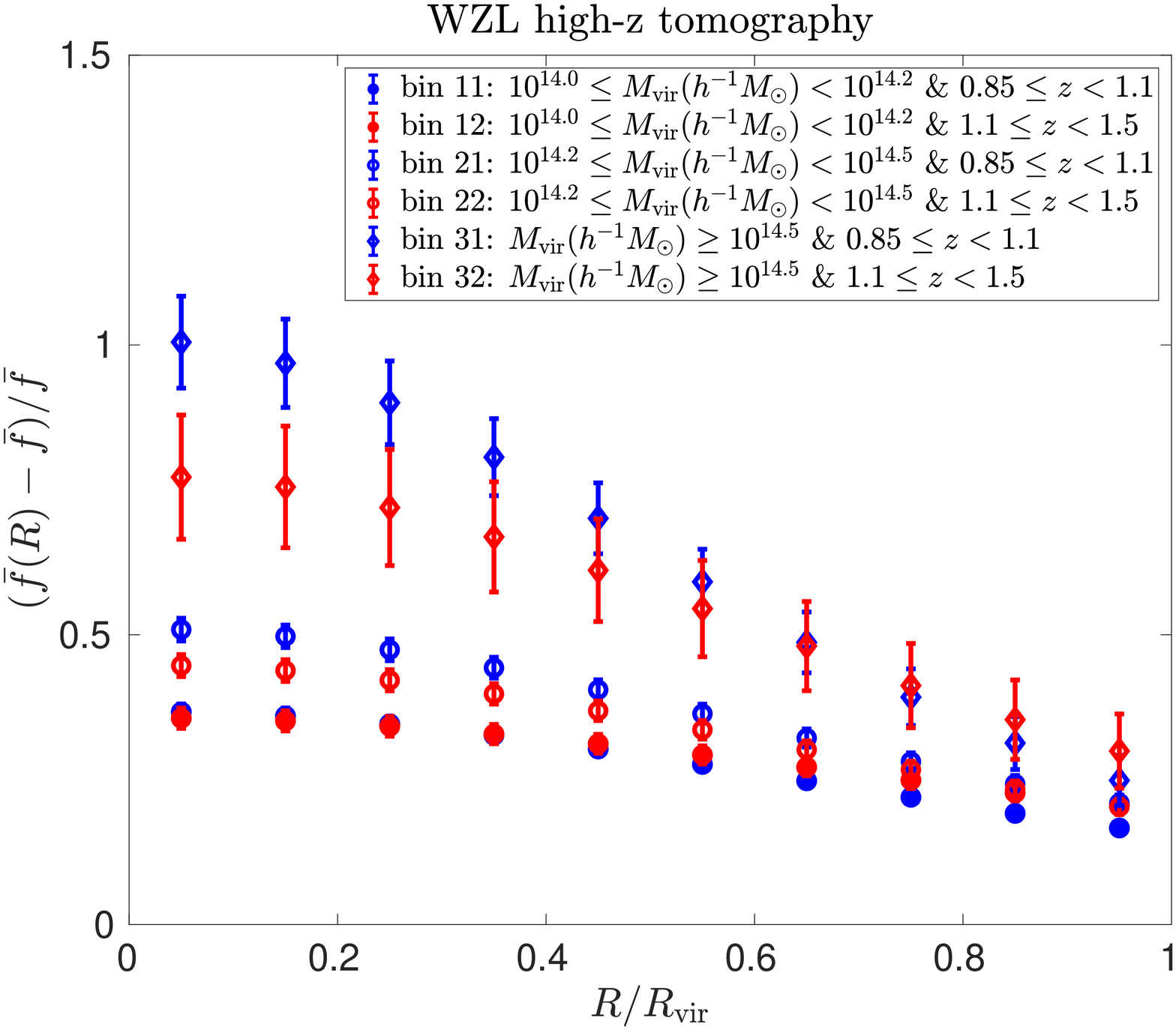}
\includegraphics[width=1.0\columnwidth]{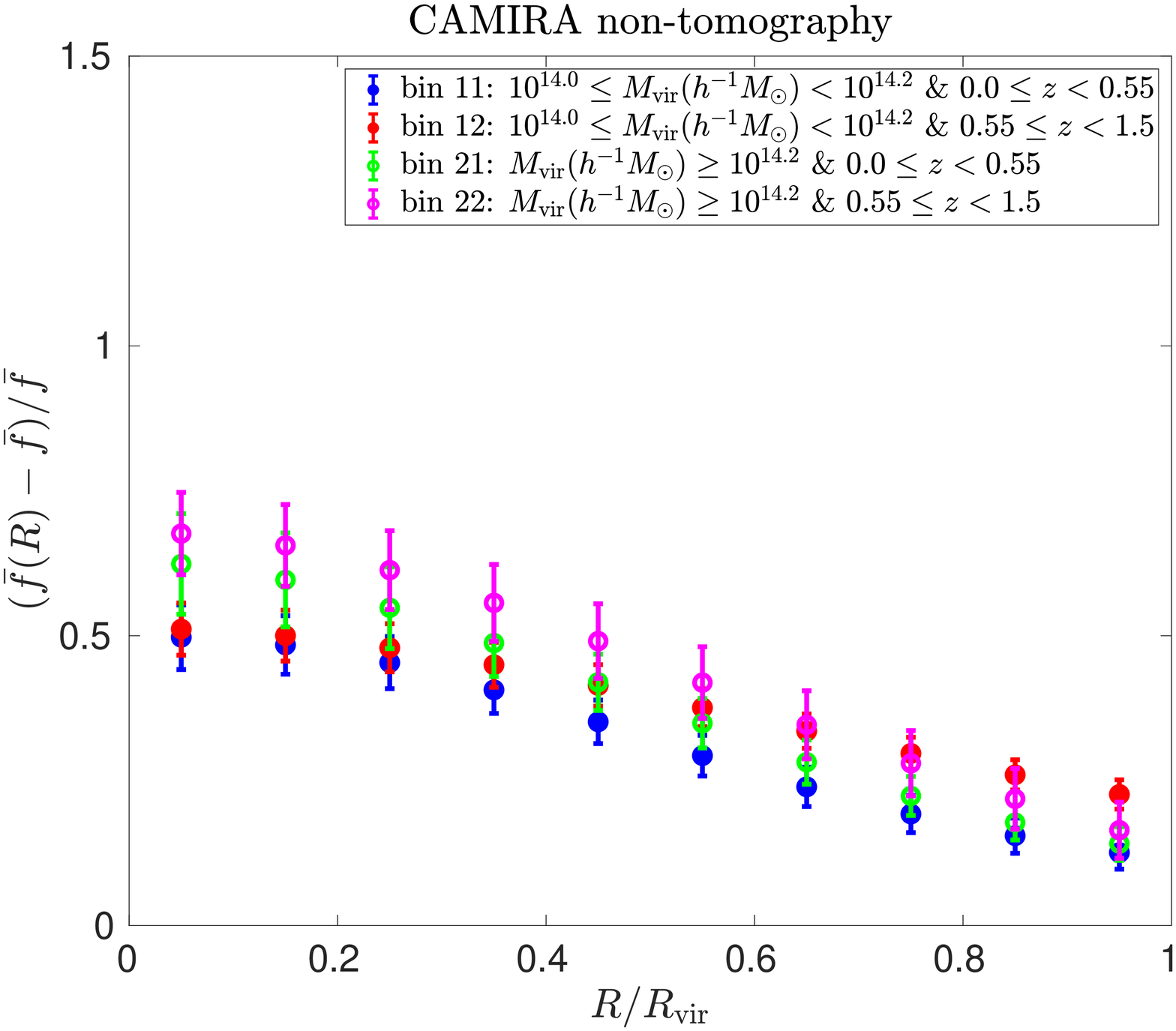}
\includegraphics[width=1.0\columnwidth]{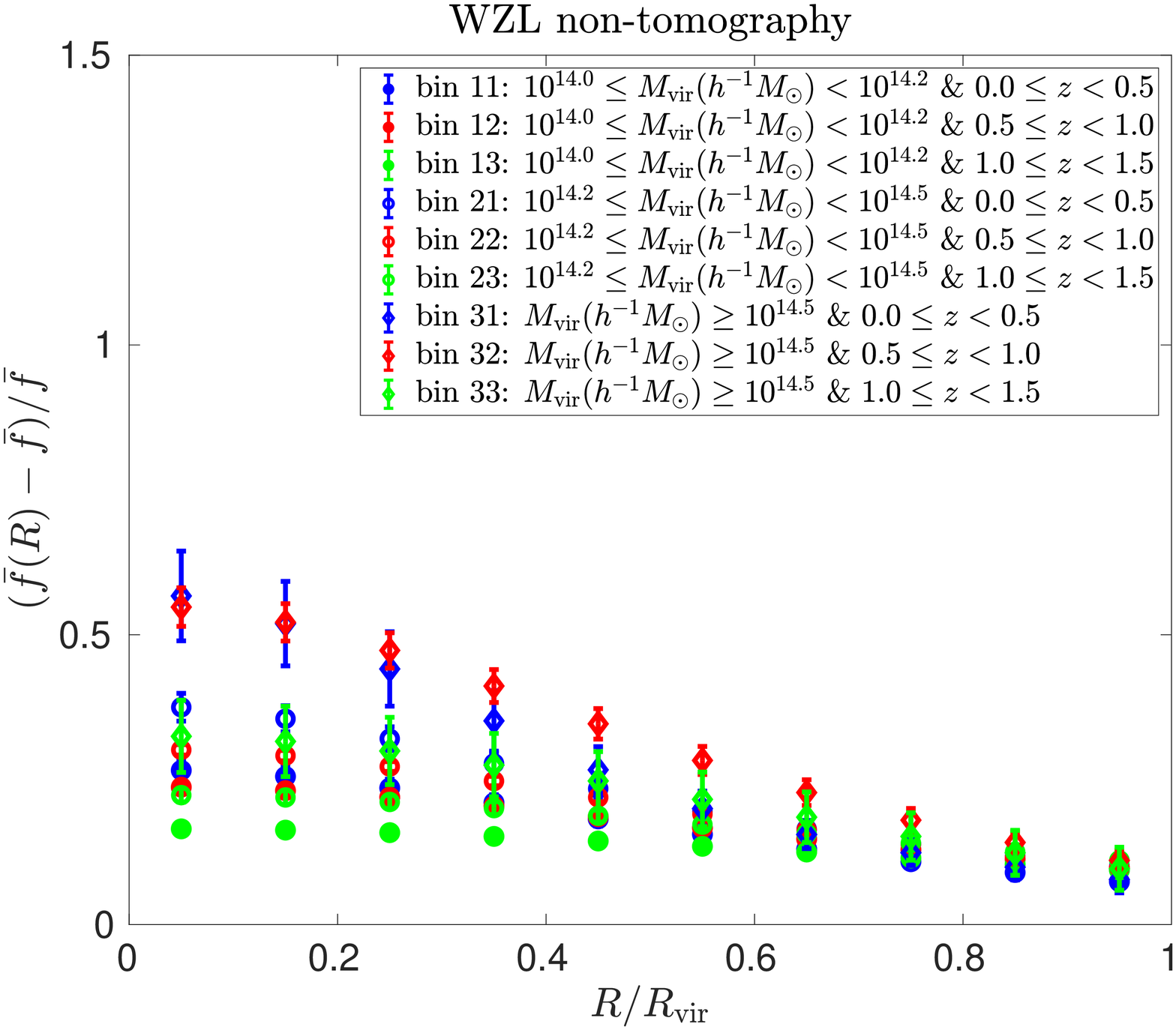}
\caption{Filling factor distributions around cluster candidates with the meaning of different symbols explained in the inserts. The error bars are estimated from a bootstrap analysis. Left snd right panels are respectively from CAMIRA and WZL clusters, and from top to bottom are the results of the low-z, high-z and non-tomographic cases.
}
\label{fig:excessfilling}
\end{figure*}

\section{Systematics} \label{sec:systematics}
In this section, we evaluate the potential impacts of different systematics on our analyses.

\begin{table*}
\caption{The cluster samples in different mass and redshift bins used in the boost factor measurement.}
\label{tab:boost}
\begin{center}
\leavevmode
\begin{tabular}{c c c c c c c  c} \hline
  &  & bin & mass range & $z$ range & dilution factor & excess fraction & Num of clusters\\
\hline
\specialrule{0em}{0.5pt}{0.5pt}
\multirow{9}*{CAMIRA} & \multirow{4}*{low-z tomography}& bin11 & $10^{14.0}\le M_\mathrm{vir}(h^{-1}{\rm M}_{\odot})<10^{14.2}$ & $0.0\le z<0.45$ & $1/1.227$ & 1.278 & 16 \\
  &  & bin12 & $10^{14.0}\le M_\mathrm{vir}(h^{-1}{\rm M}_{\odot})<10^{14.2}$ & $0.45\le z<0.85$ & $1/1.208$ & 1.566 & 17\\
  &  & bin21 & $M_\mathrm{vir}(h^{-1}{\rm M}_{\odot})\ge 10^{14.2}$ & $0.0\le z<0.45$ & $1/1.381$ & 1.438 & 17\\
  &  & bin22 & $M_\mathrm{vir}(h^{-1}{\rm M}_{\odot})\ge 10^{14.2}$ & $0.45\le z<0.85$ & $1/1.219$ & 1.513 & 10\\
\cline{2-8}
\specialrule{0em}{1pt}{1pt}
  & {high-z tomography} & bin11 & $M_\mathrm{vir}(h^{-1}{\rm M}_{\odot})\ge 10^{14.0}$ & $0.85\le z<1.5$ & $1/1.149$ & 1.678 & 10\\
\cline{2-8}
\specialrule{0em}{1pt}{1pt}
  & \multirow{4}*{non-tomography} & bin11 & $10^{14.0}\le M_\mathrm{vir}(h^{-1}{\rm M}_{\odot})<10^{14.2}$ & $0.0\le z<0.55$ & $1/1.182$ & 1.245 & 21\\
  &  & bin12 & $10^{14.0}\le M_\mathrm{vir}(h^{-1}{\rm M}_{\odot})<10^{14.2}$ & $0.55\le z<1.5$ & $1/1.098$ & 1.328 & 21\\
  &  & bin21 & $M_\mathrm{vir}(h^{-1}{\rm M}_{\odot})\ge 10^{14.2}$ & $0.0\le z<0.55$ & $1/1.232$ & 1.290  & 20\\
  &  & bin22 & $M_\mathrm{vir}(h^{-1}{\rm M}_{\odot})\ge 10^{14.2}$ & $0.55\le z<1.5$ & $1/1.131$ & 1.340 & 8\\
\hline
\specialrule{0em}{0.5pt}{0.5pt}
\multirow{21}*{WZL} & \multirow{6}*{low-z tomography}& bin11 & $10^{14.0}\le M_\mathrm{vir}(h^{-1}{\rm M}_{\odot})<10^{14.2}$ & $0.0\le z<0.4$ & $1/1.155$ & 1.190 & 148\\
  &  & bin12 & $10^{14.0}\le M_\mathrm{vir}(h^{-1}{\rm M}_{\odot})<10^{14.2}$ & $0.4\le z<0.85$ & $1/1.080$ & 1.204 & 645\\
  &  & bin21 & $10^{14.2}\le M_\mathrm{vir}(h^{-1}{\rm M}_{\odot})<10^{14.5}$ & $0.0\le z<0.4$ & $1/1.196$ & 1.213 & 120\\
  &  & bin22 & $10^{14.2}\le M_\mathrm{vir}(h^{-1}{\rm M}_{\odot})<10^{14.5}$ & $0.4\le z<0.85$ & $1/1.115$ & 1.247 & 454\\
  &  & bin31 & $M_\mathrm{vir}(h^{-1}{\rm M}_{\odot})\ge 10^{14.5}$ & $0.0\le z<0.4$ & $1/1.282$ & 1.233 & 24\\
  &  & bin32 & $M_\mathrm{vir}(h^{-1}{\rm M}_{\odot})\ge 10^{14.5}$ & $0.4\le z<0.85$ & $1/1.227$ & 1.366 & 57\\
\cline{2-8}
\specialrule{0em}{1pt}{1pt}
  & \multirow{6}*{high-z tomography}& bin11 & $10^{14.0}\le M_\mathrm{vir}(h^{-1}{\rm M}_{\odot})<10^{14.2}$ & $0.85\le z<1.1$ & $1/1.061$ & 1.242 & 360\\
  &  & bin12 & $10^{14.0}\le M_\mathrm{vir}(h^{-1}{\rm M}_{\odot})<10^{14.2}$ & $1.1\le z<1.5$ & $1/1.053$ & 1.265 & 252\\
  &  & bin21 & $10^{14.2}\le M_\mathrm{vir}(h^{-1}{\rm M}_{\odot})<10^{14.5}$ & $0.85\le z<1.1$ & $1/1.096$ & 1.316 & 212\\
  &  & bin22 & $10^{14.2}\le M_\mathrm{vir}(h^{-1}{\rm M}_{\odot})<10^{14.5}$ & $1.1\le z<1.5$ & $1/1.069$ & 1.293 & 221\\
  &  & bin31 & $M_\mathrm{vir}(h^{-1}{\rm M}_{\odot})\ge 10^{14.5}$ & $0.85\le z<1.1$ & $1/1.185$ & 1.490 & 15\\
  &  & bin32 & $M_\mathrm{vir}(h^{-1}{\rm M}_{\odot})\ge 10^{14.5}$ & $1.1\le z<1.5$ & $1/1.145$ & 1.470 & 9\\
\cline{2-8}
\specialrule{0em}{1pt}{1pt}
   & \multirow{9}*{non-tomography}& bin11 & $10^{14.0}\le M_\mathrm{vir}(h^{-1}{\rm M}_{\odot})<10^{14.2}$ & $0.0\le z<0.5$ & $1/1.095$ & 1.132 & 259\\
  &  & bin12 & $10^{14.0}\le M_\mathrm{vir}(h^{-1}{\rm M}_{\odot})<10^{14.2}$ & $0.5\le z<1.0$ & $1/1.050$ & 1.146 & 779\\
  &  & bin13 & $10^{14.0}\le M_\mathrm{vir}(h^{-1}{\rm M}_{\odot})<10^{14.2}$ & $1.0\le z<1.5$ & $1/1.025$ & 1.122 & 367\\
  &  & bin21 & $10^{14.2}\le M_\mathrm{vir}(h^{-1}{\rm M}_{\odot})<10^{14.5}$ & $0.0\le z<0.5$ & $1/1.140$ & 1.173 & 169\\
  &  & bin22 & $10^{14.2}\le M_\mathrm{vir}(h^{-1}{\rm M}_{\odot})<10^{14.5}$ & $0.5\le z<1.0$ & $1/1.067$ & 1.165 & 554\\
  &  & bin23 & $10^{14.2}\le M_\mathrm{vir}(h^{-1}{\rm M}_{\odot})<10^{14.5}$ & $1.0\le z<1.5$ & $1/1.039$ & 1.153 & 284\\
  &  & bin31 & $M_\mathrm{vir}(h^{-1}{\rm M}_{\odot})\ge 10^{14.5}$ & $0.0\le z<0.5$ & $1/1.215$ & 1.183 & 28\\
  &  & bin32 & $M_\mathrm{vir}(h^{-1}{\rm M}_{\odot})\ge 10^{14.5}$ & $0.5\le z<1.0$ & $1/1.137$ & 1.239 & 64\\
  &  & bin33 & $M_\mathrm{vir}(h^{-1}{\rm M}_{\odot})\ge 10^{14.5}$ & $1.0\le z<1.5$ & $1/1.065$ & 1.178 & 13\\
\hline
\end{tabular}
\end{center}
\end{table*}

\subsection{Shear Measurement Bias and Photo-z Errors}
In order not to significantly bias the cosmological constraints, the shear measurements need to be accurate enough. For S16A, the requirements for the systematic uncertainty of the shear measurement bias are discussed from the point of view of galaxy-galaxy (g-g) lensing and shear two-point correlations \citep{Mandelbaum2018a}. Here we present the requirement on the systematic uncertainty of the multiplicative bias $\delta m$ from the peak count statistics accordingly. 

In our analyses, we reconstruct the convergence fields based on the shear fields generated from S16A data using Eq.(\ref{eq:smoothing}), where the multiplicative shear measurement biases are corrected for. If the corrections are not perfect with a residual $\delta m$, our theoretical model calculations without accounting for the hidden $\delta m$ will lead to biases of $(\delta m){\rm d}N^{(p)}/{\rm d}m$ in the predicted peak counts. 
They then propogate into biases in the cosmological parameter constraints. 
Based on this consideration, we can derive the requirement on $\delta m$ by setting the theoretical bias to be less than half of the statistical uncertainty. 
Specifically, we follow the methodology in \citet{Mandelbaum2018a} to use the total $\mathrm{S/N}|_{\rm total}$ for a conservative estimate, i.e.,  
\begin{equation}
|(\delta m)\frac{{\rm d}N^{(p)}_{\rm peak}(\nu_{\mathrm{N},i})}{{\rm d}m}|<0.5\frac{N^{(p)}_{\rm peak}(\nu_{\mathrm{N},i})}{S/N|_{\rm total}}
\label{bias}
\end{equation}
where the theoretical peak count derivatives are calculated at $(\Omega_{\rm m}, \sigma_8)=(0.332, 0.799)$ and $m=0$, and
\begin{equation}
\mathrm{S/N}|_{\rm total}=\bigg [\sum_{i,j=1}^{N_{\rm bin}}{N^{(p)}_{\rm peak}(\nu_{\mathrm{N},i})}(\widehat{C_{ij}^{-1}}){N^{(p)}_{\rm peak}(\nu_{\mathrm{N},j})}\bigg ]^{1/2}.
\label{totalSN}
\end{equation}
For our peak analyses here with the effective area of $\sim 58\deg^2$, Eq.(\ref{bias}) and (\ref{totalSN}) give rise to the requirement on $|\delta m|<0.041, 0.024, 0.015, 0.010, 0.026, 0.017, 0.012, 0.009$ for the $4\times 2$ bins in the tomographic case. The average is $|\delta m|<0.02$.  

For the HSC-SSP S16A shear catalog, extensive internal null tests and image simulation calibrations have been performed to quantify the shear measurement biases \citep{Mandelbaum2018a, Mandelbaum2018b}. It shows that the multiplicative biases of the shear measurement in S16A have the accuracy at 1\% level, which are 
within the range of the statistical requirement on $|\delta m|$ derived above.

To further quantitively evaluate the cosmological influence from $\delta m$, we carry out analyses including $\delta m$ in theoretical model predictions as a nuisance parameter to derive cosmological constraints from the observed peak counts. 
The results are shown explicitly in \S \ref{subsec:realresults}. 
No distinguishable differences are seen comparing to the fiducial constraints without considering $\delta m$. We thus conclude that the effect from the inaccuracy $\delta m$ is insignificant for our studies.

For the additive shear bias, in all the fields, it is at the level of $\lesssim 0.1$\% \citep{Mandelbaum2018a}. Considering our bin width of $\Delta \nu_\mathrm{N}=0.5$ in peak counting, 
the effect of the additive bias should be negligible.

For the photo-z measurements, in \citet{Tanaka2018}, they use six independent codes, including template-fitting, empirical-fitting, and machine-learning techniques, and compare their performance. To the first order, all the codes perform consistently well. They show that photo-zs are most accurate in the range of $0.2\le z_{\rm p}\le 1.5$ with $\sigma[\Delta z_{\rm p}/(1+z_{\rm p})]\sim0.05$ and an outlier rate of about $15\%$ for galaxies down to $i=25$. For the DEmP photo-z catalog used in our study, it has a generally flat probability integral transform (PIT) distribution and a small mean continuous ranked probability score (CRPS), indicating a well calibrated full redshift probability distribution function $p(z)$ \citep{Polsterer2016, Tanaka2018}. 

To quantify the effects from photo-z errors on cosmological constraints, we consider the possible existence of systematic bias of photo-z in the two redshift bins, $\Delta z_\mathrm{L}$ and $\Delta z_\mathrm{H}$, and include them as free parameters in deriving cosmological constraints. 
The results are also shown in \S \ref{subsec:realresults}. It is seen that with the two wide redshift bins used in our tomographic analyses and the relatively small effective area, the impact of the photo-z errors on our analyses is negligible.

\begin{figure*}
\centering
\includegraphics[width=1.0\columnwidth]{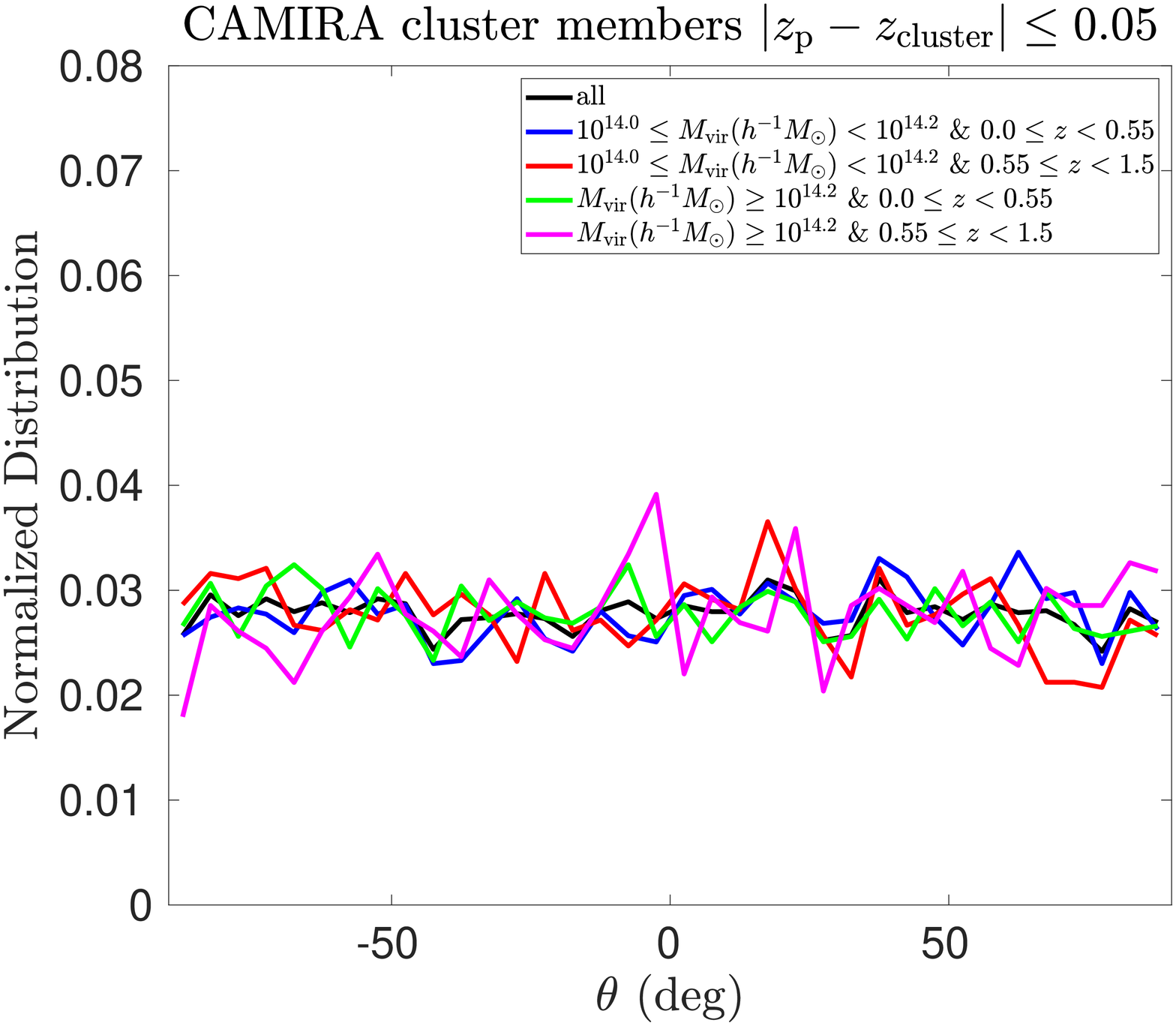}
\includegraphics[width=1.0\columnwidth]{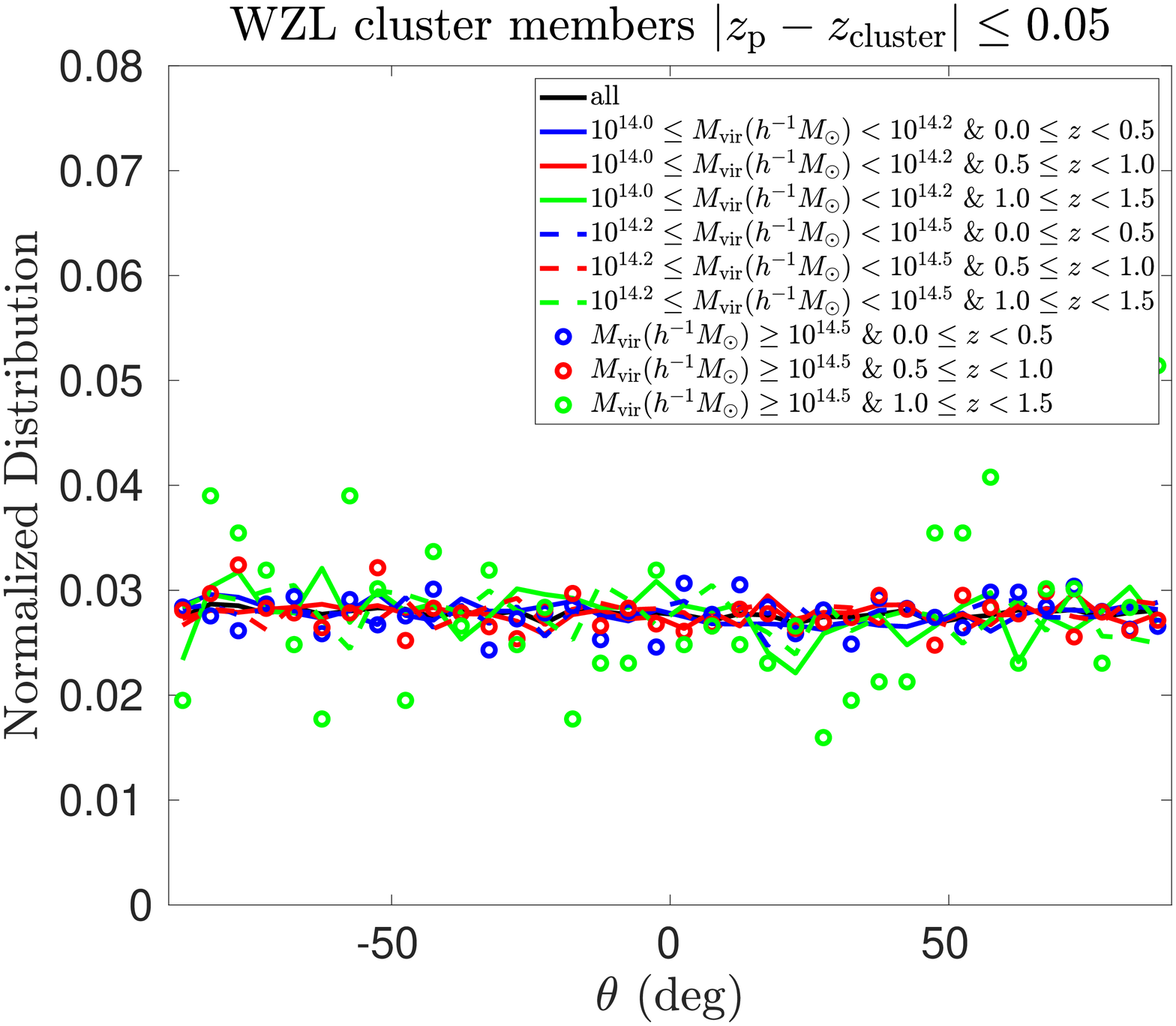}
\caption{Distributions of the angle between the projected major axis orientation of a satellite galaxy and its projected radial direction to the center of its host cluster in different mass and redshift bins with the left and right from CAMIRA and WZL, respectively.
}
\label{fig:IA}
\end{figure*}

\begin{figure}
\centering
\includegraphics[width=1.0\columnwidth]{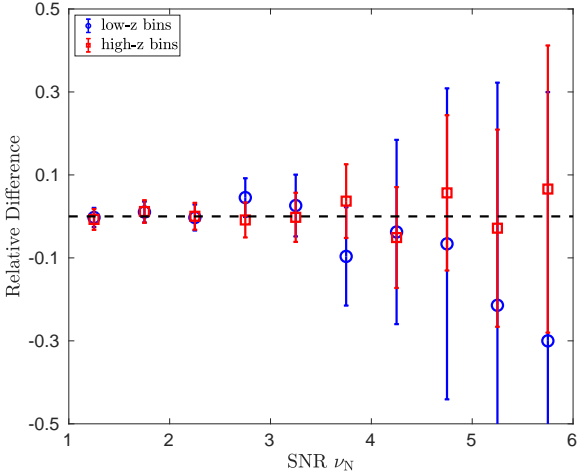}
\caption{Relative differences of the peak counts with IA and without IA for the low-z and high-z bins, considering a case with the alignment angle of satellites toward their central galaxies to be $\theta_{\rm {IA}}=30^\circ$. The error bars are Poisson errors with a survey area $\sim 58\deg^2$, about the same area as we used in the study here. 
}
\label{fig:RD_IA}
\end{figure}

\subsection{Boost Factor and Dilution Effect} \label{subsec:boost}
It is seen from Figure \ref{fig:massREC_example} that cluster member galaxies in a shear sample can lead to significant effects on peak statistics, particularly on high peaks, primarily due to two aspects. One is that the concentration of the member galaxies of a cluster alters the local redshift distribution of the source galaxies around the cluster, resulting in a dilution of the cluster lensing signal \citep{Mandelbaum2006, Miyatake2015, Dvornik2017}. Secondly, the clustering of the member galaxies increases the local number density of source galaxies and reduces the local shape noise \citep{Shan2018}. The enhancement of source galaxies around a cluster is termed as the boost factor \citep{Clowe2001,Kacprzak2016, Zrcher2022}. 

Here we follow the approach of \citet{Shan2018} to correct for the dilution resulting from the boost factor in our theoretical peak model. To obtain the boost factor around clusters of galaxies, we analyze the filling factor of source galaxies in the regions of known clusters or cluster candidates in the HSC-SSP fields. Two cluster catalogs are used separately, the CAMIRA clusters from \citet{Oguri2018Cluster}, and the WZL clusters from \citet{Wen2021}. For CAMIRA clusters, we adopt the richness-mass relation in \citet{Oguri2018Cluster} to obtain $M_{\rm 200c}$ for each cluster, and then the corresponding $M_{\rm vir}$ using the Navarro-Frenk-White (NFW) halo density profile \citep{NFW1996, NFW1997} with the same mass-concentration relation given in \citet{Duffy2008} as in our peak model calculations. Here $M_{\rm 200c}$ is the mass of a cluster within the region with the average density of 200 times of the critical density of the Universe. For WZL clusters, we use directly $M_{500{\rm c}}$ listed in their cluster catalog to obtain $M_{\rm vir}$. In accord with our analyses, we consider clusters with $M_\mathrm{vir}\ge 10^{14}h^{-1}M_\odot$ and redshift $z\le1.5$. Clusters with masks within their virial radius are excluded in boost factor analyses to avoid 
bias from masks. We divide the remaining clusters into different mass and redshift bins as detailed in Table \ref{tab:boost} for different cases, and calculate the boost factor by estimating the excess filling factor (galaxy number overdensity relative to the mean) distribution around cluster candidates. The results are shown in Figure \ref{fig:excessfilling}. The mean excess fractions for different bins are explicitly listed in the second to last column of Table \ref{tab:boost}. The choices of the mass and redshift bins for clusters primarily are based on the balance between the number of bins and the number of clusters in each bin. 
The two cluster catalogs, CAMIRA and WZL, contain different numbers of cluster candidates, and cover slightly different ranges of mass and redshift. We thus adopt different binnings as listed in Table \ref{tab:boost}. 
The consistency of the cosmological constraints derived from CAMIRA and WZL boost corrections shown in \S \ref{subsec:realresults} demonstrates that the cosmological results here are not very sensitive to the specific choices of mass and redshift bins.

With the information of boost factor, we then evaluate how it affects the WL signal in halo regions. For that, we perform simulation analyses by picking out a halo with a typical mass and redshift within each bin. We model the halo with the NFW profile, and center it in a $1.2\times 1.2 \deg^2$ field. By randomly sampling source galaxies using the corresponding number densities and global redshift distributions of low-z, high-z and non-tomography samples, we mimic the case without the boost effect. For these mock source galaxies, we add the reduced shear signals from the central dark matter halo, resulting in the `no boost' mocks. Based on these mocks, we then construct `boost' mocks by resampling source galaxies following the excess galaxy number density profile shown in Figure \ref{fig:excessfilling}. For these cluster member galaxies, no shear signals are added because they are not subject to the lensing effect from their own halo \citep{Sifon2015}. For each halo, we generate 1000 noiseless `no boost' and the corresponding `boost' mocks, and reconstruct the convergence field for each. The ratio between the convergence value of the central peak with and without the boost effect is calculated for each halo, and the mean ratio is obtained by averaging over the 1000 realizations.
These mean ratios give rise to the average dilution factors in each bin for different cases, and are listed also in Table \ref{tab:boost}. They are used in our model calculations.

For the effect of boost factor on the shape noise, it is noted that the number density of source galaxies is enhanced in cluster regions, and consequently reduced in the field regions comparing to the overall average density. Denoting $n_{\rm g}$, $n_{\rm g}^{\rm halo}$ and $n_{\rm g}^{\rm field}$ as the source number density of global average, in halo regions and in the field regions, respectively, we have 
\begin{equation}
n_{\rm g}S_{\rm eff}=\sum n_{\rm g}^{\rm halo}{S_{\rm eff}^{\rm halo}} + n_{\rm g}^{\rm field}S_{\rm eff}^{\rm field},
\end{equation}
where $S_{\rm eff}$, $S_{\rm eff}^{\rm halo}$ and $S_{\rm eff}^{\rm field}$ are the total effective area, the 
area occupied by halos, and the left-over field area with $S_{\rm eff}^{\rm field}=S_{\rm eff}-\sum S_{\rm eff}^{\rm halo}$, respectively. Both $\sum S_{\rm eff}^{\rm halo}$
and $S_{\rm eff}^{\rm field}$ are cosmology dependent. From $n_{\rm g}^{\rm halo}$ and $n_{\rm g}^{\rm field}$,
we can compute the shape noise levels $\sigma_\mathrm{N,0}^{\rm halo}$ and $\sigma_\mathrm{N,0}^{\rm field}$, respectively, for model calculations.

With the above ingredients, we modify our model calculations including the boost effect in confronting with real observed peak counts (See \S\ref{subsec:realresults}) as follows: 
(1) Halos in different mass and redshift bins are considered separately, and the corresponding average dilution factor is included in modifying the convergence field of the halo. 
The noise level $\sigma_\mathrm{N,0}^{\rm halo}$ is also adjusted accordingly. Then Eq.(\ref{eq:npk_c}) is used to calculate the number of peaks in halo regions.
(2) With the modified noise level $\sigma_\mathrm{N,0}^{\rm field}$ in the field region, we use Eq.(\ref{eq:npk_n}) to calculate peaks in field.

\subsection{Intrinsic Alignments} \label{subsec:IA}
Affected by the same environment of large-scale structures, the orientation of nearby galaxies can be correlated, which is referred to as the intrinsic alignment (IA) \citep{Joachimi2015, Kiessling2015}. The IA signals contain rich information of galaxy formation and evolution. On the other hand, they constitute a major contamination to WL analyses \citep{Troxel2015}. For cosmic shear correlation studies, they need to be included in deriving cosmological constraints \citep{Hildebrandt2017, Hikage2019, Asgari2021, Amon2022}.

The influences of IA on peak statistics include an additional contribution to the shape noise variance, i.e., $\sigma_{\mathrm{N},0}^2=\sigma_{\mathrm{N},0,\rm ran}^2+\sigma_{\mathrm{N},0,\rm  corr}^2$, 
where $\sigma_{\mathrm{N},0,\rm ran}$ is the shape noise contributed from the randomly oriented source galaxies, and $\sigma_{\mathrm{N},0,\rm corr}$ denotes the additional term from IA \citep{Fan2007}.  
In the low-z, high-z and non-tomographic cases, we have $\sigma_{\mathrm{N},0,\rm ran}^2 \approx 5.29\times 10^{-4}$, $8.41\times 10^{-4}$ and $3.24\times 10^{-4}$. 
To estimate $\sigma_{\mathrm{N},0,\rm  corr}^2$, we adopt the commonly used nonlinear tidal alignment model \citep{Hirata2004, Bridle2007} with the IA amplitude $A_{\rm IA}\sim 0.91$ from \citet{Hamana2020} and the smoothing scale $\theta_\mathrm{G}=1.5$ arcmin. 
For the II contributions to $\sigma_{\mathrm{N},0,\rm  corr}^2$, we obtain $\sim 1.6\times 10^{-6}$, $2.5\times 10^{-7}$, and $7.0\times 10^{-7}$ for the three cases, respectively. For the GI contributions, the estimated values are
$-1.5\times 10^{-5}$, $-3.7\times 10^{-6}$, and $-1.5\times 10^{-5}$ where the negative sign is due to the anti-correlations of the GI term. All of them are at least 20 times smaller than the corresponding $\sigma_{\mathrm{N},0,\rm ran}^2$. Taking into account 
also the relatively large statistical uncertainties of the peak counts in our analyses, the IA contributions to the noise are negligible.  

Besides the noise contributions, IA of satellite galaxies in a shear sample can bias the WL peak signals of their host clusters, therefore affecting peak statistics, particularly high peaks \citep{Kacprzak2016, HarnoisDeraps2022}. 
This bias depends on the level of satellite IA. With certain tidal models, it can be as large as $\sim 30\%$ \citep{HarnoisDeraps2022}. On the other hand, there are observational evidences showing no significant satellite IA signals \citep{Chisari2014, Sifon2015}.

To evaluate the effect of satellite IA on our peak analyses, we first attempt to measure the satellite orientation distributions in the S16A shear sample using the cluster catalogs of CAMIRA and WZL.   
We define satellite galaxies of a cluster to be the ones with their projected distances to the center of their host less than the cluster virial radius and $|z_{\rm p}-z_{\rm cluster}|\le 0.05$ considering the photo-z scatters \citep{Wen2021}. The angle between the projected major axis orientation of a satellite galaxy and its projected radial direction to the center of its host cluster is then calculated, where the angle is positive/negative if the major axis of a satellite is within/beyond $90$ degrees counterclockwise with respect to it radial direction. 
Figure \ref{fig:IA} shows the angle distributions in different mass and redshift bins for CAMIRA (left) and WZL (right) clusters. It is seen that in all the cases, the distributions are nearly flat, indicating no detectable satellite IA signals in the S16A shear sample. 

Because of the photo-z errors, the above analyses may suffer from non-member contaminations and thus underestimate the satellite alignment signals. We therefore further test the potential IA effects based on simulation analyses. In \citet{Zhang2022}, we perform systematic studies about the IA impacts on peak statistics using large simulations including semi-analytical galaxy formation, 
which give rise naturally central and satellite galaxies \citep{Wei2018b}. There we pay particular attention to how satellite IA affects high peak statistics by constructing different satellite IA samples with different alignment angles. 
For the purpose of evaluating the potential IA effects on our studies here using HSC-SSP S16 data, 
we construct shear samples from this set of simulations with the redshift distributions in accord with those of low-z and high-z samples used in our analyses. For IA settings, the early-type central galaxies follow their host halo orientations. 
For late-type centrals, their angular momentum directions trace that of their host halos \citep{Wei2018b}. For satellite galaxies, \cite{Wei2018b} considered two cases of alignments, perfectly radially aligned toward their central galaxies and 
purely random. They show that in comparison with the observed cosmic shear power spectra, the first satellite IA setting predicts too high signals at small scales while the second model is more consistent with the observational results. 
Here to evaluate the IA impacts, we consider a case in between the two with the alignment angle of satellites toward their central galaxies to be $\theta_{\rm {IA}}=30^\circ$.

In Figure \ref{fig:RD_IA}, we show the relative differences of the peak counts with IA and without IA for the low-z and high-z bins, respectively. The error bars are for a survey area $\sim 58\deg^2$, similar with that used in this study. It is seen that the IA effects are smaller for the high-z bins than that for the low-z bins. This is 
due to the less number of satellite galaxies of foreground clusters in the high-z shear sample than in its low-z counterpart, in accord with our diluation effect analyses shown in \S \ref{subsec:boost}. For the peaks we considered, the central values of the IA induced relative differences 
are $-9.64\%$, $-3.76\%$, $-6.62\%$ and $-21.43\%$ in the low-z case for the 4 bins of $\nu_{\rm N}$, respectively. They are $+3.68\%$, $-5.09\%$, $+5.67\%$ and $-2.84\%$ for the high-z samples. For both bins, the IA impacts are within $1\sigma$.

We employ these ratios to evaluate the IA impacts on our cosmological constraints. Specifically, we apply the ratios to adjust our model predictions, and then compare with the observed peak counts. 
The derived cosmological constraints are shown in \S \ref{subsec:realresults}. It is seen that the IA impacts are within $1\sigma$ for our analyses.

We note that with the increase of the sky coverage and depth and thus the decrease of the statistical errors, the IA effects need to be carefully treated in future peak analyses. 
Our studies in Zhang et al. (2022, submitted) investigate systematically the dependence of IA impacts on satellite alignments and on redshift distributions of shear samples.   
The results can potentially be implemented in our theoretical model for high peaks, which in turn can allow us to derive unbiased cosmological constraints and the IA information simultaneously from observed peaks using future large WL surveys.

\begin{figure*}
\centering
\includegraphics[width=1.0\columnwidth]{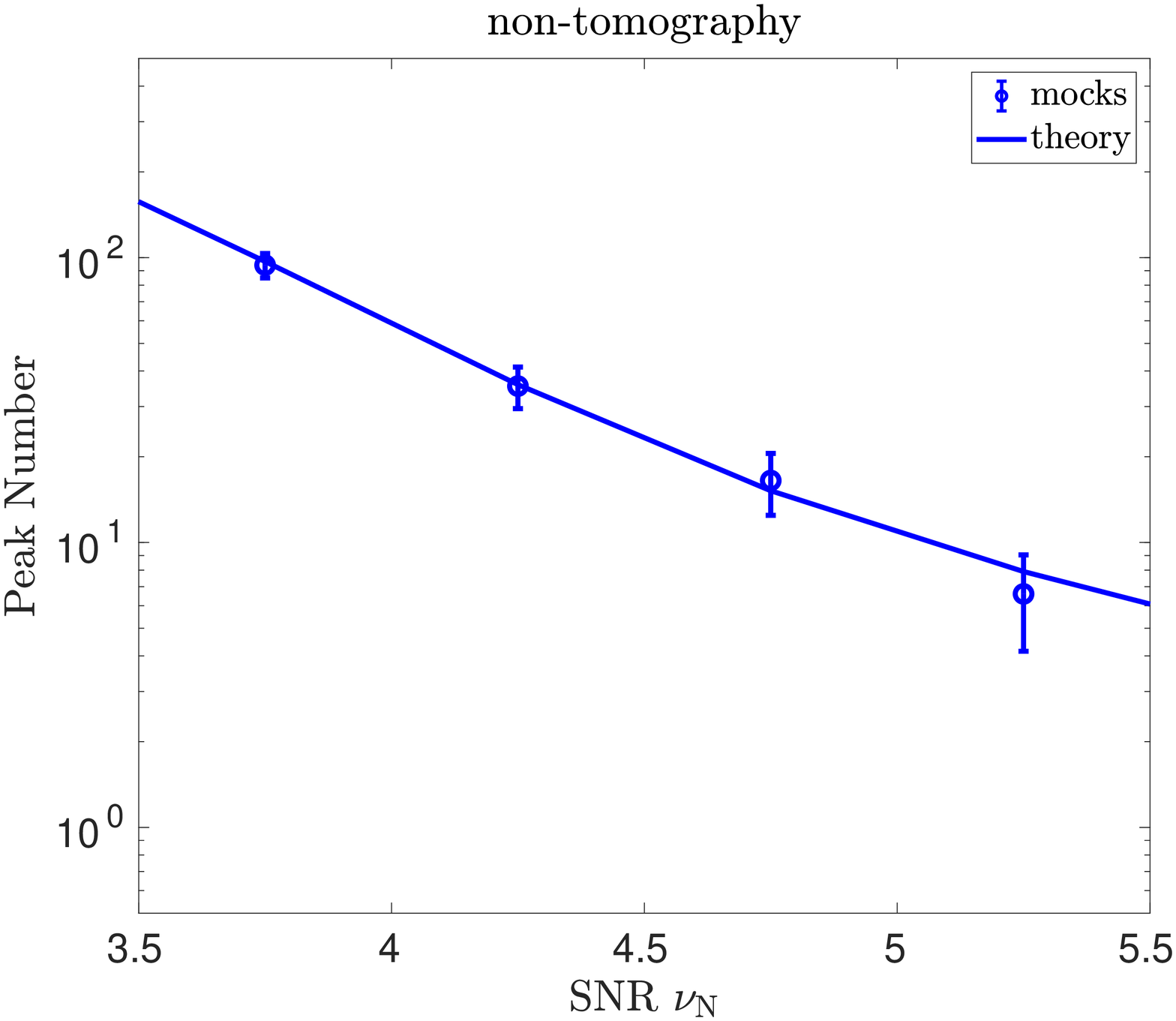}
\includegraphics[width=1.0\columnwidth]{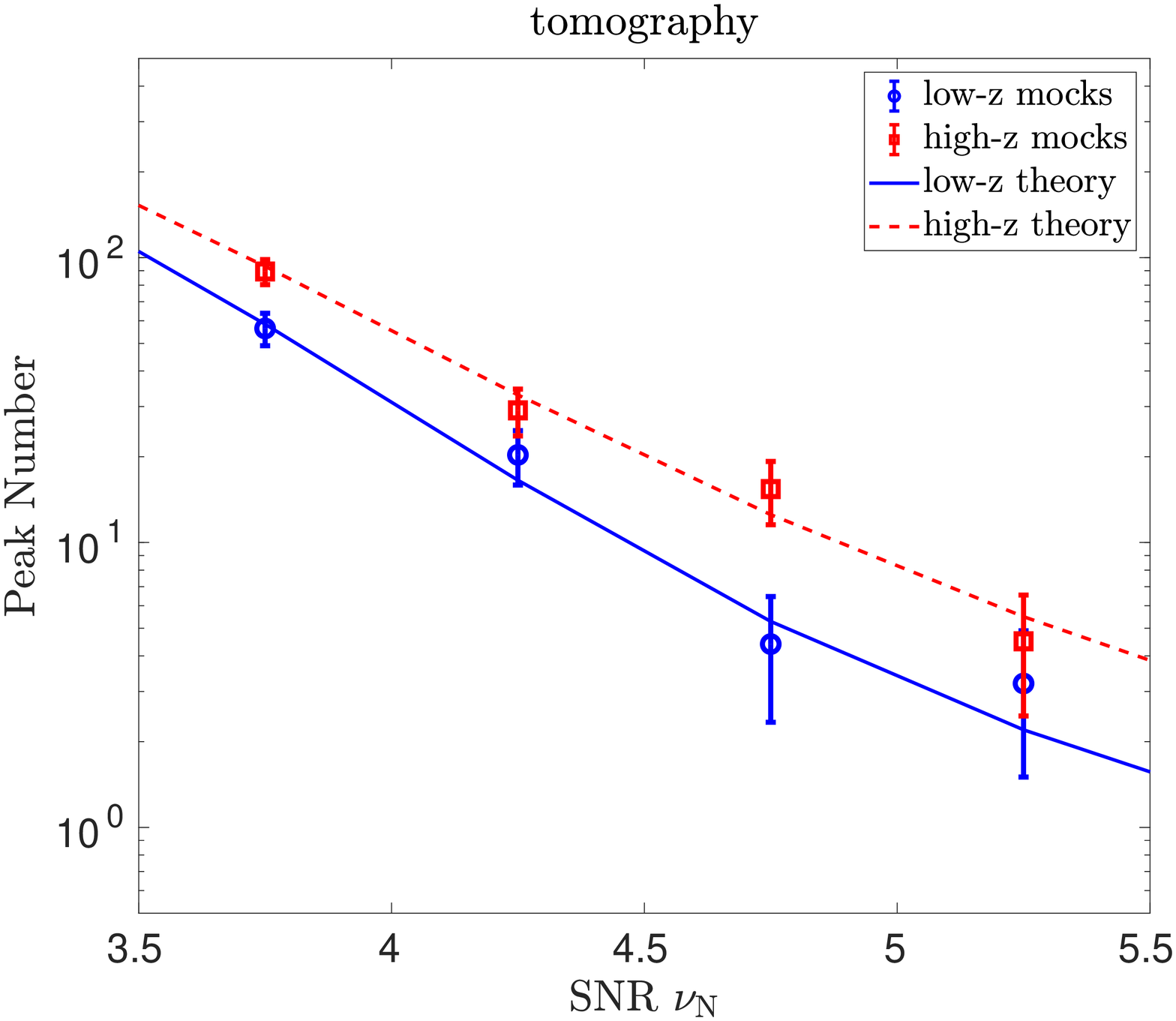}
\includegraphics[width=1.0\columnwidth]{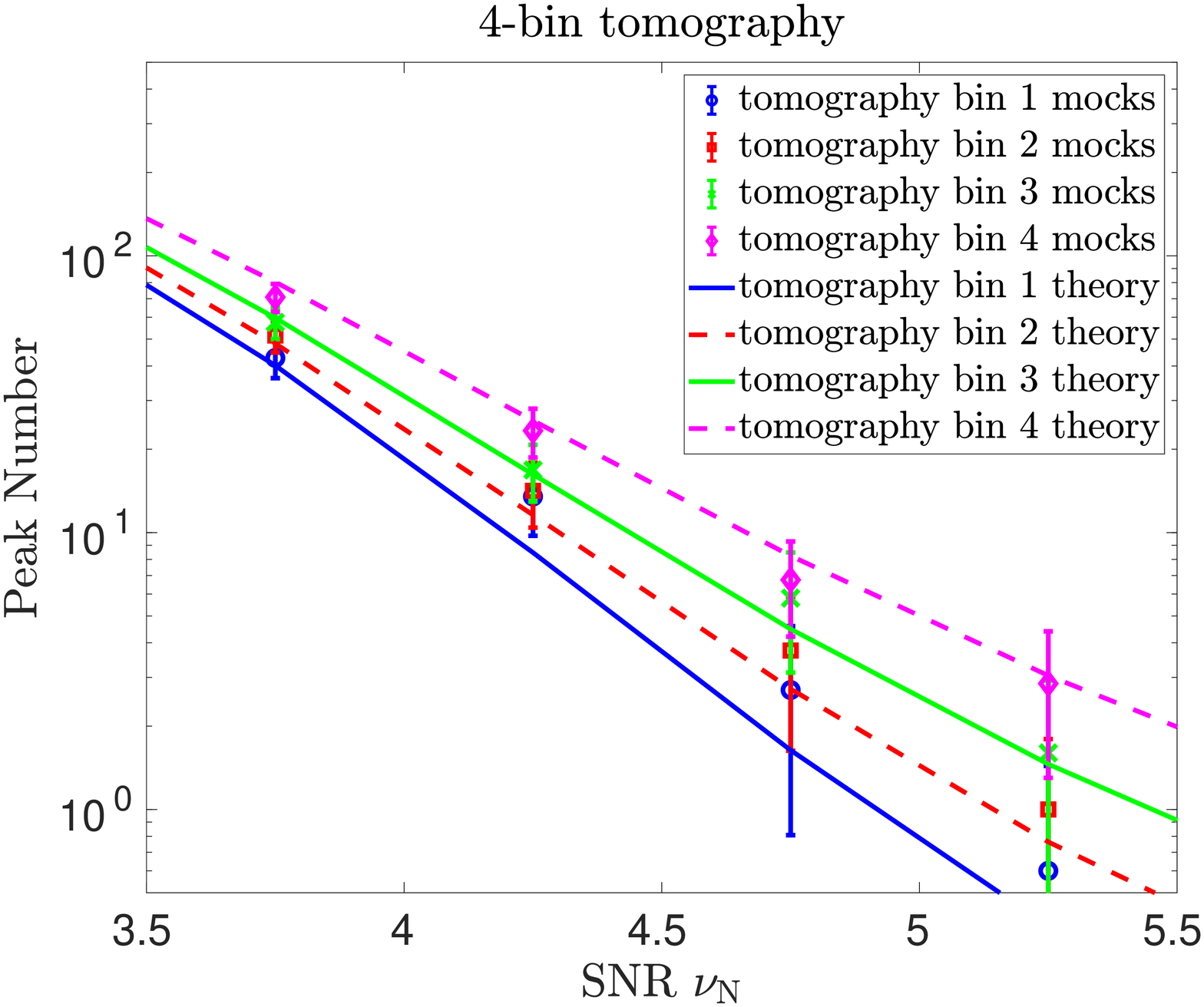}
\includegraphics[width=1.0\columnwidth]{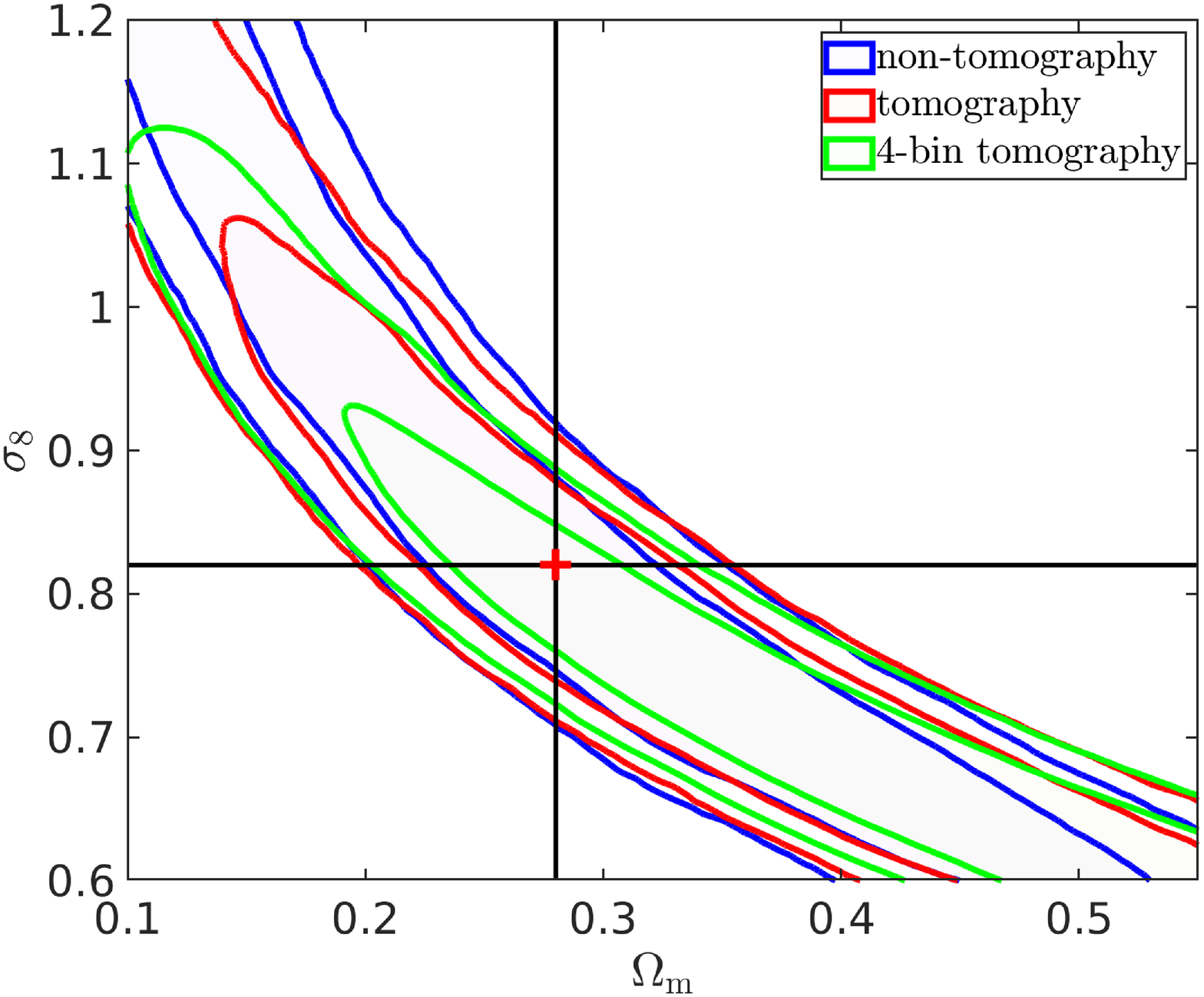}
\includegraphics[width=1.0\columnwidth, height=0.6\columnwidth]{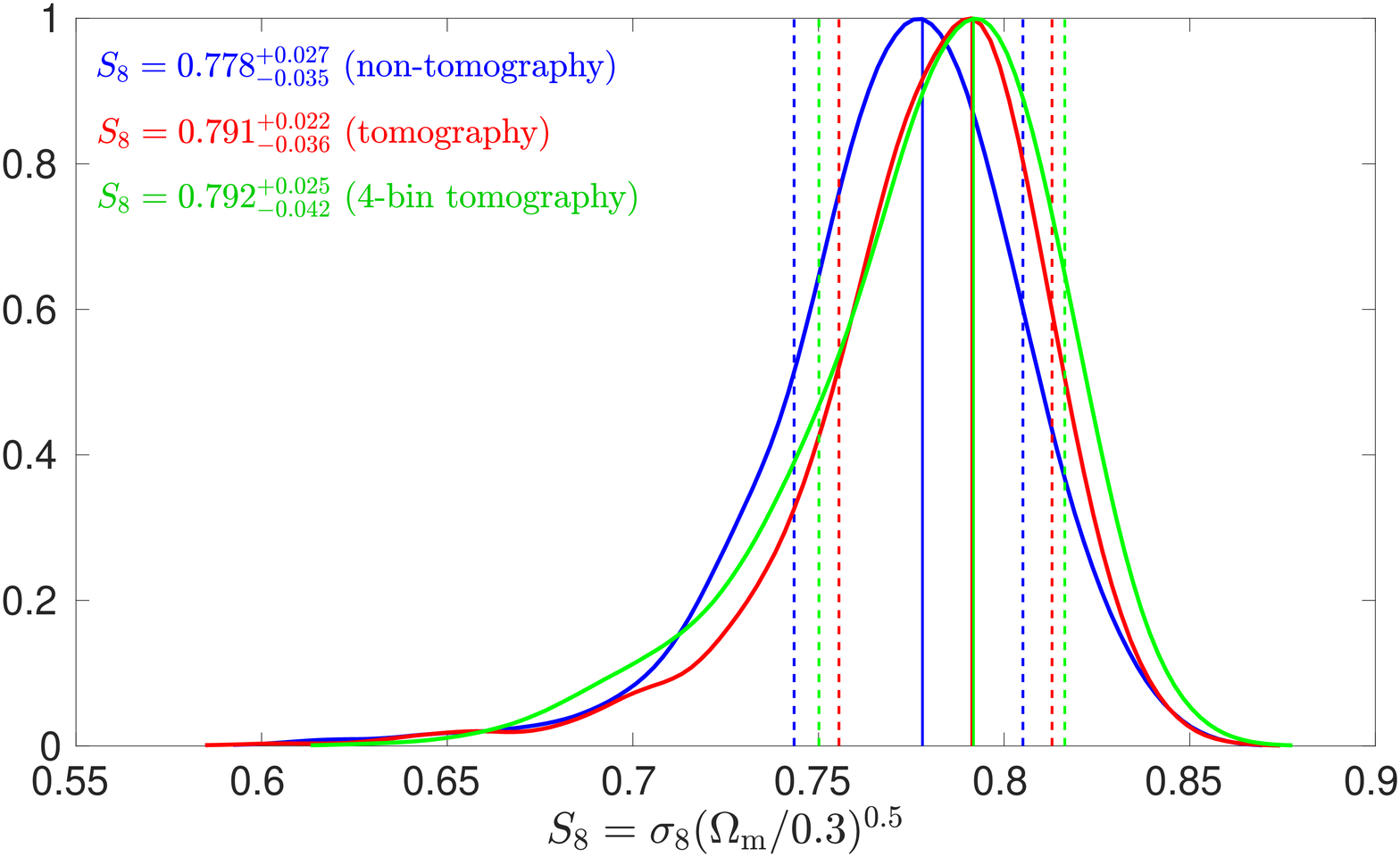}
\includegraphics[width=1.0\columnwidth, height=0.6\columnwidth]{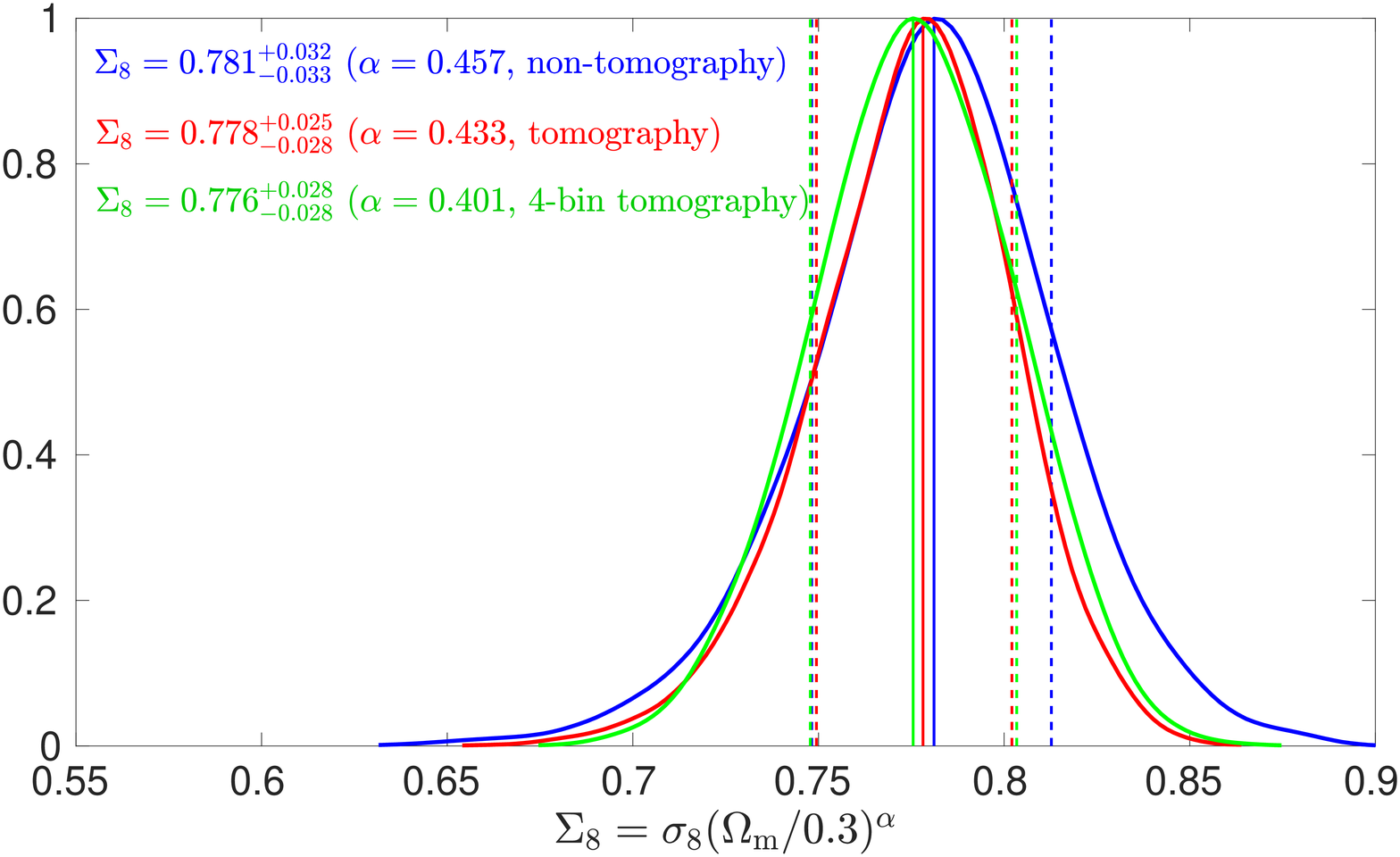}
\caption{Mock validation results. The first three panels are the peak number distributions for the non-tomographic (upper-left), 2-bin tomographic (upper-right) and 4-bin tomographic (middle-left) cases, respectively. The theoretical predictions are shown by the corresponding lines. The middle-right panel is for the corresponding cosmological constraints with the blue, red and green regions for the non-tomographic, 2-bin and 4-bin tomographic cases, respectively, and the red `+' symbol denotes the input cosmological parameters 
with ($\Omega_{\rm m}$, $\sigma_8$)=($0.28, 0.82$). In the last row, the derived $S_8$ (left) and $\Sigma_8$ (right) distributions of the non-tomographic (blue), 2-bin (red) and 4-bin tomographic (green) cases are presented explicitly for comparison.
}
\label{fig:mockresults}
\end{figure*}

\subsection{Baryonic Effects} 
Baryonic physics beyond gravity can affect the mass distribution of clusters of galaxies, and thus the WL high peak studies. The effects are however complex, including heating, cooling and feedbacks from stars and active galactic nuclei (AGN). Some of them have counter impacts on the mass distributions. Thus the results from various studies have not reached a consensus about the baryonic effects on WL peak studies, which depend sensitively on the detailed physics considered. 

By manually steepening the halo density profile to mimic the baryonic cooling and concentration effects, \citet{Yang2013} claim that baryonic effects could result in an increase for high peak counts. A few recent studies find that the overall baryonic effects may suppress $\sim 5-10\%$ in peak counts for both positive and negative tails, depending on the strength of the feedbacks implemented in the hydrodynamic simulations \citep{Coulton2020, Osato2021, Martinet2021b}. 
The bias from the baryonic effects is within $1\sigma$ for the current surveys, and will become significant for Stage IV surveys \citep{Shan2018, Martinet2021b}. 

In \citet{Weiss2019}, they adopt a theoretical model to modify the total mass distribution of halos to investigate the potential impact of baryonic effects on WL peak statistics. Following \citet{Schneider2019}, the total density distribution of a halo includes
the collisionless components of dark matter and satellite galaxies, the gas component and the central galaxy contribution. For the collisionless part, they are modeled as the truncated NFW profile corrected by the adiabatic relaxation. The gas distribution
is taken to be consistent with X-ray observations and modeled using a power-law decrease profile with the power index depending on the halo mass and a truncation at some gas ejection radius. For the central galaxy component, it is described by a power-law distribution with an exponential 
cut off, which only affects the very central region of the halo and has little impact on WL peaks \citep{Weiss2019}. For WL peaks, in general, the modified profiles of halos lead to decreases of the numbers of high peaks and increases of low peaks.

Comparing to the two survey configurations, DES and {\it Euclid}, considered in \citet{Weiss2019}, the HSC-SSP data used here has an effective galaxy number density in between of the two with $n_{\rm g}\sim 18.5 \hbox{ arcmin}^{-2}$. 
The galaxy redshift distribution is close to that of {\it Euclid}, but the effective area in our peak analyses is $\sim 58\hbox { deg}^2$, about 300 times smaller than $20000 \hbox { deg}^2$ for {\it Euclid} used in \citet{Weiss2019}. Consequently, 
the parameter $s_{\rm bcm}=(N_{\rm bcm}-N_{\rm dmo})/\sigma_{\rm dmo}$ adopted to measure the baryonic effects on peak statistics would decrease by a factor of $\sim \sqrt{300}\sim 17$ in our case comparing to that of {\it Euclid} without taking into account the difference of $n_{\rm g}$.
Here $N_{\rm bcm}$ and $N_{\rm dmo}$ are the peak numbers with and without modified halo profiles, respectively, and $\sigma_{\rm dmo}$ is the statistical error of the peak count $N_{\rm dmo}$ and $\sigma_{\rm dmo}\approx \sqrt{N_{\rm dmo}}$ under the Poisson error approximation.
From the results shown in Figure 7 of \citet{Weiss2019}, we see that decreasing $s_{\rm bcm}$ by a factor of $\sim 17$ would make it smaller than unity over the entire peak range. Considering the smaller $n_{\rm g}$ than that of {\it Euclid}, we expect even smaller $s_{\rm bcm}$.
Thus the baryonic effects on our peak analyses should be insignificant.

Additionally, assuming the baryonic effects mainly affect the density profile of halos, we also evaluate their impacts by allowing the amplitude of the mass-concentration (M-c) relation of dark matter halos to vary simultaneously 
with the cosmological parameters in the fitting \citep{Shan2018}. In this work, we adopt the M-c relation from \citet{Duffy2008}, given as 
\begin{equation}
c_\mathrm{vir}=\frac{A}{(1+z)^{0.7}}\left(\frac{M_{\mathrm{vir}}}{10^{14}h^{-1}M_\odot}\right)^{-0.081}.
\end{equation}
As presented in \S \ref{subsec:realresults}, the cosmological parameter constraints obtained by including $A$ as a free parameter show little differences with the results of our fiducial analyses, demonstrating further the negligible baryonic effects on our studies here.

\begin{figure*}
\centering
\includegraphics[width=1.0\columnwidth]{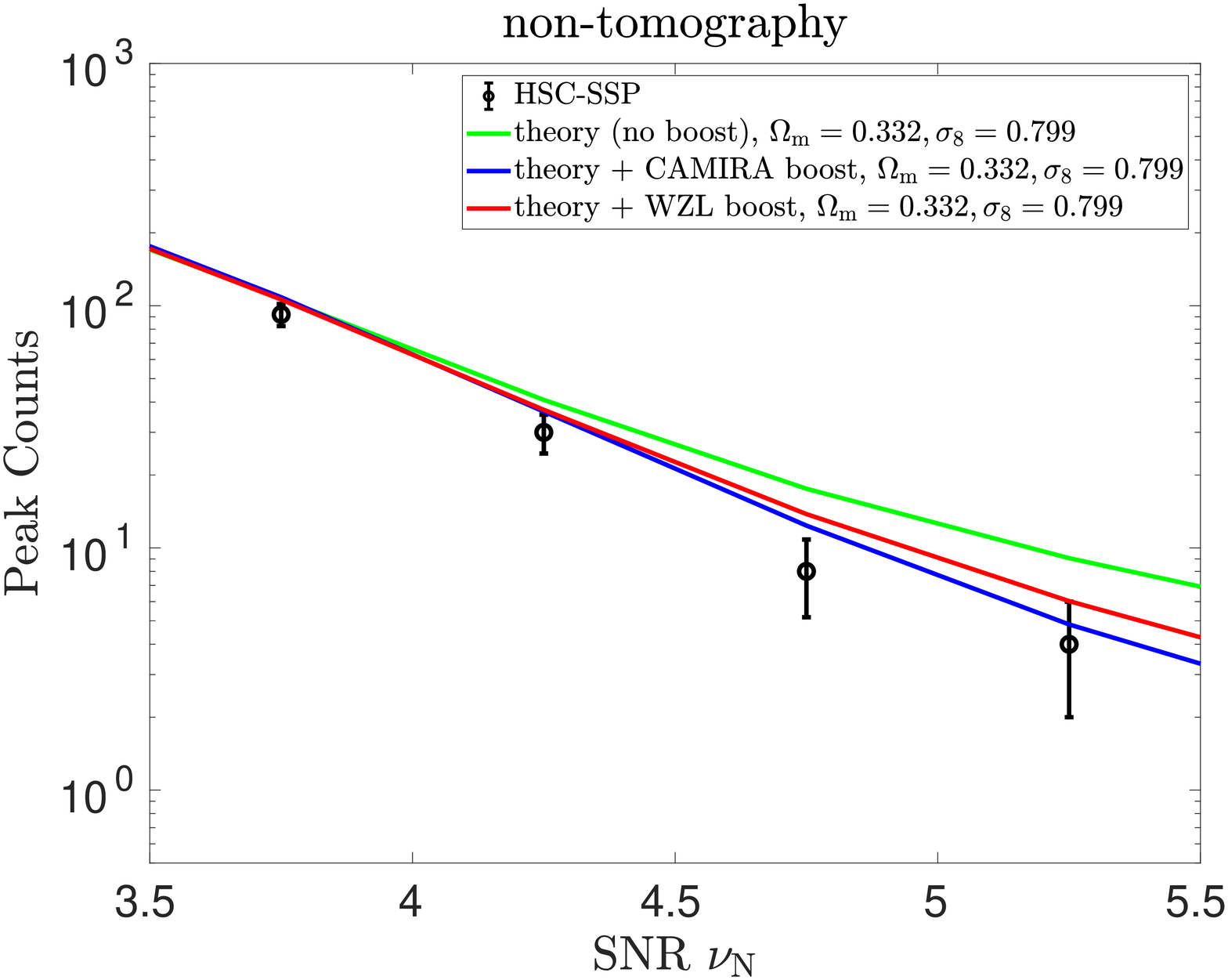}
\includegraphics[width=1.0\columnwidth]{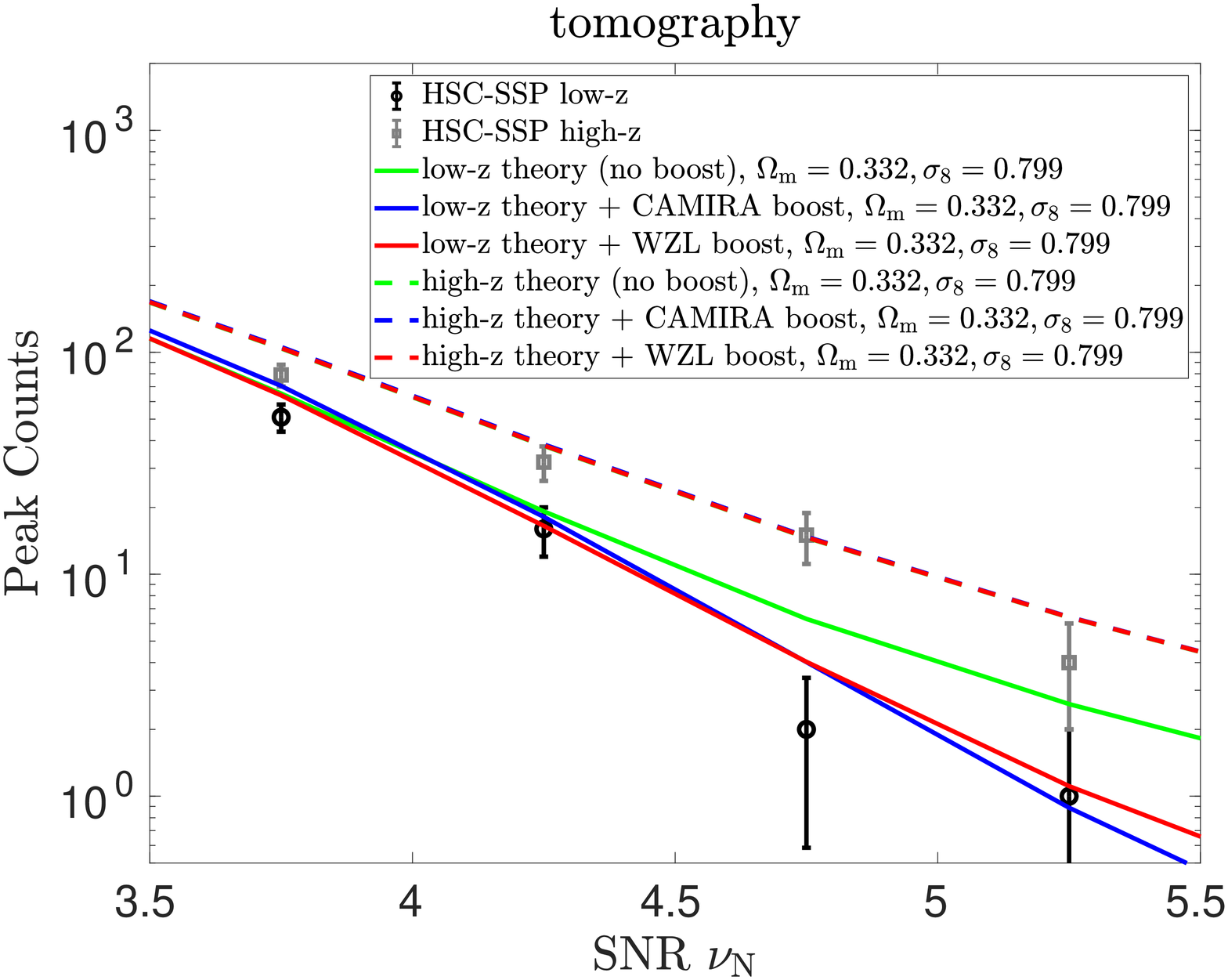}
\includegraphics[width=1.0\columnwidth]{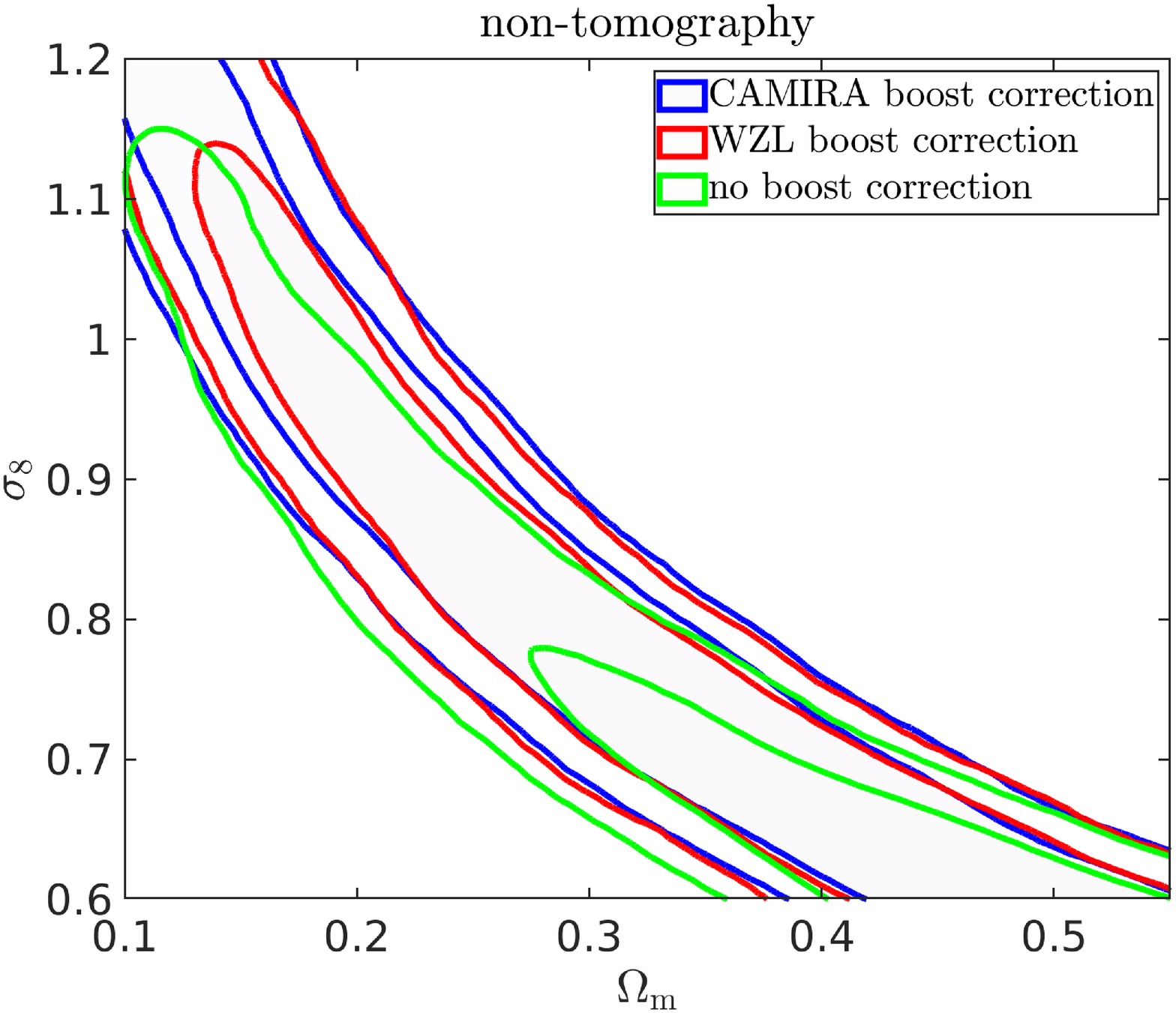}
\includegraphics[width=1.0\columnwidth]{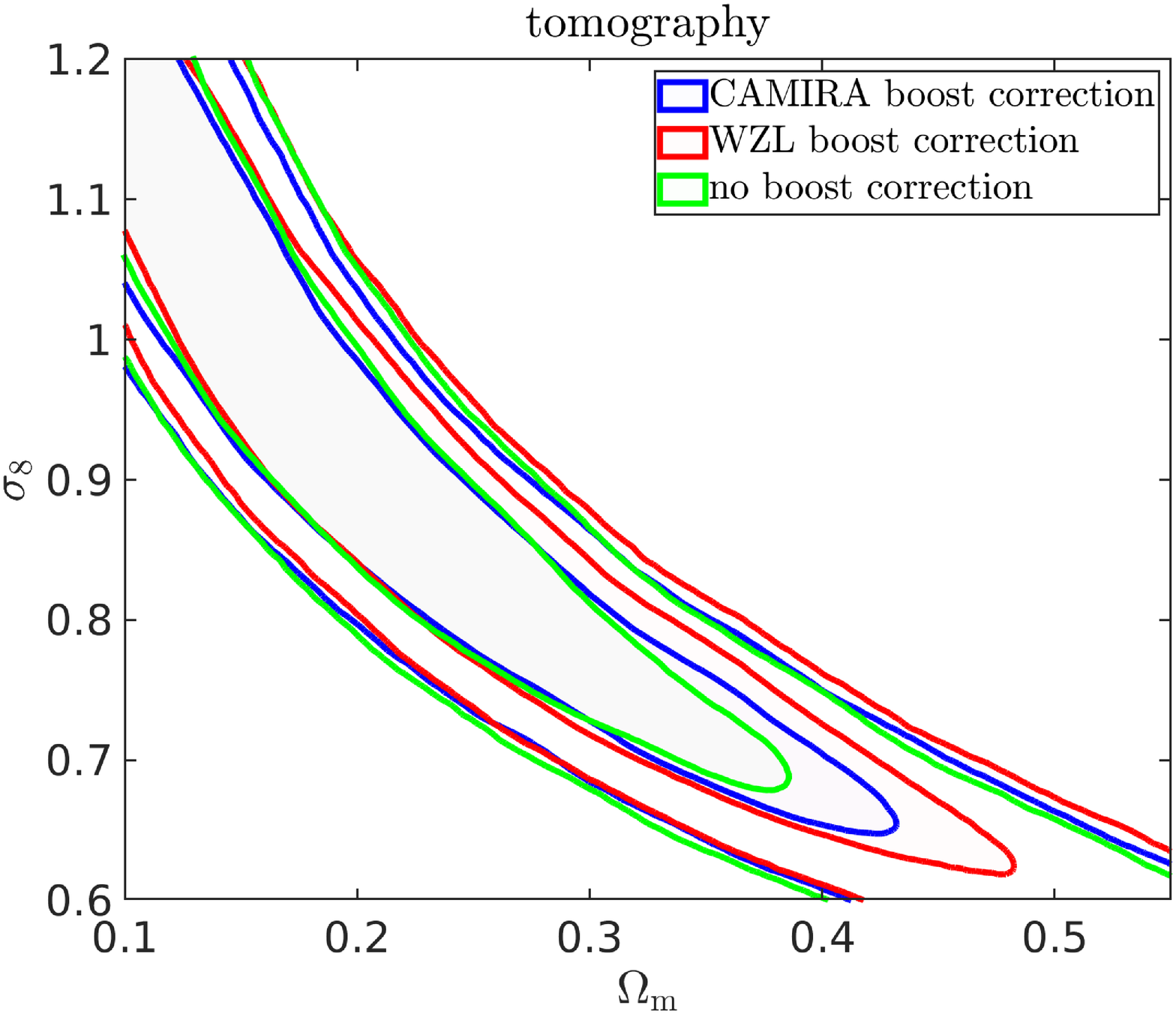}
\caption{Observational results. The error bars in the upper panels are from the scaled covariance matrix as explained in \S\ref{subsec:fitting}. Theoretical results with and without boost corrections at the cosmological model with ($\Omega_{\rm m}$, $\sigma_8$)=($0.332, 0.799$) from the marginalized constraints of the HSC shear correlation analyses in \citet{Hamana2020, Hamana2022} are shown in colored lines. The bottom panels are the corresponding cosmological constraints with CAMIRA (blue), WZL (red) boost correction and no boost correction (green).
}
\label{fig:obsresults}
\end{figure*}

\begin{figure}
\centering
\includegraphics[width=1.0\columnwidth]{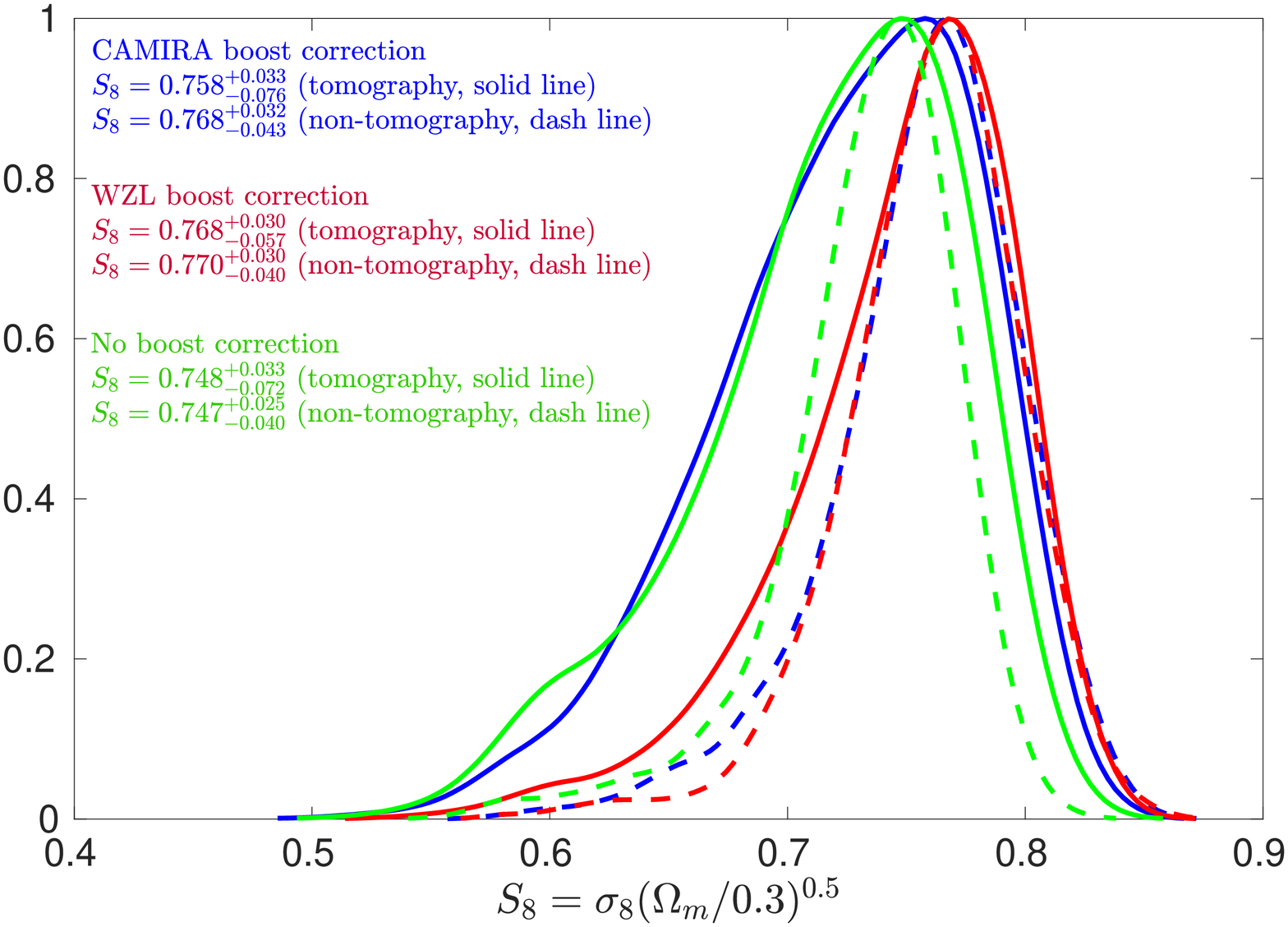}
\caption{The marginalized probability distribution of $S_8$ for tomographic (solid lines) and non-tomographic (dash lines) cases, with blue, red and green representing the constraints with CAMIRA, WZL boost correction and no boost correction, respectively.
}
\label{fig:S8}
\end{figure}

\begin{figure*}
\centering
\includegraphics[width=1.0\columnwidth, height=0.6\columnwidth]{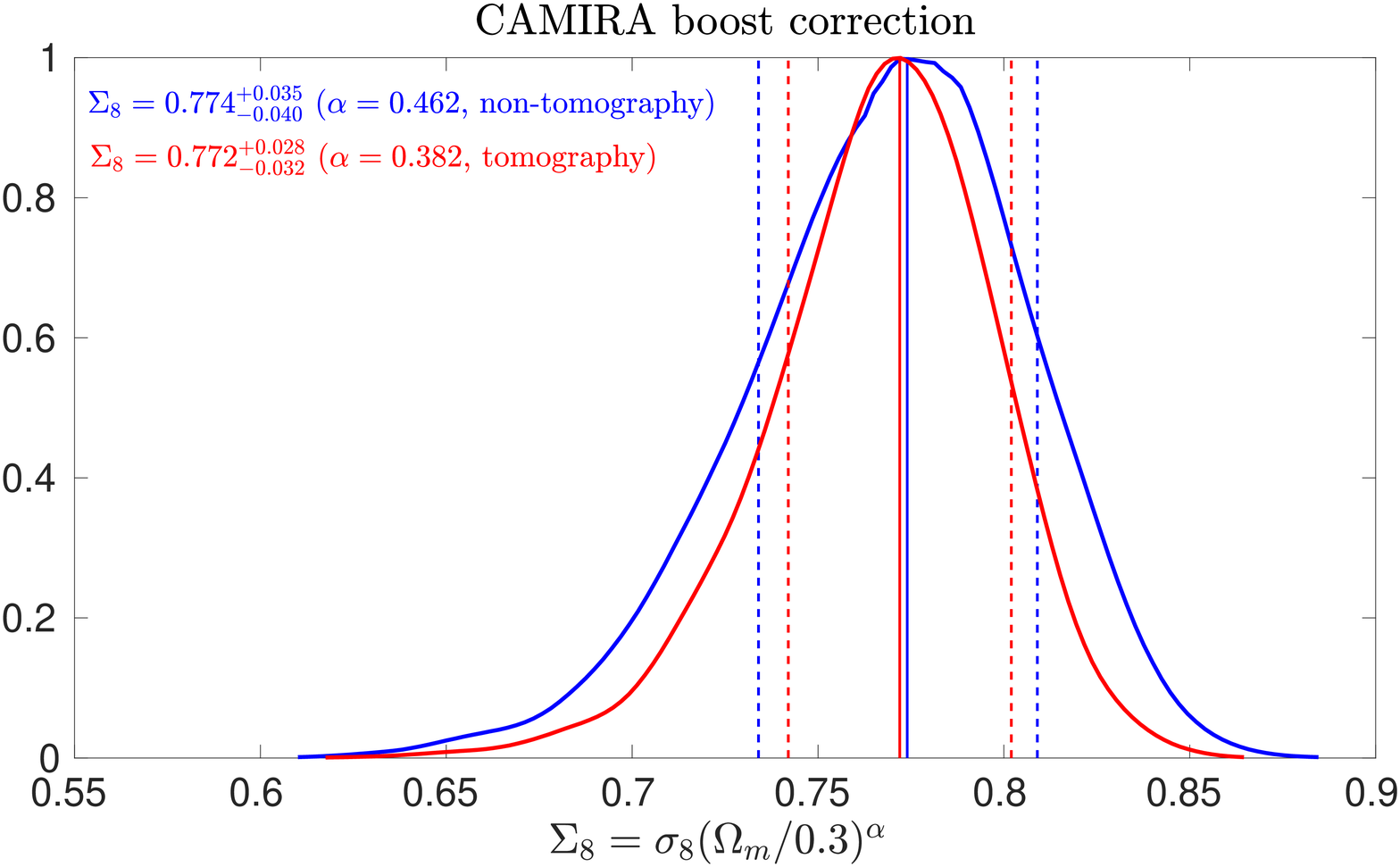}
\includegraphics[width=1.0\columnwidth, height=0.6\columnwidth]{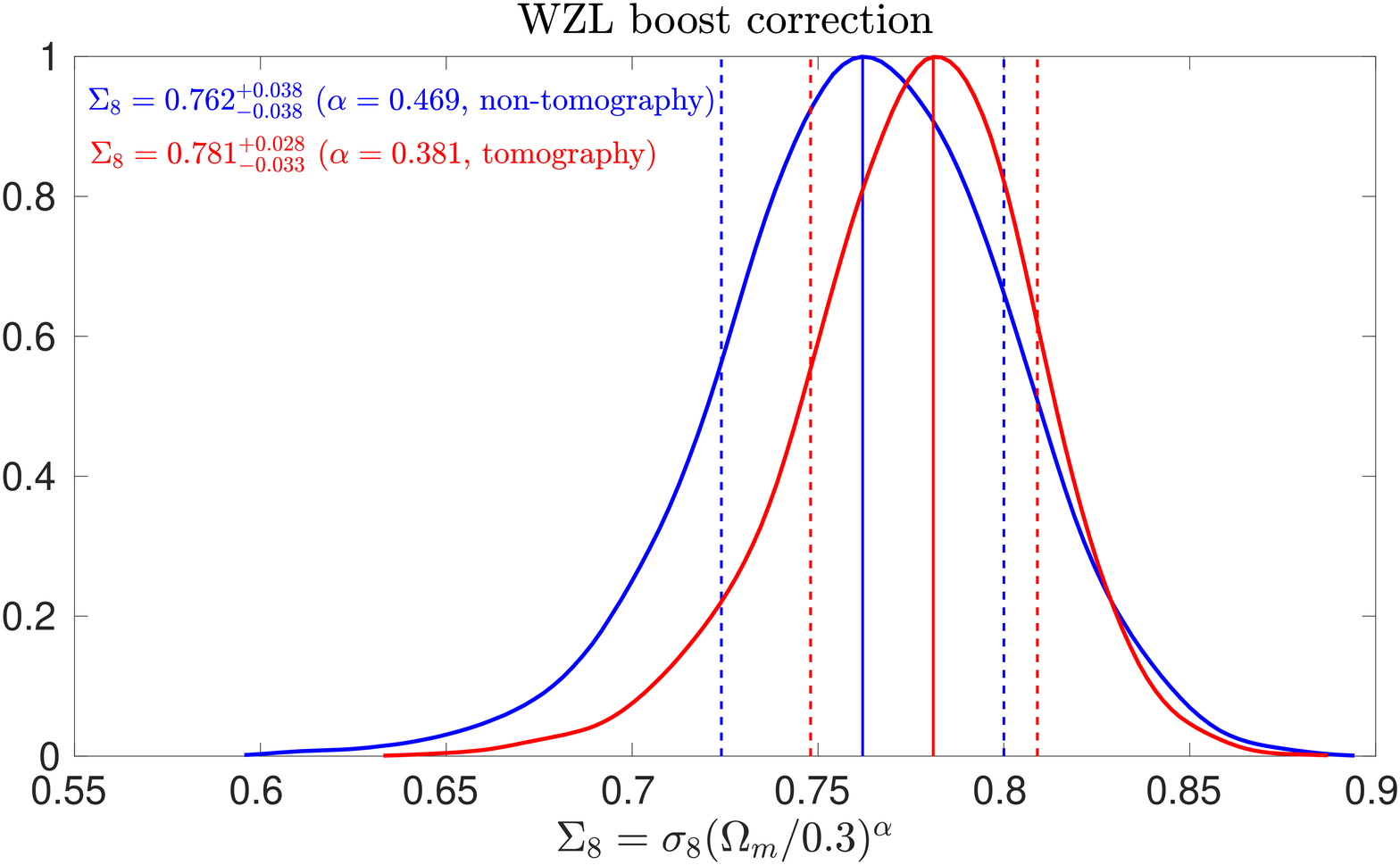}
\caption{The marginalized probability distribution of $\Sigma_8$ from CAMIRA (left) and WZL (right) boost correction. Blue and red lines are the results for non-tomographic and tomographic cases, with best-fit index parameter $\alpha\sim0.38$ and $0.46$, respectively.
}
\label{fig:C8}
\end{figure*}

\begin{table*}
\caption{Summary for $S_8$ and $\Sigma_8$ constraints of mock simulation analyses and real observations.}
\label{tab:summary}
\begin{center}
\leavevmode
\begin{tabular}{c c c c c } \hline
  &  & $S_8$ & $\Sigma_8$ & $\alpha$\\
\hline
\specialrule{0em}{0.5pt}{0.5pt}
\multirow{3}*{MOCK} & non-tomography & $0.778_{-0.035}^{+0.027}$ & $0.781_{-0.033}^{+0.032}$ & $0.457\pm 0.014$ \\
\specialrule{0em}{0.5pt}{0.5pt}
 & tomography & $0.791_{-0.036}^{+0.022}$ & $0.778_{-0.028}^{+0.025}$ & $0.433\pm 0.013$ \\
 \specialrule{0em}{0.5pt}{0.5pt}
  & 4-bin tomography & $0.792_{-0.042}^{+0.025}$ & $0.776_{-0.028}^{+0.028}$ & $0.401\pm 0.014$ \\
  \specialrule{0em}{0.5pt}{0.5pt}
 \hline
\specialrule{0em}{1pt}{1pt}
\multirow{4}*{OBS} & non-tomography (CAMIRA) & $0.768_{-0.043}^{+0.032}$ & $0.774_{-0.040}^{+0.035}$ & $0.462\pm 0.013$ \\
\specialrule{0em}{0.5pt}{0.5pt}
 & tomography (CAMIRA) & $0.758_{-0.076}^{+0.033}$ & $0.772_{-0.032}^{+0.028}$ & $0.382\pm 0.013$ \\
 \specialrule{0em}{0.5pt}{0.5pt}
 & non-tomography (WZL) & $0.770_{-0.040}^{+0.030}$ & $0.762_{-0.038}^{+0.038}$ & $0.469\pm 0.015$ \\
 \specialrule{0em}{0.5pt}{0.5pt}
 & tomography (WZL) & $0.768_{-0.057}^{+0.030}$ & $0.781_{-0.033}^{+0.028}$ & $0.381\pm 0.012$ \\
 \specialrule{0em}{0.5pt}{0.5pt}
\hline
\end{tabular}
\end{center}
\end{table*}

\begin{figure*}
\centering
\includegraphics[width=1.0\columnwidth]{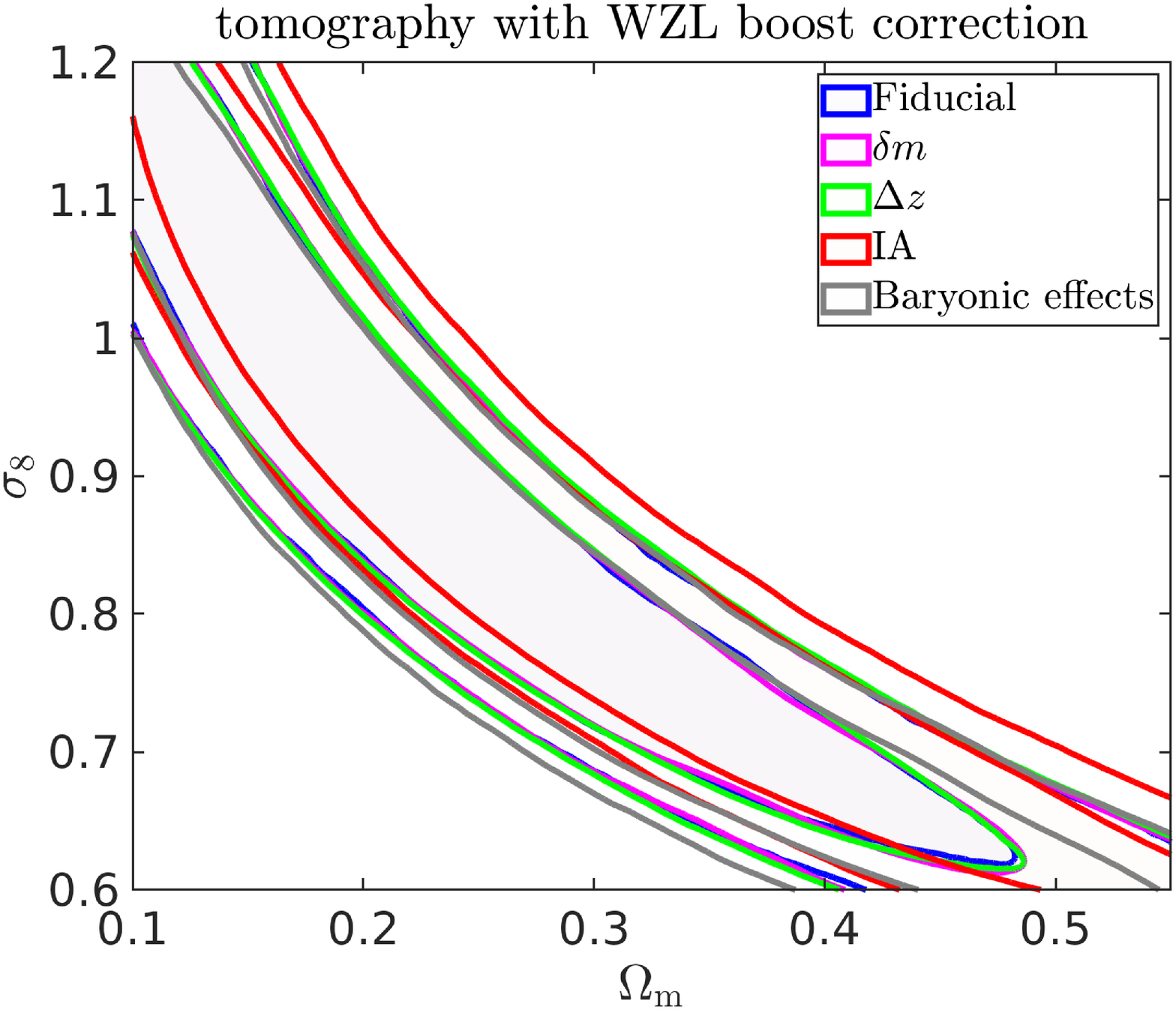}
\includegraphics[width=1.0\columnwidth, height=0.77\columnwidth]{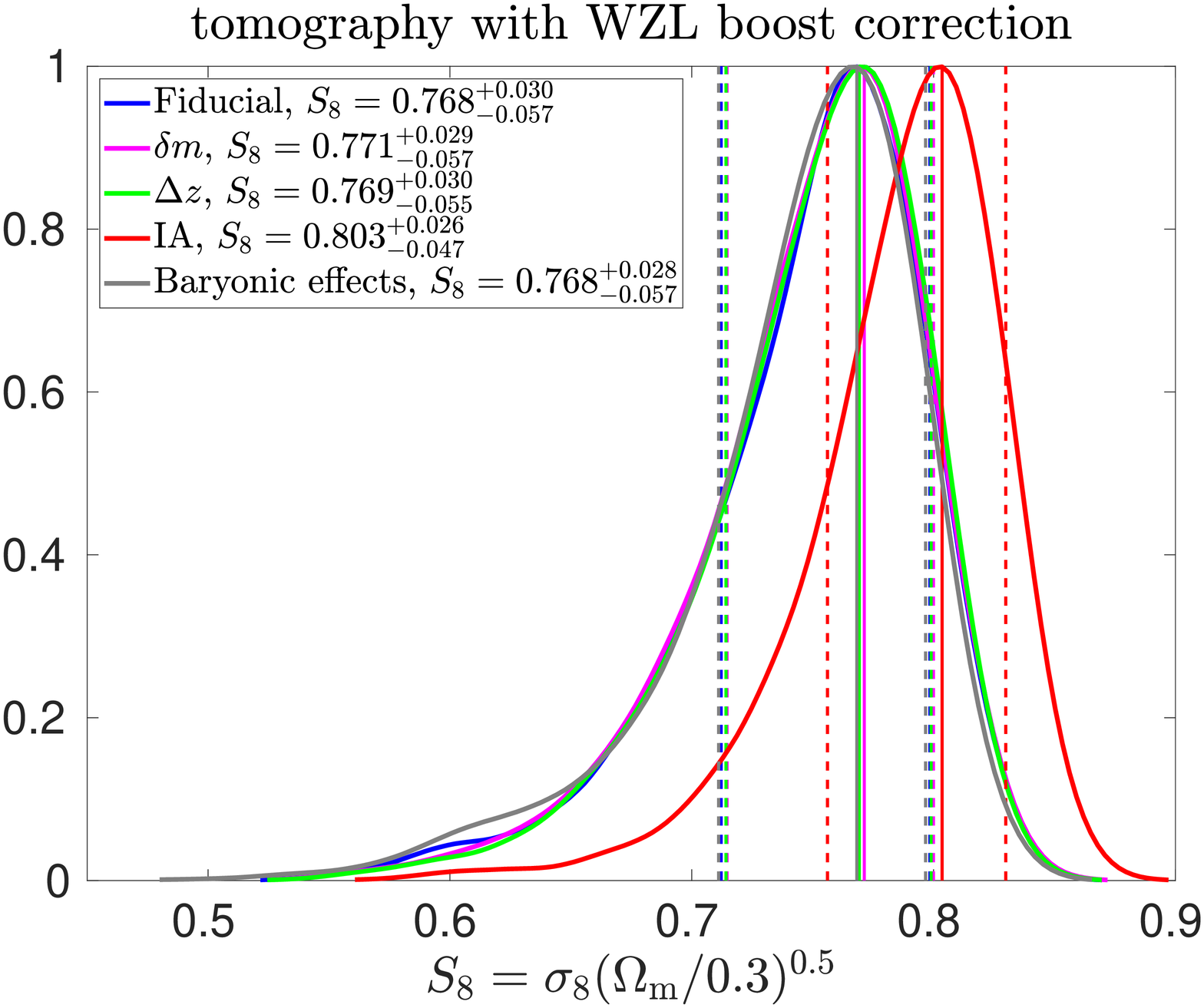}
\caption{The comparison of the derived cosmological constraints (Left) and the marginalized probability distribution of $S_8$ (Right) among different systematic cases.
}
\label{fig:systematics}
\end{figure*}

\begin{figure}
\centering
\includegraphics[width=1.0\columnwidth, height=0.9\columnwidth]{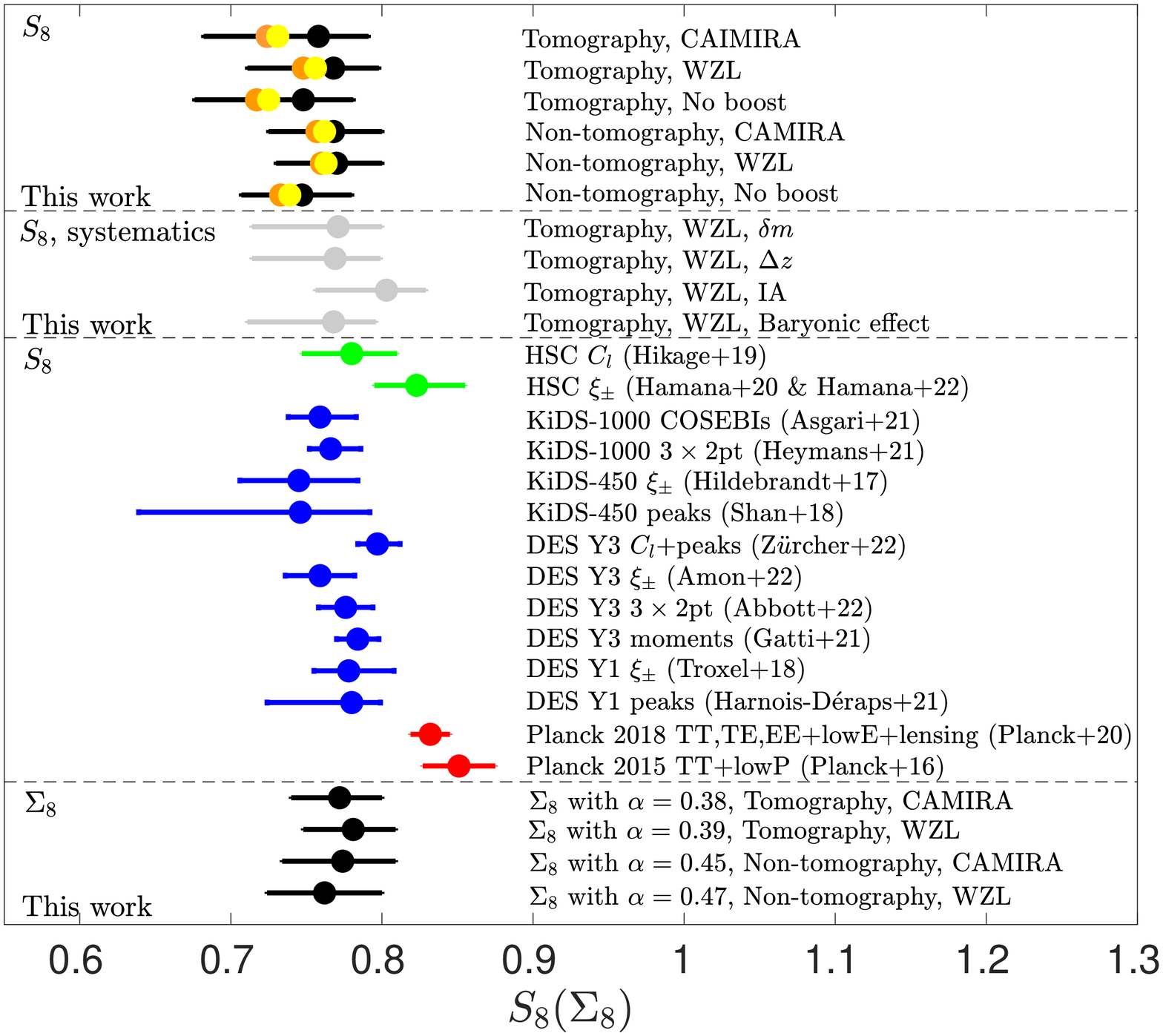}
\caption{Comparison of $S_8$ ($\Sigma_8$) from our WL peak analyses with the results from other studies. Black and gray data points are for the results in this work. For $S_8$, besides the best-fit value (black filled circles), we also show the mean (orange) and the median (yellow) values of our studies. Green points are for the constraints from HSC-SSP cosmic shear power spectra and two-point correlation functions. Blue points are the results from other WL surveys, including KiDS-450, KiDS-1000, DES Y1 and DES Y3, and two red points are for the Planck 2018 \citep{Planck2020} and 2015 \citep{Planck2016} results, respectively.}
\label{fig:S8compare}
\end{figure}

\section{Cosmological Constraints From HSC-SSP Tomographic Peak Analyses} \label{sec:constraints}
In this section, we present HSC-SSP cosmological WL tomographic peak analyses and the non-tomographic case as a comparison.

\subsection{Mock Validation} \label{subsec:mocktest}
We first generate mocks from ray-tracing simulations detailed in \citet{LiuX2015} to validate the analyses. Briefly, 
for the cosmological model with $(\Omega_\mathrm{m}, \Omega_\Lambda, \Omega_\mathrm{b}, h, \sigma_8, n_{\rm s})=(0.28, 0.72, 0.046, 0.7, 0.82, 0.96)$, we run a large set of N-body simulations with independent initial random seeds, and pad these simulation boxes to generate light cones to redshift $z=3$. For each light cone, from redshift $z=0$ to $z=1$, we fill with 8 simulation boxes each with a size of $320 h^{-1}$Mpc, and 4 larger simulations with a size of $600 h^{-1}$Mpc are used from $z=1$ to $z=3$. For both small and large sized simulations, 
we use $640^3$ particles and thus the mass resolution is $\sim 9.7\times 10^9 h^{-1}M_\odot$ and $\sim6.4\times 10^{10} h^{-1}M_\odot$, respectively. We then use 59 lens planes for ray-tracing calculations. In total we obtain 96 sets of convergence 
and shear maps with each set consisting of 59 lens-plane outputs. The individual map size is $3.5\times 3.5\deg^2$ sampled on $1024\times1024$ pixels, and thus the total simulated area is $1176\deg^2$.

Using these maps, we then produce HSC-SSP mocks following the steps in \citet{Oguri2018Mass}. Specifically,
(1) We randomly rotate galaxies ($\epsilon_\mathrm{ran}$) in the S16A catalog to remove their original lensing signals while keeping their positions and redshifts $z_{\rm p}$ fixed to the observed values to generate the mock galaxy sample. For each galaxy, a randomized intrinsic ellipticity is derived as
\begin{equation}
\epsilon_\mathrm{int}=\frac{\sigma_\mathrm{int}}{\sqrt{\sigma_\mathrm{int}^2+\sigma_\mathrm{sta}^2}}\epsilon_\mathrm{ran}
\end{equation}
where $\sigma_\mathrm{int}$ is an estimate of the intrinsic rms ellipticity, and $\sigma_\mathrm{sta}$ is the measurement error from the S16A catalog.
(2) These mock galaxies are painted onto our simulated maps, and their lensing signals are calculated by interpolating, in the dimensions of redshift and angular position, the grid values of the lensing maps. 
The lensed ellipticities are then constructed in accord with S16A as follows 
\begin{equation}
\epsilon_\mathrm{lens}=\frac{\epsilon_\mathrm{int}+2g+g^2\epsilon_\mathrm{int}^*}{1+|g|^2+2\mathrm{Re}[g\epsilon_\mathrm{int}^{*}]}.
\end{equation}
Adding the measurement errors, the ellipticity for a galaxy is given by 
\begin{equation}
\epsilon_\mathrm{mock}=\epsilon_\mathrm{lens}+(N_1+iN_2)
\end{equation}
where $N_i~(i=1,2)$ is a random value drawn from a normal distribution with a standard deviation of $\sigma_\mathrm{sta}$.
Finally, the mock `observed' shear catalog is made with the shear bias from S16A included \citep{Oguri2018Mass}.
(3) With the mock catalog, the same convergence reconstruction, peak identification and statistics are done as for the observational data. 
(4) For the mock galaxies of each of the 52 HSC-SSP fields, we perform 20 different paintings of (2) and thus create 20 mocks. 
(5) We randomly select one from the 20 for each field and compose one bootstrap set of HSC-SSP mock with 52 fields. A total of 1000 bootstrap sets are generated.  
(6) The mean peak counts and covariance matrices are estimated from these 1000 bootstrap mocks. 
It is noted that the mock galaxy distribution is the same as that of the observation. However, the observed galaxy distribution is not necessarily the same as the large-scale structures in our simulations. 
Thus the cluster-member boost effects do not show in our mocks.

In Figure \ref{fig:mockresults}, we show the peak counts for the non-tomographic (left) and 2-bin tomographic (right) cases in the upper panels. The error bars and the lines are the rms of the diagonal elements of the covariance matrix and our theoretical model predictions under the simulated cosmology. The corresponding cosmological constraints from our mock data are shown in the middle-right panel, with the blue and red contours are for the non-tomographic and 2-bin tomographic results, respectively. The red `+' symbol denotes the input cosmological parameters. We see that in both cases, the input cosmology is recovered excellently, demonstrating the robustness of our analyses. 

In addition, we also carry out analyses using 4 tomographic redshift bins with the redshift ranges of $0.2\le z_{\rm p}\le 0.51$, $0.51\le z_{\rm p}\le 0.75$, $0.75\le z_{\rm p}\le 1.02$ and $1.02\le z_{\rm p}\le 1.5$, respectively. Such a division leads to about the same number density 
of source galaxies in each bin. The peak results are shown in the middle-left panel of Figure \ref{fig:mockresults}. For comparison,  the cosmological constraints from the 4-bin tomographic case are shown as green contours in the middle-right panel. In the last row, we present the derived $S_8=\sigma_8(\Omega_{\rm m}/0.3)^{0.5}$ in the left.
We can see that the asymmetry of the $S_8$ distribution increases somewhat with the increase of the number of redshift bins, reflecting the somewhat different degeneracy directions of the constrained $(\Omega_{\rm m}, \sigma_8)$. 
To better characterise the constraining power, we thus also calculate $\Sigma_8=\sigma_8(\Omega_{\rm m}/0.3)^{\alpha}$ with $\alpha$ derived separately for each case. Following a similar procedure done in \citet{Kilbinger2013} and \citet{Fu2014}, we first construct the 2D nomalized posterior probability distribution for $(\Omega_\mathrm{m}, \sigma_8)$ from the corresponding MCMC chains with optimal bin numbers for estimating the posterior density \citep{Scott1979}. It corresponds to an optimal bin width with $h_N=3.49 sN^{-1/3}$, where $s$ is an estimate of the sample standard deviation, and $N$ denotes the sample size. We then perfrom non-linear least-squares fits by Levenberg-Marquardt method to obtain $\alpha$ and its $68\%$ confidence interval. The results are shown in the right panel of the last row. It is seen that 
$\alpha$ does change, and the values are $\sim 0.457\pm 0.014$, $0.433\pm 0.013$ and $0.401\pm 0.014$ for the non-tomographic, 2-bin and 4-bin cases, respectively. For the $\Sigma_8$ uncertainties, they are smaller by about a factor of 1.3 in the 2-bin case than that of the non-tomographic analyses.
No further improvements are seen from 2-bin to 4-bin studies because the increase of the shape noise in each bin dilutes the cosmological gain from the 4-bin case. We should note that here we do not include cross-peaks \citep{Martinet2021a}, which could bring extra cosmological information in the 4-bin case comparing to the 2-bin case. 

The specific $S_8$ and $\Sigma_8$ constraints from mock simulation analyses are presented in the upper part of 
Table \ref{tab:summary}. For the following observational analyses, we focus on the 2-bin case.   

\subsection{Constraints from Real Observations} \label{subsec:realresults}

We show the observational results in Figure \ref{fig:obsresults}. As discussed previously, for the real observational analyses, we need to take into account the boost effect from cluster member contaminations in the shear samples. This effect is not apparent in the simulation mocks because of the mismatch of massive halos in the simulations and those in the Universe.   

We use the boost information obtained in \S\ref{subsec:boost} from CAMIRA and WZL cluster catalogs, respectively. 
The revised theoretical model and the scaled covariance matrix as described in \S\ref{subsec:boost} and \S\ref{subsec:fitting} are used to derive cosmological constraints from observed peak counts. 
To illustrate the boost effect, in the upper panels of Figure \ref{fig:obsresults}, on top of the observational data, we show the theoretical results with and without boost corrections at the cosmological model with $(\Omega_{\rm m}, \sigma_8)=(0.332, 0.799)$ 
from the marginalized constraints of the HSC shear correlation analyses in \cite{Hamana2020, Hamana2022}. We can see that for the low-z peak counts the boost effect is significant, while for the high-z case it is minimal. This is consistent with that 
shown in Figure \ref{fig:massREC_example} where the association between peaks and clusters is stronger in the low-z case than that in the high-z case. The theoretical predictions with the boost correction from CAMIRA and WZL are in good agreements.
The observed peak counts are slightly lower than the model predictions with $(\Omega_{\rm m}, \sigma_8)=(0.332, 0.799)$ from \citet{Hamana2022}. 
The cosmological constraints from our peak analyses are shown in the bottom panels of Figure \ref{fig:obsresults}. 

In Figure \ref{fig:S8}, we show the constraints on $S_8=\sigma_8(\Omega_{\rm m}/0.3)^{0.5}$. It is noted that the power index $\alpha=0.5$ is a suitable one from cosmic shear two-point correlations (power spectra) studies. 
As discussed in the mock analyses, high peak statistics lead to somewhat different degeneracy directions of $(\Omega_{\rm m}, \sigma_8)$, and thus the asymmetric distributions of $S_8$ in Figure \ref{fig:S8}. 
The asymmetry is more apparent in the tomographic case than that of non-tomographic analyses. 
We then use $\Sigma_8=\sigma_8(\Omega_{\rm m}/0.3)^{\alpha}$ and fit $\Sigma_8$ and the index $\alpha$ simultaneously following the same fitting methods as described in \S\ref{subsec:mocktest}. 
We can see that for the tomographic case, $\alpha\sim 0.382\pm 0.013$ (CAMIRA) and $0.381\pm 0.012$ (WZL), which is more deviated from $\alpha=0.5$ than the non-tomographic cases with $\alpha\sim 0.462\pm 0.013$ and $0.469\pm0.015$. 
With the best-fit $\alpha$, the constraints on $\Sigma_8$ are plotted in Figure \ref{fig:C8} with the left and right panels for the results from CAMIRA and WZL boost corrections, respectively. 
For $\Sigma_8$, we have the constraints of $\Sigma_8=0.772^{+0.028}_{-0.032}$ and $\Sigma_8=0.774^{+0.035}_{-0.040}$ with CAMIRA boost correction from the tomographic and non-tomographic peak counts, respectively. 
The corresponding results with WZL boost correction are $\Sigma_8=0.781^{+0.028}_{-0.033}$ and $\Sigma_8=0.762^{+0.038}_{-0.038}$, well consistent with that from CAMIRA correction. For clarity, we also summarize these $S_8$ and $\Sigma_8$ constraints 
from real observations in the lower part of Table \ref{tab:summary}. The $1\sigma$ uncertainty of $\Sigma_8$ from 2-bin tomographic peak counts 
is about 1.3 times smaller than that of the non-tomographic peak anlayses.   

For the systematics discussed in \S\ref{sec:systematics}, we perform additional analyses to evaluate their possible impacts on cosmological constraints. Specifically, we consider the following cases.

1) Shear multiplicative bias correction $\delta m$. We add $\delta m$ in our theoretical model as an additional nuisance parameter to in the cosmological fitting, adopting a Gaussian prior with $(\mu,\bm{\sigma})=(0,\bm{0.01})$ where $\mu$ and $\sigma$ 
are the mean value and the standard deviation of the prior, respectively.

2) Photo-z bias. We introduce two nuisance parameters $\Delta z_\mathrm{L}$ and $\Delta z_\mathrm{H}$ to describe the overall redshift shifts of the low-z and high-z tomographic bins, each with a flat prior of $[-0.01, 0.01]$ and perform 4-parameters constraints.
We also test using the prior of $[-0.02, 0.02]$, and the results are about the same.

3) IA effects. As discussed in \S \ref{subsec:IA}, we adopt the simulation results with $\theta_{\rm IA}= 30^\circ$ and scale the model predicted peak counts with the corresponding ratios shown in Figure \ref{fig:RD_IA}.
We then re-perform cosmological fitting to the observed peak counts to illustrate the potential IA impacts. 

4) Baryonic effects. We perform 3-parameter fitting allowing the amplitude $A$ of the adopted halo M-c relation to vary in deriving cosmological constraints. A wide flat prior of $[0,20]$ on $A$ is applied. 

The test results including the systematics are shown in Figure \ref{fig:systematics} for the 2-bin tomographic case with WZL boost corrections. The left panel shows the derived $(\Omega_{\rm m}, \sigma_8)$ constraints and
the right panel is for the corresponding $S_8$, marginalized over the nuisance parameters. It is seen that adding $\delta m$, $\Delta z_\mathrm{L}$ and $\Delta z_\mathrm{H}$, and $A$ in the fitting results in no significant changes of the cosmologcial constraints. 
For the IA effects with the test case of $\theta_{\rm IA}= 30^{\circ}$ for satellites, $S_8$ shifts to a higher value but is still within the statistical uncertainty. Thus we do not expect significant IA bias in our analyses. 
For future studies with larger areas and thus the reduced statistical errors, IA effects need to be taken into account carefully.

Figure \ref{fig:S8compare} presents the comparison of $S_8$ ($\Sigma_8$) from our peak analyses with that from other studies. For $S_8$, besides the best-fit value (black filled circles), we also show the mean (orange) and the median (yellow) values of our studies. 
It is seen that our results are in good accordance with other WL studies \citep{Asgari2021, Heymans2021, Hildebrandt2017, Shan2018, Zrcher2022, Amon2022, Abbott2022, Gatti2021, Troxel2018, HarnoisDeraps2021}. Within the HSC-SSP analyses, our constraints are in excellent agreements with that from cosmic shear power spectra measurements in \citet{Hikage2019}. They are
about $1.3-1.8\sigma$ smaller than the update results from 2-point correlation studies of \citet{Hamana2022}. Comparing to that from Planck 2018 \citep{Planck2020}, our $S_8$ values are about $2.0\sigma$ smaller, which is calculated by subtracting our best-fit $S_8$ value from the Planck best-fit and then dividing by the square root of the quadratic sum of our $S_8$ error on its higher side and that from Planck.
We also note that without accounting for the boost correction in our theoretical modelling, $S_8$ is biased to a lower value. For the current analyses, this bias is already close to $1\sigma$.

In \citet{Shan2018}, similar high peak analyses are done using KiDS-450 data without redshift tomography. There the effective number density of galaxies is $n_{\rm g}\sim 7\hbox{ arcmin}^{-2}$ and the effective area used in peak statistics is $\sim 300 \deg^2$. In comparison,
we have $n_{\rm g}\sim 18.5\hbox{ arcmin}^{-2}$ and the effective area of $\sim 58 \deg^2$. Our constraints on $S_8$ and $\Sigma_8$ are about a factor of $1.6$ tighter although the total number of galaxies used here 
is about half of that in \citet{Shan2018}. It is noted that for WL analyses, the redshift distribution of the shear sample plays an important role. In \citet{Shan2018}, the photo-z range is limited to $0.1<z_{\rm p}\le 0.9$, while in our studies $0.2\le z_{\rm p}\le 1.5$.
The higher redshift part enhances the cosmological information gain in our analyses. Additionally, the projection effects from large-scale structures are more significant for HSC-SSP than that for KiDS-450 also because of the higher redshift range in the former. The projection effects 
are cosmology dependent and thus further boost the constraining power of our peak analyses here.  

In comparison with the HSC-SSP results of \citet{Hikage2019} and \citet{Hamana2020, Hamana2022}, we use a smaller effective area because of the necessary mask and boundary exclusions. Our $S_8$ constraints have larger uncertainties than theirs. For $\Sigma_8$, 
our uncertainties are comparable to their $S_8$. We note that in their analyses, they include IA model parameters in the fitting, while we do not here. From Figure \ref{fig:systematics}, we see that IA can shift the constraints from peak statistics. We thus 
expect that adding IA parameters as nuisance parameters properly in the fitting would broaden the constraining range in comparison to our results here. As one of the future efforts, we will improve our peak model to
include IA effects. Therefore their impacts on both the cosmological parameter bias and the uncertainty estimates can be better understood. 

\section{Conclusions} \label{sec:summary}
Using HSC-SSP S16A shear data, we carry out tomographic peak analyses, focusing on high peaks with a physical origin primarily due to individual massive halos along lines of sight. To derive cosmological constraints, we use our theoretical peak model including the effects of random shape noise and the projection of large-scale structures \citep{Yuan2018},
which is well validated by our simulation mocks. With the cluster (candidates) catalogs of CAMIRA and WZL, we carefully exam the dilution resulting from the boost factor of cluster member galaxies in the shear samples, and include the information in our model calculations. We also evaluate the satellite IA impacts by employing simulations with semi-analytical galaxy formation. Considering a test case with the alignment angle of satellite galaxies being $\theta_{\rm IA}=30^{\circ}$ with respect to the radial directions toward their central galaxies, the derived $S_8$ is higher than that without including the IA impacts, but the difference is within $1\sigma$ for our analyses here. The possible systematics from the shear bias and photo-z errors and the baryonic effects are discussed, and they do not affect our studies significantly given the effective area of $\sim 58\deg^2$.   

Our results find $S_8=0.758^{+0.033}_{-0.076}$ and $0.768^{+0.030}_{-0.057}$ from the 2-bin tomographic peak counts using the CAMIRA and WZL boost corrections, respectively. They are consistent with the cosmic shear power spectra measurements using the same HSC-SSP S16A data, as well as using other WL surveys. A deviation of $\sim 1.5\sigma$ and $\sim 2.0\sigma$ from the updated HSC-SSP 2-point correlation and the Planck result is noted, respectively. 

We fit the degeneracy direction of ($\Omega_{\rm m}, \sigma_8$) from our peak analyses, and obtain, respectively, $\alpha\approx 0.38$ and $0.46$ for the tomographic and non-tomographic cases. Comparing to $\alpha\approx 0.5$ from cosmic shear correlations, the constraints from high peak analyses exhibit a different degeneracy direction, which is particularly apparent for the case of tomographic peak counts. The different degeneracies reveal important complementarity between peak statistics and 2PCF. Tighter constraints can be obtained from their joint analyses. With the fitted $\alpha$, the tomographic peak results are $\Sigma_8=0.772^{+0.028}_{-0.032}$ (CAMIRA) and $\Sigma_8=0.781^{+0.028}_{-0.033}$ (WZL) with the $1\sigma$ error about a factor of 1.3 smaller than that of the non-tomographic case.  

For the first-year HSC-SSP shear catalog, the sky area is about $100\deg^2$ and it is further 
reduced to about $58\deg^2$ in our peak analyses after mask and boundary exclusions. In this case, we test and show 
that the 4-bin tomography does not significantly increase the cosmological information gain comparing to the 2-bin analyses. 
For the three-year HSC-SSP shear catalog \citep{Li2021}, however, the area is increased by $\sim 4$ times.
Future surveys can reach the sky coverage of $\sim 15000\deg^2$ with the similar or deeper depth than HSC-SSP. Thus the number 
of high peaks will expectedly increase by orders of magnitude, and tomographic peak analyses with more redshift bins
and also including cross peaks are achievable.

In this paper, we focus on applying tomographic high peak analyses to observational data.
We show the different $(\Omega_{\rm m}, \sigma_8)$ degeneracy direction compared with that from the cosmic shear correlations. 
In our future studies, we will further explore the joint constraints by combining the two statistics.  

For high precision peak studies, systematic controls are challenging. Within the framework of our theoretical high peak model, it is potentially possible to include the astrophysics-related systematics as additional ingredients in the modelling, 
such as the IA and the baryonic effects. This can allow us to extract those astrophysical information 
together with the cosmological ones using high peak statistics from future WL surveys.

\section*{Acknowledgements}
The Hyper Suprime-Cam (HSC) collaboration includes the astronomical communities of Japan and Taiwan, and Princeton University. The HSC instrumentation and software were developed by the National Astronomical Observatory of Japan (NAOJ), the Kavli Institute for the Physics and Mathematics of the Universe (Kavli IPMU), the University of Tokyo, the High Energy Accelerator Research Organization (KEK), the Academia Sinica Institute for Astronomy and Astrophysics in Taiwan (ASIAA), and Princeton University. Funding was contributed by the FIRST program from Japanese Cabinet Office, the Ministry of Education, Culture, Sports, Science and Technology (MEXT), the Japan Society for the Promotion of Science (JSPS), Japan Science and Technology Agency (JST), the Toray Science Foundation, NAOJ, Kavli IPMU, KEK, ASIAA, and Princeton University. 

This paper makes use of software developed for the Large Synoptic Survey Telescope. We thank the LSST Project for making their code available as free software at  http://dm.lsst.org.

The Pan-STARRS1 Surveys (PS1) have been made possible through contributions of the Institute for Astronomy, the University of Hawaii, the Pan-STARRS Project Office, the Max-Planck Society and its participating institutes, the Max Planck Institute for Astronomy, Heidelberg and the Max Planck Institute for Extraterrestrial Physics, Garching, The Johns Hopkins University, Durham University, the University of Edinburgh, Queen’s University Belfast, the Harvard-Smithsonian Center for Astrophysics, the Las Cumbres Observatory Global Telescope Network Incorporated, the National Central University of Taiwan, the Space Telescope Science Institute, the National Aeronautics and Space Administration under Grant No. NNX08AR22G issued through the Planetary Science Division of the NASA Science Mission Directorate, the National Science Foundation under Grant No. AST-1238877, the University of Maryland, and Eotvos Lorand University (ELTE) and the Los Alamos National Laboratory.

The analyses of this paper are based on data collected at the Subaru Telescope and retrieved from the HSC data archive system, which is operated by Subaru Telescope and Astronomy Data Center at National Astronomical Observatory of Japan.

The calculations of this study are partly done on the Yunnan University Astronomy Supercomputer. Z.H.F. and X.K.L. are supported by NSFC of China under Grant No. 11933002 and No. U1931210, and a grant from CAS Interdisciplinary Innovation Team. X.K.L. also acknowledges the supports from NSFC of China under Grant No. 11803028 and No. 12173033, YNU Grant No. C176220100008. X.K.L. and S.Y. acknowledge the supports by the research grants from the China Manned Space Project with No. CMS-CSST-2021-B01. Z.H.F. also acknowledges the supports from NSFC of China under Grant No. 11653001, and the research grants from the China Manned Space Project with No. CMS-CSST-2021-A01. We also acknowledge the support from ISSI/ISSI-BJ International Team Program - Weak gravitational lensing studies from space missions. 

\section*{Data Availability}

The data used in this article will be shared on reasonable request to the authors.



\bibliographystyle{mnras}
\bibliography{ms} 




\appendix

\section{The Theoretical Peak Model} \label{sec:appendix}


In this appendix, we summarise the high peak model that is applied in our analyses to derive cosmological constraints. 

We divide the reconstructed convergence $K$ into three parts of contributions from massive halos $K_\mathrm{H}$, projection effects of large-scale structures $K_{\rm LSS}$, and the shape noise $N$, i.e.,
\begin{equation}
K=K_{\mathrm{H}}+K_{\mathrm{LSS}}+N.\label{kall}
\end{equation}
Here the term $K_{\mathrm{H}}$ is from individual massive halos with $M_{\rm vir}\ge M_*$ that dominate the lensing signals of high peaks. We take $M_*=10^{14}h^{-1}M_\odot$ in our analyses \citep{Yuan2018, Wei2018a}. 

For $K_\mathrm{LSS}$, it is cosmology dependent, and includes the contribution from large-scale structures excluding the massive halo part, which is already explicitly split out as $K_\mathrm{H}$. 
For high peaks, typically, $|K_\mathrm{LSS}|\ll |K_\mathrm{H}|$, and $K_\mathrm{LSS}$ shows up as 
small additive contributions \citep{Yuan2018}. Comparing with the overall convergence field, $K_\mathrm{LSS}$ 
should be more Gaussian because of the exclusion of the heavily non-Gaussian massive halos. Therefore, in our model, 
we approximate $K_{\rm LSS}$ as a Gaussian random field. We test its validity for high peak modelling extensively using 
simulations \citep{Yuan2018}. For the studies here, we also perform tests with simulated mock data.

The moments of $K_\mathrm{LSS}$ can be calculated by 

\begin{equation}
\sigma_{\mathrm{LSS},i}^{2}=\int_{0}^{\infty}\frac{\ell\mathrm{d}\ell}{2\pi}\ell^{2i}C_{\ell}^{\mathrm{LSS}};(i=0,1,2).
\end{equation}
We calculate the power spectrum $C_{\ell}^{\mathrm{LSS}}$ by
\begin{equation}
\ensuremath{C_{\ell}^{\mathrm{LSS}}=\frac{9H_{0}^{4}\Omega_{\rm m}^{2}}{4c^4}\int_{0}^{\chi_{H}}\mathrm{d}\chi^{\prime}\frac{g^{2}\left(\chi^{\prime}\right)}{a^{2}\left(\chi^{\prime}\right)}P_{\delta}^{\mathrm{LSS}}\left(\frac{\ell}{f_{K}\left(\chi^{\prime}\right)},\chi^{\prime}\right)}\label{eq:WLsignal}
\end{equation}
with the lensing efficiency factor ${g\left(\chi^{\prime}\right)}$,
\begin{equation}
{g\left(\chi^{\prime}\right)=\int_{\chi^{\prime}}^{\chi_{H}}\mathrm{d}\chi p_{\mathrm{s}}(\chi)\frac{f_{K}\left(\chi-\chi^{\prime}\right)}{f_{K}(\chi)}}.\label{eq:WLeff}
\end{equation}
For the 3D power spectrum $P_{\delta}^{\mathrm{LSS}}$, it is computed by subtracting the one halo term of massive halos with mass $M\ge M_{*}$ from the full nonlinear power spectrum \citep{Yuan2018},
\begin{equation}
P_{\delta}^{\mathrm{LSS}}[k,\chi(z)]=P_{\delta}[k,\chi(z)]-\left.P_{\delta}^{1\mathrm{H}}\right|_{M\geqslant M_{*}}[k,\chi(z)],
\end{equation}
and \citep{CS2002}
\begin{equation}
P_{\delta}^{\mathrm{1H}} \Bigm| _{M\geqslant M_{*}}[k,\chi(z)]=\frac{4\pi}{\bar{\rho}^{2}}\int_{M_{*}}^{\infty}\mathrm{d}M\,M^{2}W^{2}(k,M)n(M,z),\label{eq:p-1h}
\end{equation}
\begin{equation}
W(k,M)=\frac{1}{M}\int_{0}^{R_{\mathrm{vir}}}\mathrm{d}r\,\frac{\sin(kr)}{kr}4\pi r^{2}\rho(r,M).
\end{equation}
Here $\bar{\rho}$ is the mean matter density of the universe, $n(M,z)$ is the halo mass function, $\rho(r,M)$ is the spherically symmetric halo density profile, and $R_\mathrm{vir}$ is the halo virial radius. 
In accord with the peak count calculations shown in below, we adopt the \citet{Watson2013} halo mass function and the NFW halo density profile with the mass-concentration relation given in \citet{Duffy2008} in  
computing $\left.P_{\delta}^{1\mathrm{H}}\right|_{M\geqslant M_{*}}$. For the full nonlinear power spectrum $P_{\delta}[k,\chi(z)]$, it is calculated using CAMB \citep{Lewis2000} based on \citet{Takahashi2012}.   

For the smoothed shape noise field $N$, because of the central limit theorem, it can be well aproximated as a Gaussian random field with the moments 
\begin{equation}
\sigma_{\mathrm{N},i}^{2}=\int_{0}^{\infty}\frac{\ell\mathrm{d}\ell}{2\pi}\ell^{2i}C_{\ell}^{\mathrm{N}};(i=0,1,2).\label{eq:3moments}
\end{equation}
where $C_{\ell}^{\mathrm{N}}$ is the power spectrum of the smoothed noise field. These moments can be computed directly from the noise maps described in \S\ref{subsec: MapPeak}. 
We have $\sigma_{\mathrm{N},0}\approx 0.023$, $0.029$, $0.018$, $\sigma_{\mathrm{N},1}\approx 0.021$, $0.027$, $0.017$, and $\sigma_{\mathrm{N},2}\approx 0.028$, $0.036$, $0.022$ for the low-z, high-z and non-tomographic cases, respectively.

With the above information, we can calculate the total number density of convergence peaks by \citep{Fan2010}
\begin{equation}
n_{\mathrm{peak}}(\nu)\mathrm{d}\nu=n_{\mathrm{peak}}^{\mathrm{c}}(\nu)\mathrm{d}\nu+n_{\mathrm{peak}}^{\mathrm{n}}(\nu)\mathrm{d}\nu,\label{npksum}
\end{equation}
where $\nu=K/\sigma_{0}$ with $\sigma_{0}^{2}=\sigma_\mathrm{LSS,0}^{2}+\sigma_\mathrm{N,0}^{2}$ being the total variance of the sum of the two Gaussian fields $K_\mathrm{LSS}+N$, which is also Gaussian.
The term $n_{\mathrm{peak}}^{\mathrm{c}}(\nu)$ is the number density of peaks per unit $\nu$ at $\nu$ within virial radii of massive halos of $M_\mathrm{vir}\ge M_*$. 
The second term in the right $n_{\mathrm{peak}}^{\mathrm{n}}(\nu)$ is the number density of peaks in the field regions outside the massive halos. 

For $n_{\mathrm{peak}}^{\mathrm{c}}(\nu)$, we have
\begin{multline}
n_{\mathrm{peak}}^{\mathrm{c}}(\nu)=\int\mathrm{d}z\:\dfrac{\mathrm{d}V(z)}{\mathrm{d}z\mathrm{d\Omega}}\int_{M_{*}}^{\infty}\mathrm{d}M\:n(M,z) \\ \times \int_{0}^{\theta_{\mathrm{vir}}}\mathrm{d}\theta(2\pi\theta)\hat{n}^{\mathrm{c}}_{\mathrm{peak}}(\nu,M,z,\theta),\label{eq:npk_c}
\end{multline}
where $\mathrm{d}V(z)$ and $\mathrm{d}\Omega$ are, respectively, the cosmic volume element at redshift $z$ and the solid angle element. The angular virial radius of a halo is $\theta_{\mathrm{vir}}=R_{\mathrm{vir}}(M,z)/D_{\mathrm{A}}(z)$ 
with $D_{\mathrm{A}}$ being the angular diameter distance to the halo. In the integrand, $\hat{n}_{\mathrm{peak}}^{\mathrm{c}}(\nu,M,z,\theta)$ is the number density of peaks at $\theta$, 
canculated from the theory of Gaussian random field modulated by the halo profile $K_{\rm H}$ \citep{BBKS1986, Bond1987}. It is given by,
\begin{eqnarray}
 &  & \hat{n}_{\mathrm{peak}}^{\mathrm{c}}(\nu,M,z,\theta)=\exp\bigg[-\frac{(K_\mathrm{H}^{1})^{2}+(K_\mathrm{H}^{2})^{2}}{\sigma_{1}^{2}}\bigg]\nonumber \\
 &  & \times\bigg[\frac{1}{2\pi\theta_{*}^{2}}\frac{1}{(2\pi)^{1/2}}\bigg]\exp\bigg[-\frac{1}{2}\bigg(\nu-\frac{K_\mathrm{H}}{\sigma_{0}}\bigg)^{2}\bigg]\nonumber \\
 &  & \times\int_{0}^{\infty}dx\bigg\{\frac{1}{[2\pi(1-\gamma^{2})]^{1/2}}\nonumber \\
 &  & \times\exp\bigg[-\frac{[{x+(K_\mathrm{H}^{11}+K_\mathrm{H}^{22})/\sigma_{2}-\gamma(\nu-K_\mathrm{H}/\sigma_{0})}]^{2}}{2(1-\gamma^{2})}\bigg]\nonumber \\
 &  & \times F(x)\bigg\}\label{nchat}
\end{eqnarray}
and 
\begin{eqnarray}
 &  & F(x)=\exp\bigg[-\frac{(K_\mathrm{H}^{11}-K_\mathrm{H}^{22})^{2}}{\sigma_{2}^{2}}\bigg]\times\nonumber \\
 &  & \int_{0}^{1/2} \mathrm{d}e ~\hbox{}8(x^{2}e)x^{2}(1-4e^{2})\exp(-4x^{2}e^{2})\times\nonumber \\
 &  & \int_{0}^{\pi}\frac{d\psi}{\pi}\hbox{}\exp\bigg[-4xe\cos(2\psi)\frac{(K_\mathrm{H}^{11}-K_\mathrm{H}^{22})}{\sigma_{2}}\bigg],\nonumber \\
\label{fx}
\end{eqnarray}
where $\theta_{*}^{2}=2\sigma_{1}^{2}/\sigma_{2}^{2}$, $\gamma=\sigma_{1}^{2}/(\sigma_{0}\sigma_{2})$ and $\sigma_{i}^{2}=\sigma_{\mathrm{LSS},i}^{2}+\sigma_{\mathrm{N},i}^{2}$ ($i=0,1,2$). 

For the smoothed halo quantities, we have
\begin{equation}
K_\mathrm{H}(\bm{\theta})=\int \mathrm{d}\bm{\theta'}W_{\theta_\mathrm{G}}(\bm{\theta'}-\bm{\theta})\kappa(\bm{\theta'}),
\label{kH}
\end{equation}
\begin{equation}
K_\mathrm{H}^{i}(\bm{\theta})=\partial_{i}K_\mathrm{H}(\bm{\theta})=\int \mathrm{d}\bm{\theta'}\partial_iW_{\theta_\mathrm{G}}(\bm{\theta'}-\bm{\theta})\kappa(\bm{\theta'}),
\label{kHi}
\end{equation}
and
\begin{equation}
K_\mathrm{H}^{ij}(\bm{\theta})=\partial_{ij}K_\mathrm{H}(\bm{\theta})=\int \mathrm{d}\bm{\theta'}\partial_{ij}W_{\theta_\mathrm{G}}(\bm{\theta'}-\bm{\theta})\kappa(\bm{\theta'}),
\label{kHij}
\end{equation}
where $\kappa(\bm{\theta})$ is modeled using the NFW halo density profile with the mass-concentration relation given in \citet{Duffy2008}.

For $n_{\mathrm{peak}}^{\mathrm{n}}(\nu)$, we have 
\begin{equation}
 \begin{split}
n_{\mathrm{peak}}^{\rm n}(\nu)=\frac{1}{{\rm d}\Omega}\Big\{n_{\mathrm{ran}}(\nu)\Big[{\rm d}\Omega-\int {\rm d}z\frac{{\rm d}V(z)}{{\rm d}z}\\
   \times\int_{M_{*}}^{\infty} {\rm d}M\,n(M,z)\,(\pi \theta_{\rm vir}^{2})\Big]\Big\},
\end{split}
\label{eq:npk_n}
\end{equation}
where $n_{\mathrm{ran}}(\nu)$ is the number density of peaks from the random field $K_\mathrm{LSS}+N$ without foreground massive halos. It can be calculated from Eq.(\ref{nchat}) by setting all the halo quantities to be zero.


\bsp	
\label{lastpage}
\end{document}